\documentclass[manuscript,screen]{acmart}

\usepackage{enumitem}
\usepackage{booktabs}
\usepackage{colortbl}
\usepackage{multirow}
\usepackage{bigstrut}
\usepackage{xspace}
\usepackage{hyperref}
\usepackage{tabularx}
\usepackage{soul}
\usepackage{xcolor}
\usepackage{graphicx}
\usepackage{threeparttable}
\usepackage{siunitx}
\usepackage{longtable}

\sethlcolor{gray!12!white}
\newcommand{\yellowhl}[1]{\texttt{#1}}

\newcommand{\up}{\includegraphics[scale=0.01]{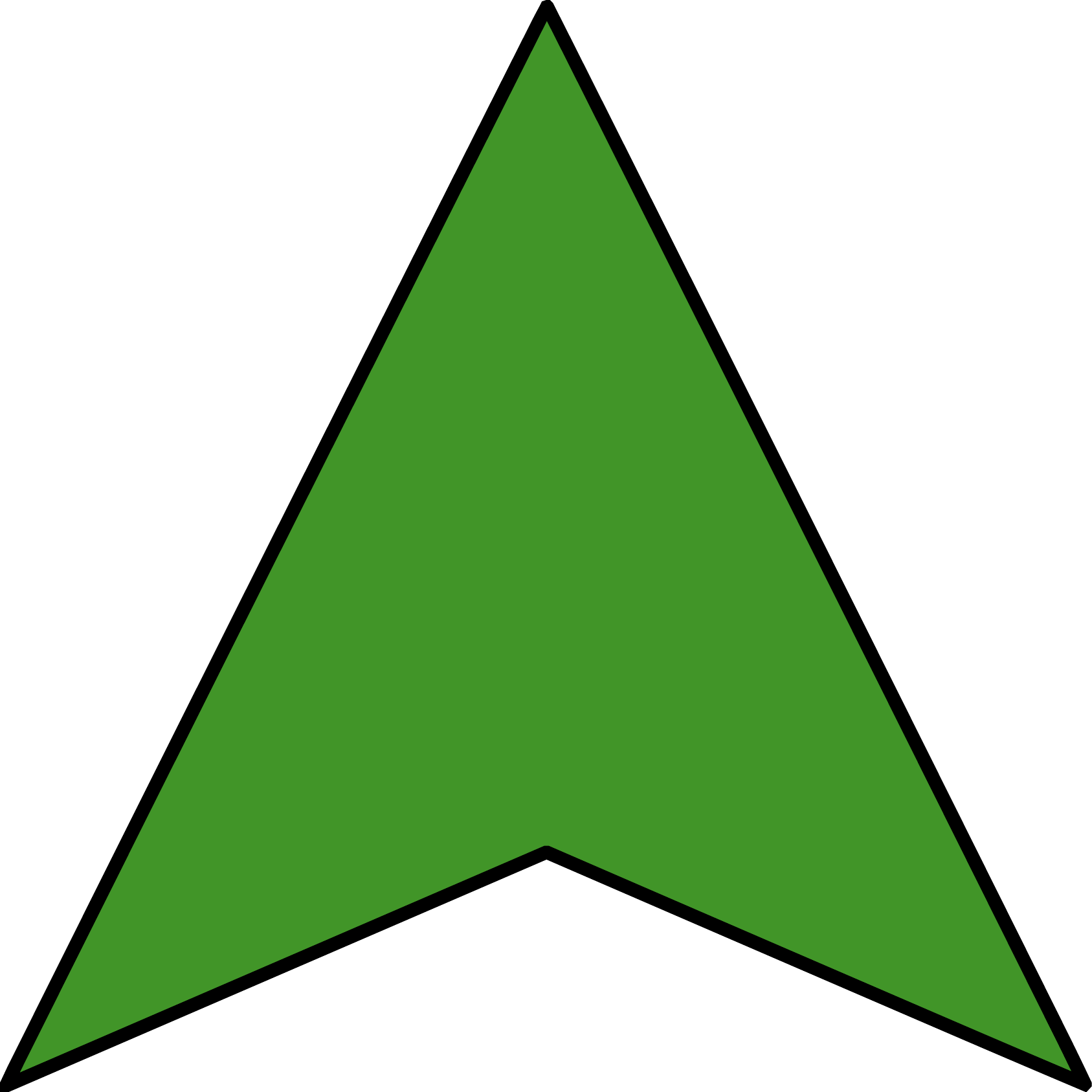}\xspace}
\newcommand{\down}{\includegraphics[scale=0.01]{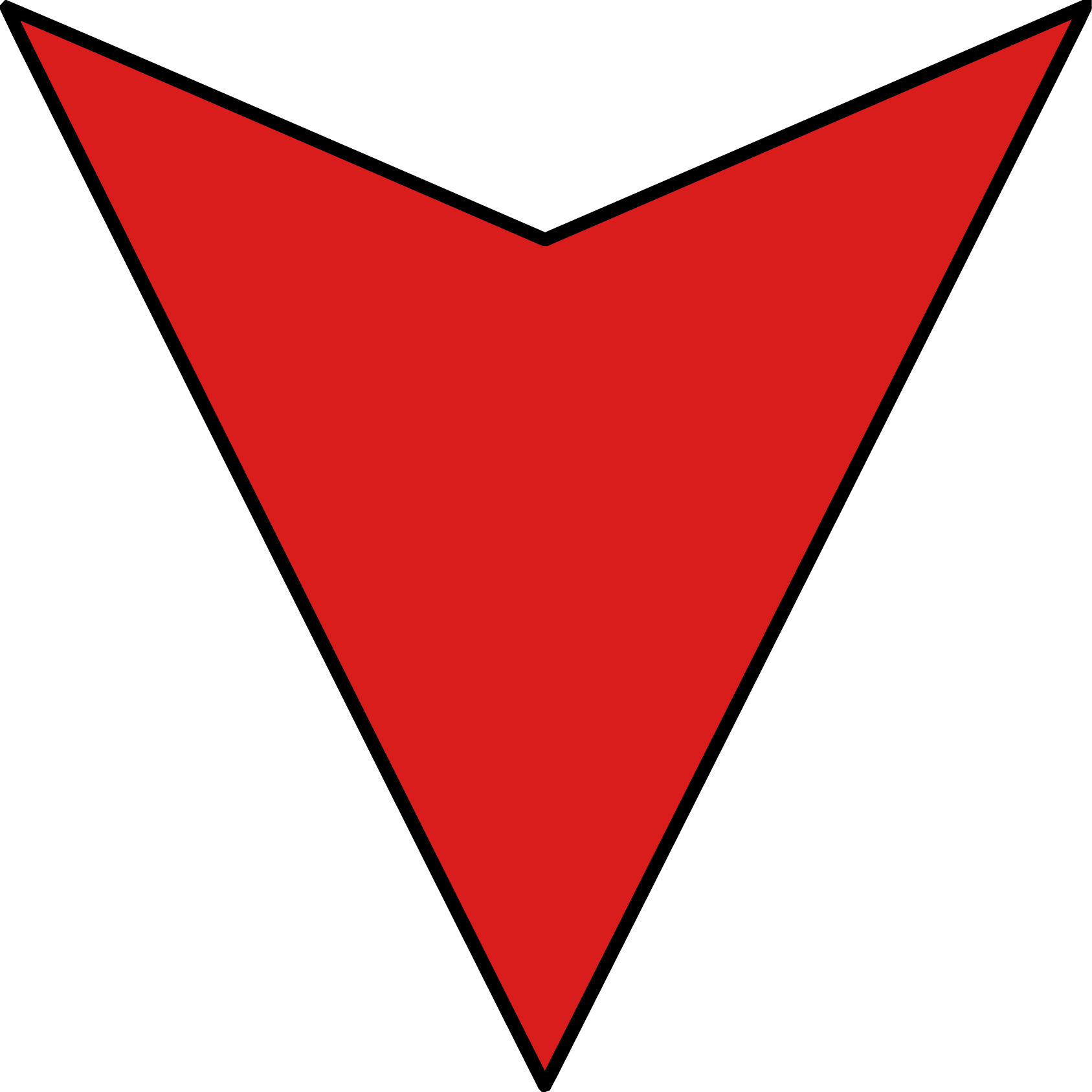}\xspace}

\newcommand{\rev}[1]{{\color{black} #1}} 
\newcommand{\rever}[1]{{\color{black} #1}} 

\AtBeginDocument{%
  \providecommand\BibTeX{{%
    \normalfont B\kern-0.5em{\scshape i\kern-0.25em b}\kern-0.8em\TeX}}}

\setcopyright{acmlicensed}
\copyrightyear{2018}
\acmYear{2018}
\acmDOI{XXXXXXX.XXXXXXX}

\acmConference[Conference acronym 'XX]{Make sure to enter the correct
  conference title from your rights confirmation emai}{June 03--05,
  2018}{Woodstock, NY}
\acmISBN{978-1-4503-XXXX-X/18/06}

\begin{document}

\title{Perceiving and Countering Hate: The Role of Identity in Online Responses}

\author{Kaike Ping}
\email{pkk@vt.edu}
\affiliation{%
  \institution{Virginia Tech}
  \streetaddress{220 Gilbert St}
  \city{Blacksburg}
  \state{VA}
  \country{USA}
  \postcode{24060}
}
\orcid{0000-0002-7520-7503}

\author{James Hawdon}
\email{hawdonj@vt.edu}
\affiliation{%
  \institution{Virginia Tech}
  \streetaddress{495 Old Turner Street}
  \city{Blacksburg}
  \state{VA}
  \country{USA}
  \postcode{24060}
}

\author{Eugenia Rho}
\authornote{Corresponding author.}
\affiliation{%
  \institution{Virginia Tech}
  \streetaddress{495 Old Turner Street}
  \city{Blacksburg}
  \state{VA}
  \country{USA}
  \postcode{24060}
}
\email{eugenia@vt.edu}








\begin{abstract}
This study investigates how online counterspeech, \rev{defined as direct responses to harmful online content with the intention of dissuading the perpetrator from further engaging in such behavior,} is influenced by the match between a target of the hate speech and a counterspeech writer's identity. Using a sample of 458 English-speaking adults who responded to online hate speech posts covering race, gender, religion, sexual orientation, and disability status, our research reveals that the match between a hate post’s topic and a counter-speaker’s identity (topic-identity match, or TIM) shapes perceptions of hatefulness and experiences with counterspeech writing. Specifically, TIM significantly increases the perceived hatefulness of posts related to race and sexual orientation. TIM generally boosts counter-speakers’ satisfaction and perceived effectiveness of their responses, and reduces the difficulty of crafting them, with an exception of gender-focused hate speech. In addition, counterspeech that displayed more empathy, was longer, had a more positive tone, and was associated with higher ratings of effectiveness and perceptions of hatefulness. Prior experience with, and openness to AI writing assistance tools like ChatGPT, correlate negatively with perceived difficulty in writing online counterspeech. Overall, this study contributes insights into linguistic and identity-related factors shaping counterspeech on social media. The findings inform the development of supportive technologies and moderation strategies for promoting effective responses to online hate.
\end{abstract}

\ccsdesc[500]{Human-centered computing}
\ccsdesc[300]{Collaborative and social computing}
\ccsdesc{Empirical studies in collaborative and social computing}

\keywords{Empirical Methods, Mixed Methods ; Social Networking Site Design and Use ; Computer Mediated Communication ; Gender/Identity ; Social Media/Online Communities ; Empirical study that tells us about people ; Method ; Quantitative Methods ; Survey}


\maketitle

\section{Introduction}

The escalation of online hate speech presents a significant threat to individuals and society \cite{chandrasekharanYouCanStay2017, mathewThouShaltNot2019}. With the proliferation of social media, people now have access to a vast audience to disseminate harmful content that attacks individuals or groups based on their \rev{race \cite{matamoros-fernandezRacismHateSpeech2021a, costelloSocialGroupIdentity2019, mungerTweetmentEffectsTweeted2017}, gender \cite{henryTechnologyFacilitatedSexualViolence2018, cowanEmpathyWaysKnowing2003, stroudVarietiesFeministCounterspeech2018b}, religion \cite{obermaierLlBeThere2023, bonottiReligionHateSpeech2017, castano-pulgarinInternetSocialMedia2021}, sexual orientation \cite{costelloWeDonYour2019, cowanHeterosexualsAttitudesHate2005, goyalYourToxicityMy2022}, or disability status \cite{sherryDisabilityHateSpeech2019, sherryDisablistHateSpeech2019, vedelerHateSpeechHarms2019}. These topics represent some of the most common targets of online hate speech \cite{pazHateSpeechSystematized2020}.} The United Nations characterizes hate speech as any communication that vilifies individuals or groups based on aspects such as religion, ethnicity, nationality, race, color, descent, gender, or other identity factors \cite{unitednationsUnitedNationsStrategy2023}. Unlike generally offensive language, hate speech specifically targets core aspects of an individual's or a group's inherent identity - in essence, who they are. For this reason, hate speech is particularly insidious as it targets fundamental aspects of a person's or group's identity, exacerbating social divisions and often prompting discrimination \cite{beneschDefiningDiminishingHate2014}.

\rev{The harm inflicted by such speech can have profound impacts on the targeted individuals and groups. For instance, Schmid et al. (2024) found in their qualitative study that being confronted with hate speech can have similar consequences to traumatic events, causing frustration, fear, and anger, and inducing psychological stress or even depression, particularly for targeted groups such as women who tend to perceive such incivility as more severe \cite{schmidHowSocialMedia2024}. Exposure to online hate speech has been linked to experieincing mood swings, fear, and anger \cite{costelloHateSpeechOnline2020, keipiHarmadvocatingOnlineContent2017}. Exposure is also related to diminshed levels of trust \cite{nasiExposureOnlineHate2015} and adopting discriminatory attitudes \cite{foxmanViralHateContaining2013}. The gravest concern regarding encounters with hate material on the Internet is its potential to radicalize. Indeed, there are numerous instances linking exposure to online hate to violence, including mass violence and even terrorism \cite{holtLonersColleaguesPeers2019, frissenInternetGreatRadicalizer2021, hollewellRadicalizationSocialMedia2022}. Evidence suggests that exposure to online hate is widespread and frequent \cite{reichelmannHateKnowsNo2021}. Given the dangers associated with exposure, it is critically important that we find effective ways to combat it and reduce its impact.} 

One possible solution to online hate speech is online counterspeech, which is the act of responding to hateful content with the intention of stopping it, reducing its impact, or supporting the target \cite{garlandImpactDynamicsHate2022, hangartnerEmpathybasedCounterspeechCan2021, riegerHateCountervoicesInternet2018, ruthsConsiderationsSuccessfulCounterspeech2016}. Online counterspeech can take various forms (e.g., memes or pictures, written text, etc.\cite{fatimaHateSpeechSocial2023}) and use different strategies, such as using humor \cite{macavaneyHateSpeechDetection2019, mathewThouShaltNot2019}, showing empathy \cite{hangartnerEmpathybasedCounterspeechCan2021}, or warning the perpetrators \cite{macavaneyHateSpeechDetection2019, mathewThouShaltNot2019}. Research has shown that counterspeech can be effective in challenging online hate and promoting civility in online communities \cite{hangartnerEmpathybasedCounterspeechCan2021, mathewThouShaltNot2019, ziegeleJournalisticCountervoicesComment2018}. Nonetheless, crafting effective online counterspeech is complex \cite{buergerCounterspeechLiteratureReview2021} and often demands specific skills (e.g., linguistic fluency, motivation, and confidence \cite{garlandImpactDynamicsHate2022, salmivalliBystandersMatterAssociations2011}).

Another critical factor that may influence how and whether people engage in online counterspeech may be what we call in this paper, \textbf{\textit{Topic-Identity Match (TIM)}}, or the alignment between the topical focus of the hate speech and the demographic identity of the individuals responding to hate speech. For instance, hate speech directed at Asians might resonate differently with an Asian individual compared to others. Similarly, a woman countering a hateful online post against women might draw from her own personal experiences to make her response more authentic and impactful \cite{stroudVarietiesFeministCounterspeech2018b}. Studies have shown that the extent to which individuals perceive online hate speech as offensive significantly affects their likelihood of and approach to responding to it \cite{buergerIamhereCollectiveCounterspeech2021}. Thus, TIM may not only influence the intensity with which individuals perceive hate speech as offensive, as a direct match between their identity and the hate speech’s target can heighten the perceived hatefulness, but also influence \textit{how} someone engages in counterspeech. 

Hence, understanding the role of TIM is essential in evaluating the perception of hatefulness by individuals who respond to hate speech and how they write online counterspeech, as these perceptions shape their engagement strategies that indirectly contribute to the effectiveness and overall discourse quality of counterspeech. However, most prior research has primarily focused on the impact of online counterspeech on hate speakers \cite{mungerTweetmentEffectsTweeted2017, buergerIamhereCollectiveCounterspeech2021, garlandImpactDynamicsHate2022} or its overall effectiveness in reducing hate \cite{hangartnerEmpathybasedCounterspeechCan2021, bilewiczArtificialIntelligenceHate2021, saltmanNewModelsDeploying2023}. Limited attention has been given to how the alignment between the hate speech's topic and the identity of individuals responding to hate speech influences perceptions and responses to hate speech. In this paper, we address this gap by examining how TIM influences how users perceive and respond to online hate speech. In summary, we ask the following research questions in this paper:

\begin{enumerate}[label=RQ\arabic*:]
    \item How does the alignment between an individual's identity and the target of hate speech, known as \textit{Topic-Identity Match (TIM)}, shape the individual's perceived hatefulness of online hate speech? 
    \item How does \textit{TIM} influence users' subjective experience of writing a counterspeech - namely, their perceived \textit{satisfaction} with their counterspeech, their perceptions of its \textit{effectiveness} in responding to hate speech, and their perceived \textit{difficulty} in crafting online counterspeech? 
    \item Given the influence of TIM, how do specific linguistic features of participant-written counterspeech, including strategy, length, and sentiment polarity, correlate with \textit{(a) the participants' perceived hatefulness of the online hate speech they are responding to} and \textit{(b) their subjective experience of writing counterspeech}, measured in terms of \textit{satisfaction}, \textit{perceived effectiveness}, and \textit{difficulty}?
\end{enumerate}

Meanwhile, social media companies are beginning to take advantage of artificial intelligence (AI) with the intent to foster more positive online interactions while preventing harmful discourse on their platforms. For instance, Quora, a question-and-answer platform, uses AI to help users write clearer questions, in addition to providing AI-generated responses to users' questions \cite{pierceBetterChatGPTApp2023}. Instagram utilizes AI to offer suggested replies for creators in direct messages \cite{WhatNewOur2023}. Nextdoor, a neighborhood-based social network, has integrated OpenAI’s language models to recommend modifications for user posts that could potentially incite hostility \cite{NextdoorIntegratingGenerative2023}. Amidst this technological shift, several researchers advocate the use of AI to help generate online counterspeech as a strategy against online hate \cite{chungKnowledgeGroundedCounterNarrative2021, sahaSelfsupervisionControllingTechniques2023, zhuGeneratePruneSelect2021, munCounterspeakersPerspectivesUnveiling2024}. \rev{In particular, Mun et al. (2024) \cite{munCounterspeakersPerspectivesUnveiling2024} discussed the potential benefits and concerns of AI involvement in the counterspeech process, such as providing guidance on formulating effective responses and helping with emotion regulation and clear communication. In the process of our analysis, we noticed a pattern where individuals who perceived AI writing assistants as more useful also found writing counterspeech to be less challenging. Given the potential implications of AI, we found it relevant to present this observation beyond our primary three research questions. Therefore, we included an exploratory analysis to briefly investigate the relationship between the perceived usefulness of AI writing assistants like ChatGPT and the challenges people face when writing online counterspeech.}

Our study uses a survey with 458 participants who wrote counterspeech in response to three online hate posts randomly selected from a pool of 900 hate posts covering topics such as race, gender, religion, sexual orientation, or disability status. We then asked them follow-up questions to understand their perceptions of online hate speech and experience of writing counterspeech, such as their satisfaction, self-perceived effectiveness, the difficulty of their counterspeech, and their attitudes toward using AI to assist them in writing counterspeech. We used mixed-effects models to analyze the hierarchical data and capture individual and contextual effects. We investigated how the TIM between the hate post and the user's identity affected the user's perceived hatefulness of the hate post and their experience of responding to it with a counterspeech. We also examined how various linguistic characteristics of the user-written counterspeech were associated with their counterspeech writing experience, and their perceived hatefulness of the online hate post they were responding to. Finally, we investigated how the user's prior use of, and attitudes towards AI writing-assistant tools were associated with their difficulty in writing online counterspeech.  

Our results reveal that the TIM between the hate post and the identity of counterspeech writers (or \textit{counter-speakers}) influences their perception of online hate speech and their counterspeech writing experience. We found that TIM and prior exposure to hate posts (seeing more hate posts online) increased the perceived hatefulness of the hate posts, especially for race and sexual orientation topics (RQ1). Second, TIM positively influenced the satisfaction and self-perceived effectiveness of counterspeech and negatively influenced the difficulty of writing counterspeech for most topics, except for gender. We also found that counterspeech perceptions were affected by counter-speaker characteristics and behavior, such as, more frequent exposure to online hate speech, using their real name online, and higher commenting frequency. All of these factors were related to higher satisfaction and self-perceived effectiveness (RQ2). Third, linguistic characteristics of counterspeech were associated with counter-speakers’ writing experience and perceptions of hate speech. We found that the use of empathy in counterspeech was related to higher difficulty, satisfaction, and self-perceived effectiveness; longer counterspeech was related to higher satisfaction, self-perceived effectiveness, and hatefulness ratings; and more sentimentally positive counterspeech was linked to higher satisfaction and hatefulness ratings (RQ3). \rev{Finally, in an exploratory analysis, we found that prior use of ChatGPT and perceived usefulness of ChatGPT were negatively correlated with the difficulty of writing counterspeech, especially for those who found ChatGPT more useful.}

\textbf{Contributions}: We contribute to CSCW research by offering a comprehensive understanding of the various factors that shape the counter-speakers’ perception and writing of counterspeech on social media. First, \rev{we offer a theoretical lens to understand how \textit{\textbf{Topic-Identity Match (TIM)}}, counter-speakers' characteristics, and linguistic features shape the counter-speakers' writing experience. The theoretical explanation allows our work to extend to counter-extremist efforts more generally, informing the broader literature on ways to potentially thwart radicalization in online environments \cite{hawdonConfrontingOnlineExtremism2022}. Second, we extend existing research by considering a comprehensive set of counter-speakers' characteristics, including demographic factors, political views, hate speech exposure, and social media behavior. By simultaneously examining these identity factors, we address limitations in previous studies that often focus on fewer factors in isolation. Third, we contribute to the literature investigating the linguistic characteristics of online counterspeech and how this influences perceptions of writing counterspeech narratives. We offer empirical evidence on the relationship between these characteristics and perceived effectiveness, satisfaction, and difficulty in counterspeech writing. Understanding these factors can guide the development of improved moderation tools, user interfaces promoting constructive dialogue, and AI-assisted writing systems for counterspeech on social media \cite{meskeDesignPrinciplesUser2023}. Fourth, we provide a large-scale quantitative analysis of how TIM influences perception and writing of counterspeech. While prior scholarship has mostly focused on the impact of counterspeech on the hate speakers or its effectiveness \cite{blackwellClassificationItsConsequences2017, costelloWhoAreOnline2018, wachsAssociationsBystandersPerpetrators2018}, our work adds a new layer of depth by empirically validating TIM. Finally, our exploratory analysis contributes to the ongoing discourse on the role of AI in crafting online counterspeech by offering an empirical, quantitative perspective that complements the existing qualitative insights into countering hate speech on social media. Understanding how the identity of counter-speakers and the use of AI influence an individual's willingness and ability to intervene upon encountering hate speech provides valuable insights for designing effective counter-extremism strategies.}

\section{Related Work}

Online hate is a pervasive and harmful phenomenon that affects individuals and society \cite{chandrasekharanYouCanStay2017, kaakinenDidRiskExposure2018, reichelmannHateKnowsNo2020}. Researching people’s perception of online hate posts is important for understanding the causes \cite{costelloWhoAreOnline2018, kaakinenImpulsivityInternalizingSymptoms2020, wachsAssociationsBystandersPerpetrators2018}, consequences \cite{mathewHateBegetsHate2020, seglowHateSpeechDignity2016, soralExposureHateSpeech2018}, and potential solutions to this problem \cite{hangartnerEmpathybasedCounterspeechCan2021, mathewThouShaltNot2019}. For example, Soral et al. (2018) found that more exposure to online hate speech makes people less sensitive. They also found that this desensitization process results in lower evaluations of the victims and greater distancing from them, thus increasing outgroup prejudice \cite{soralExposureHateSpeech2018}. \rev{However, a broader examination of literature suggests a nuanced dynamic: Increased exposure to online hate speech has been linked with heightened awareness and a greater propensity to recognize and challenge hate speech content \cite{hawdonConfrontingOnlineExtremism2022, costelloConfrontingOnlineExtremism2017, pingCounterExploringMotivations2024}. This apparent contradiction underscores the complex interplay between individual and the contextual factors influencing responses to hate speech.} As a potential solution, Lepoutre et al. (2017) suggested counterspeech as an effective way to counteract the dilution effect of hate speech, as it can challenge, correct, or counteract the negative effects of hate speech \cite{lepoutreHateSpeechPublic2017}. Providing evidence for this argument, Hangartner et al. (2021) showed that empathetic counterspeech was particularly effective in compelling users to delete racist and xenophobic tweets in a field experiment \cite{hangartnerEmpathybasedCounterspeechCan2021}. Additional studies have explored how different factors, such as the content \cite{cohen-almagorTakingNorthAmerican2018, tontodimammaThirtyYearsResearch2020}, context \cite{chettyHateSpeechReview2018, dordevicSociocognitiveDimensionHate2020}, and source of online hate posts \cite{costelloWhoAreOnline2018} influence the perception of hatefulness by the recipients \cite{soralExposureHateSpeech2018}, bystanders \cite{buergerWhyTheyIt2022, obermaierLlBeThere2023, buergerIamhereCollectiveCounterspeech2021, wachsAssociationsBystandersPerpetrators2018}, and perpetrators \cite{costelloWhoAreOnline2018, wachsAssociationsBystandersPerpetrators2018, mungerTweetmentEffectsTweeted2017}. However, these studies have not sufficiently examined the perspective of the counter-speakers – the people who write counterspeech in response to online hate speech. The counter-speakers’ perception of online hate posts may affect their motivation, strategy, and effectiveness in countering online hate. Therefore, in this paper, we aim to address this gap by investigating how various factors influence the counter-speakers’ perception of hatefulness and their counterspeech writing experience. These factors include the counter-speakers’ demographics, social media experience, experience with AI writing tools, and the characteristics of the hate speech itself, such as the topic and TIM.

\subsection{The Effect of Hate Speech Topics}

Hate speech is a complex phenomenon that can be characterized by various features, such as the language \cite{garlandImpactDynamicsHate2022, schmidtSurveyHateSpeech2017}, tone \cite{polettoAnnotatingHateSpeech2019}, intensity \cite{ibrohimMultilabelHateSpeech2019}, and intention of the speaker \cite{holgateWhySwearAnalyzing2018}. However, as Poletto et al. (2020) noted in their systematic review of hate speech corpora, the topic of hate speech, or the protected group that is targeted by hateful or derogatory expressions, is one of its most salient features. The target can be either a group or an individual belonging to such a group, not for their individual characteristics, but for their group membership \cite{polettoResourcesBenchmarkCorpora2021}. The topic of hate speech depends on the context, culture, and ideology of the audience \cite{fantonHumanintheLoopDataCollection2021, mollasETHOSOnlineHate2022, ousidhoumMultilingualMultiAspectHate2019}. Therefore, it is crucial to examine how different topics of hate speech affect the perception of the people who encounter them, especially those who write counterspeech to challenge online hate. In this study, we categorize the topics of hate speech into five groups: \rev{race \cite{matamoros-fernandezRacismHateSpeech2021a, costelloSocialGroupIdentity2019, mungerTweetmentEffectsTweeted2017}, gender \cite{henryTechnologyFacilitatedSexualViolence2018, cowanEmpathyWaysKnowing2003, stroudVarietiesFeministCounterspeech2018b}, religion \cite{obermaierLlBeThere2023, bonottiReligionHateSpeech2017, castano-pulgarinInternetSocialMedia2021}, sexual orientation \cite{costelloWeDonYour2019, cowanHeterosexualsAttitudesHate2005, goyalYourToxicityMy2022}, or disability status \cite{sherryDisabilityHateSpeech2019, sherryDisablistHateSpeech2019, vedelerHateSpeechHarms2019}, as these topics represent some of the most common targets of online hate speech \cite{pazHateSpeechSystematized2020}.} We investigate how these topics influence the perception of counterspeech writers.

\subsection{The Effect of Counter-speaker’s Social Identity}

Besides the topic of hate speech, another key influencing factor that may affect the perception of the counterspeech writers is their social identity \cite{costelloSocialGroupIdentity2019, simpsonDignityHarmHate2013, elsheriefHateLingoTargetbased2018, roussosHateSpeechEye2018, stroudVarietiesFeministCounterspeech2018b}. The identity of the counterspeech writers refers to their social group membership. Previous research has examined how the demographic or identity of the raters, such as age \cite{wachsAssociationsBystandersPerpetrators2018}, gender \cite{cowanEmpathyWaysKnowing2003}, race \cite{celuchFactorsAssociatedOnline2022, goyalYourToxicityMy2022}, and sexual orientation \cite{costelloWeDonYour2019, cowanHeterosexualsAttitudesHate2005, goyalYourToxicityMy2022} etc., influences their perception of hatefulness in online posts. For example, Celuch et al. (2022) found significant differences in online hate acceptance levels among individuals from different countries, races, or cultural backgrounds \cite{celuchFactorsAssociatedOnline2022}. Similarly, Zhang et al. (2018) also found that the perception of hate speech and offensive language was affected by the rater’s gender and personal experience \cite{zhangHateSpeechDetection2018}. \rev{Although these studies effectively quantify the influence of social identity on perceptions and attitudes towards hate speech, they fall short in examining the interaction between the specific content of hate speech and the demographic identities of the respondents. This leaves a gap in understanding how different types of hate speech are perceived or countered by individuals from varied demographic backgrounds. In a recent study, Schmid et al. (2024) divided the cognition of hate speech into two levels: first-level (recognition) and second-level (attitudes/opinions). They found that the counter-speakers' identity influenced both levels, with indications that women had a heightened sensitivity to hate speech and perceived it as more severe compared to men \cite{schmidHowSocialMedia2024}. Such studies relied on small qualitative samples that can offer profound insights into individual experiences, but this richness lacks the large-scale quantitative validation that is particularly important in research on sensitive topics such as hate speech.} Another recent large-scale quantitative study by Obermaier et al. (2023) examined the effects of Islamophobic online hate speech on the perceived religious identity threat and the intentions to utter factual counterspeech among Muslim participants \cite{obermaierLlBeThere2023}. They found that exposure to online hate speech increased the perception of religious identity threat, which in turn enhanced the sense of personal responsibility to intervene and the willingness to engage in factual counterspeech. \rev{However, these studies often focus on specific aspects of identity. For instance, while factors like age, gender, and social media use are considered, others such as political attitudes, education level, or disability status are often overlooked. Focusing on a single identity ignores the fact that humans have multiple identities \cite{learyHandbookSelfIdentity2012}, and each of these can potentially influence how the react to hate speech exposure.} Therefore, it is important for us to consider a range of identity information. 

In this study, we use the term \textbf{\textit{Topic-Identity Match (TIM)}} to describe whether the counter-speakers’ identity aligns with the hate speech’s topic. For example, if the hate speech targets women and the counter-speaker is also a woman, then they have a TIM. \rev{Research in this growing field has collectively suggested that the social identity of the perceiver may influence how they justify or counter hate speech against different groups \cite{obermaierLlBeThere2023, stroudVarietiesFeministCounterspeech2018b, schmidHowSocialMedia2024}. However, as previously noted, these studies often focus on individual identity aspects like age, gender, and social media use, rather than a more comprehensive set of identity information such as political attitudes, education level, disability status, or their social media usage behaviors \cite{obermaierLlBeThere2023, cowanEmpathyWaysKnowing2003, wachsAssociationsBystandersPerpetrators2018}. Furthermore, while small qualitative samples can provide deep insights into individual experiences, they are often prone to social desirability bias, especially on sensitive topics like hate speech \cite{schmidHowSocialMedia2024, elsheriefHateLingoTargetbased2018}. Moreover, while these studies concentrate on perceptions and attitudes towards hate speech, they often assume topic matches that are implied rather than empirically validated \cite{costelloSocialGroupIdentity2019, simpsonDignityHarmHate2013, roussosHateSpeechEye2018}.} Therefore, it is important to consider the identity of the counter-speakers as an individual variable that may influence the perception and the behavior of the counter-speakers. \rev{In this study, we provide further empirical evidence for the role of identity through a large-scale quantitative analysis, using a multilevel linear mixed-effects model to measure the effect of TIM. This serves as a complementary and supportive evidence to the prior studies.}

\subsection{Writing Experience and Counterspeech}

Prior research indicates that the satisfaction with counterspeech efforts \cite{ hensonThereVirtuallyNo2020}, perceptions of their effectiveness in deterring hateful behavior \cite{wachsAssociationsWitnessingPerpetrating2019}, and the difficulty encountered in responding to online hate \cite{buergerIamhereCollectiveCounterspeech2021} are critical in shaping individuals' experiences and decisions to engage in counterspeech \cite{buergerCounterspeechLiteratureReview2021}. Henson et al. (2020) investigated the frequency and predictors of bystander intervention behaviors in online situations among college students. They found that satisfaction with intervention and confidence in violence prevention skills were positively associated with online bystander intervention \cite{hensonThereVirtuallyNo2020}. Wachs et al. (2019) found that beliefs about the response's impact on perpetrator behavior influence the willingness to engage \cite{wachsAssociationsWitnessingPerpetrating2019}. Buerger et al. (2021) examined a major counterspeech effort on Facebook and found that the writing experience of the counter-speakers, such as the challenge of crafting suitable and persuasive replies, influenced their motivation and confidence to engage in counterspeech \cite{buergerIamhereCollectiveCounterspeech2021}. These studies suggest that enhancing these factors to increase the writer's positive perceptions of the experience may increase the likelihood of the person intervening in online hate \cite{hensonThereVirtuallyNo2020, buergerIamhereCollectiveCounterspeech2021, wachsAssociationsWitnessingPerpetrating2019}, including writing a counterspeech \cite{buergerCounterspeechLiteratureReview2021}. However, these studies have not sufficiently explored how these writing experience factors are associated with the characteristics of the counter-speakers or the linguistic characteristics of counterspeech. We aim to fill this gap by exploring how the counter-speakers’ characteristics and the linguistic features of counterspeech are related to three key experiential factors experienced by users when composing online counterspeech: their satisfaction with their own counterspeech, their belief in the effectiveness of their counterspeech in mitigating the hate speech they are responding to, and the level of difficulty experienced when crafting the counterspeech.

\subsection{Linguistic Characteristics of Counterspeech}

Recent research examined the correlation between the linguistic characteristics of counterspeech and its effectiveness \cite{hangartnerEmpathybasedCounterspeechCan2021, mathewThouShaltNot2019, baiderAccountabilityIssuesOnline2023}. Baider et al. (2023) highlighted the predominant use of argumentative strategies in counterspeech, often accompanied by a tone of refutation, and examined how these approaches influence the outcomes of counterspeech \cite{baiderAccountabilityIssuesOnline2023}. They found that depending on the context and the audience, although using a tone of refutation could sometimes foster dialogue, it could also lead to backlash from the perpetrator and more hostile verbal exchanges \cite{baiderAccountabilityIssuesOnline2023}. This finding highlights the importance of choosing the right tone and rhetorical strategy for effective counterspeech interventions. Hangartner et al. (2021) conducted an experiment to test the effects of three counterspeech strategies — empathy, warning of consequences, and humor — on reducing xenophobic hate speech on Twitter. They discovered that only empathy-based counterspeech was effective in increasing the deletion of hate speech by the original perpetrators and in decreasing the likelihood of backlash \cite{hangartnerEmpathybasedCounterspeechCan2021}. Using a manually annotated dataset of YouTube comments, Mathew et al. (2019) examined the linguistic structure of counterspeech. They discovered that the effectiveness of counterspeech was significantly influenced by features such as tone, first-person language, and psycholinguistic categories \cite{mathewThouShaltNot2019}. 

However, only a few studies have examined the relationship between counterspeech strategies and the perceptions of counter-speakers. In their 2021 study, Buerger et al. qualitatively explored this relationship through the experiences of members in a Facebook counterspeech group \cite{buergerIamhereCollectiveCounterspeech2021}. This research demonstrates the members' perceptions of the challenges involved in crafting counterspeech that is both suitable and persuasive, underscoring the complexity of responding to hate speech in a manner that is both effective and respectful by taking into account the subtleties of tone and content of counterspeech.  Our work extends prior research by further examining such connection between the linguistic characteristics of counterspeech and the perceptions (individuals' reported satisfaction, self-perceived effectiveness, and difficulty in their writing experience; as well as perceived intensity of hate in the posts) of the counter-speakers. Common linguistic characteristics of counterspeech include strategy \cite{baiderAccountabilityIssuesOnline2023, hangartnerEmpathybasedCounterspeechCan2021}, length \cite{chungCONANCOunterNArratives2019}, use of questions and first-person language \cite{mathewThouShaltNot2019, saltmanNewModelsDeploying2023}, and sentiment polarity \cite{baiderAccountabilityIssuesOnline2023}.

\section{Methods}

To explore the role of participants’s social identity in online counterspeech, we carried out a pre-registered survey with English-speaking U.S. participants (N = 458). The participants were presented with three random examples of hate speech from a pool of 900 hateful posts that covered different topics, and they were asked to write a counterspeech in response to each one. We obtained 1374 pairs of hate posts and counterspeech from the participants’ responses. We then had six independent annotators review the pairs and remove those that were of low quality or irrelevant. This resulted in 1261 pairs of hate posts and counterspeech that were used for analysis.

\subsection{Collection of Online Hate Posts} \label{subsec:hatepost}

We obtained hateful posts from three online hate datasets that are widely used in literature: ETHOS \cite{mollasETHOSOnlineHate2022}, Multi-Target Counter Narrative \cite{fantonHumanintheLoopDataCollection2021}, and MLMA \cite{ousidhoumMultilingualMultiAspectHate2019}. We randomly sampled hateful posts from the combined corpus that covered five frequent topics of hate speech: race, gender, religion, sexual orientation, and disability. We manually checked all the sampled posts to ensure that they were relevant to the topics and balanced the number of posts for each topic. We ended up with 900 hateful posts for our survey, with five topics: race (183), gender (183), religion (182), sexual orientation (182), and disability (170).

\subsection{Survey Design and Variables}
The survey was designed using Qualtrics and consisted of (a) a consent form, (b) relevant background information about hateful speech and counterspeech, (c) 3 hateful posts and questions pertaining to them, (d) questions about past online hate speech experience, frequency of writing counterspeech online, and motivations as well as barriers to writing online counterspeech, (e) questions about prior use of ChatGPT, perceived usefulness of ChatGPT, as well as willingness of using such AI tools to aid in counterspeech writing, and finally (f) demographic and social media use questions.

The consent form informed participants that they were being invited to a study to evaluate the efficacy of counterspeech. \rev{Our study specifically focuses on the direct public replies to hateful posts on social media, with the aim of dissuading the perpetrators from further engaging in such behavior. Although counterspeech can take forms such as indirect responses that encourage bystanders to speak up \cite{buergerCounterspeechLiteratureReview2021} and one-on-one private messages \cite{wrightVectorsCounterspeechTwitter2017}, these forms are outside the scope of our current study. The focus on direct engagement represents one important type of counterspeech interaction, as proposed by Wright et al. (2017) and Benesch et al. (2016) \cite{ruthsCounterspeechTwitterField2016, wrightVectorsCounterspeechTwitter2017}, who include direct engagement as a distinct category in their classification systems for counterspeech.} Given the offensive nature of hateful speech, participants were also informed of the potential psychological risks involved in this study. Then, participants were provided with definitions of hateful speech and counterspeech, as well as examples of counterspeech. Following this, participants were shown three unique hateful posts randomly selected from the set of 900 hate posts described in \ref{subsec:hatepost}. For each hateful post, participants were prompted with “Imagine you are a user of an online group on social media. Another user (perpetrator) in the group posted the following. Do you consider this post to be hateful?” If they answered Yes, participants were also asked to rate the hatefulness of each post using a four-point scale, with the question, “How hateful do you find this post?” Response options ranged from (1) A little to (4) A great deal. Participants were then prompted to respond to the hateful post shown. The survey asked, “Please write a counterspeech to this post. The goal is to further reduce hateful behavior from the perpetrator.” Participants were then asked to rate their satisfaction, perceived effectiveness, and perceived difficulty of each counterspeech they wrote using a five-point Likert scale. Finally, participants answered questions related to motivations and barriers to writing online counterspeech, frequency of writing online counterspeech, and willingness to use ChatGPT to write counterspeech on social media. \rev{The demographic data collected in this study was based on participants' self-disclosure, which reflects their subjective identification with the social identity groups.}

\subsection{Recruitment}
We used Prolific to recruit U.S.-based, English-speaking adults who had approval ratings above 95\%. We informed the participants about the possibility of encountering harmful content in the survey. We initially had 536 respondents, but we excluded those who did not pass attention checks or did not finish the survey. The final sample consisted of 458 participants. The participants took an average of 15 minutes to complete the survey and received a compensation rate of \$12/hour.

\subsection{Data Annotation} \label{sec:anno}

We first conducted a quality and relevance check of the hate posts and counterspeech pairs that were collected from the survey. We hired six independent annotators to review the pairs and remove those that were of low quality or irrelevant. Low-quality pairs were those that had incomplete, incomprehensible, or inappropriate hate posts or counterspeech. Irrelevant pairs were those where hate posts and counterspeech did not match. The annotators removed 113 pairs out of 1374, resulting in 1261 pairs that were used for further analysis. We also calculated the inter-rater reliability (IRR) of the annotators using Cohen’s kappa coefficient. The IRR was 0.882 (95\% CI, 0.806 to 0.958), which indicates a very high level of agreement among the annotators \cite{wanKappaCoefficientPopular2015}. This suggests that the quality and relevance of the hate posts and counterspeech pairs were consistently evaluated.

We then annotated the hate posts and counterspeech pairs on two dimensions: TIM and the linguistic characteristics of the counterspeech (including strategy, use of first-person language, use of questions, and sentiment polarity). Topic-Identity Match (TIM) is a binary variable that indicates whether the topic of the hate post matches the identity of the counterspeech writer. For example, if the hate post targeted women and the participant who wrote a counterspeech in response was also a woman, then the topic matched. The same logic applied to other topics, such as race. If the hate post targeted African Americans and the participant who wrote a counterspeech was white, then the topic did not match.
Strategy is a categorical variable that indicates the type of strategy that the counterspeech writer uses to write their counterspeech. We had five types of strategy: empathy, humor,  warning of consequence, refutal, and other. Empathy is when the counterspeech writer shows empathy or compassion to the target, such as by expressing support or understanding. Humor is when the counterspeech writer uses humor or sarcasm to mock or ridicule the hate speech, such as by making jokes or irony. Warning of consequence is when the counterspeech writer warns the hate speaker of the potential consequences of their hate speech, such as legal action or social backlash. Refutal is when the counterspeech writer refutes or challenges the hate speech with facts or logic, such as by providing evidence or counterarguments. \rever{Table A1 in the supplementary materials shows examples of different counterspeech strategies.} First-person language is used when the counterspeech writer uses “I” or “we” to express their opinion or experience.     
The sentiment polarity in counterspeech reflects the extent of its negativity or positivity, categorizing responses based on their emotional tone. To determine the polarity of each counterspeech instance, we utilized an automated sentiment analysis tool \cite{vidgenLearningWorstDynamically2021}, which evaluates the emotional tone based on specific linguistic markers and context.

\subsection{Analysis}

\subsubsection{Data Structure}

\begin{figure}[htbp]
  \centering
  \includegraphics[width=\linewidth]{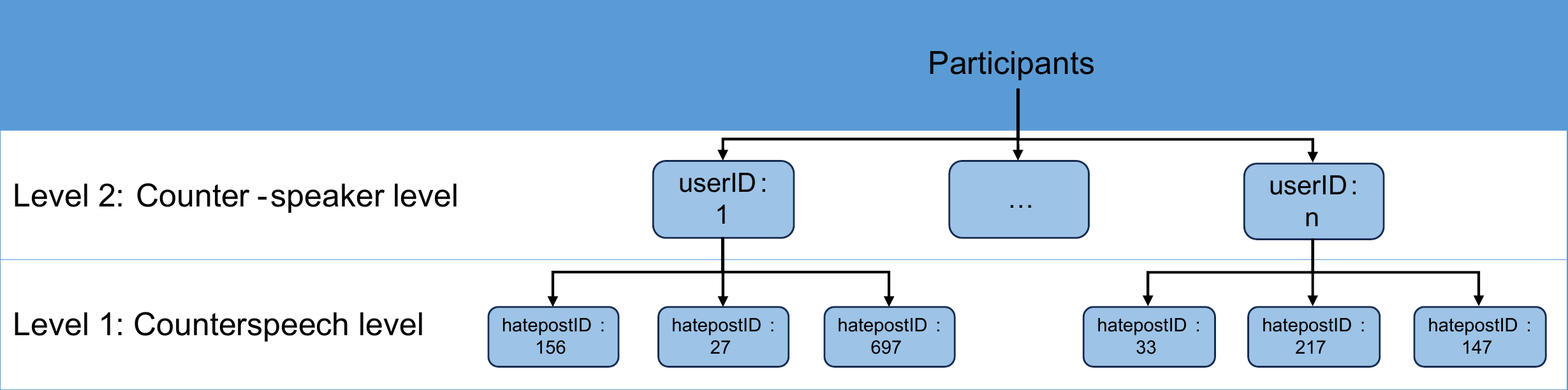}
  \caption{\textbf{A Two-Level Hierarchical Study of Counterspeech.} The figure shows the data structure of our study, where counterspeech responses are nested within participants. Each participant responded to three random hate posts, each with a different topic (race, religion, gender, disability, or sexual orientation).}
    \label{fig:level}
\end{figure}

Our data were multilevel in nature, as the responses of the participants were nested within the hate posts they responded to. Each participant responded to three hate posts. Therefore, we had two levels of analysis: the counter-speaker level (level 2) and the hate post level (level 1). We used multilevel linear mixed models (LMMs) to account for the dependency of the observations within each level and to examine the effects of both level-1 and level-2 predictors on the outcome variables. Table \ref{tab:varlist} lists all variables included in our analysis, and Figure \ref{fig:level} shows the structure of levels. The text in this paper uses a \texttt{typewriter} \texttt{font} to highlight the variable names.

\begin{table}[htbp]
  \centering
  \caption{Variables and Effects Used in the Multilevel Linear Mixed Model Analysis. The demographic variables were used to match participants with hate posts by topic.}
    \begin{tabular}{|c|p{19em}|p{12.5em}|}
    \hline
    \rowcolor[rgb]{ .933,  .925,  .882} \textbf{Levels} & \textbf{Variables} & \textbf{Effects} \bigstrut\\
    \hline
    \multicolumn{1}{|c|}{\multirow{14}[28]{*}{2: Counter-speaker level}} & userID & Random intercept \bigstrut\\
\cline{2-3}          & Frequency of encountering online hate speech & \multirow{3}[6]{*}{Social media control variable} \bigstrut\\
\cline{2-2}          & Use of real name on social media & \multicolumn{1}{l|}{} \bigstrut\\
\cline{2-2}          & Social media commenting frequency & \multicolumn{1}{l|}{} \bigstrut\\
\cline{2-3}          & Gender & \multirow{5}[10]{*}{\parbox{12em}{Demographic control variable (used to match hate posts by topic)}} \bigstrut\\
\cline{2-2}          & Ethnicity & \multicolumn{1}{l|}{} \bigstrut\\
\cline{2-2}          & Sexual orientation & \multicolumn{1}{l|}{} \bigstrut\\
\cline{2-2}          & Religion & \multicolumn{1}{l|}{} \bigstrut\\
\cline{2-2}          & Disability & \multicolumn{1}{l|}{} \bigstrut\\
\cline{2-3}          & Age   & \multirow{3}[6]{*}{Demographic control variable} \bigstrut\\
\cline{2-2}          & Education level & \multicolumn{1}{l|}{} \bigstrut\\
\cline{2-2}          & Political view & \multicolumn{1}{l|}{} \bigstrut\\
\cline{2-3}          & Prior use of chatgpt & Fixed slope \bigstrut\\
\cline{2-3}          & Perceived usefulness of chatgpt & Fixed slope \bigstrut\\
    \hline
    \multicolumn{1}{|c|}{\multirow{12}[24]{*}{1: Counterspeech level}} & hatepostID & Random intercept \bigstrut\\
\cline{2-3}          & Perceived hatefulness rating & Dependent variable \bigstrut\\
\cline{2-3}          & Satisfaction & Dependent variable \bigstrut\\
\cline{2-3}          & Effectiveness & Dependent variable \bigstrut\\
\cline{2-3}          & Difficulty & Dependent variable \bigstrut\\
\cline{2-3}          & Hate post topic & Fixed slope \bigstrut\\
\cline{2-3}          & Topic-Identity Match (TIM) & Fixed slope \bigstrut\\
\cline{2-3}          & Strategy & Fixed slope \bigstrut\\
\cline{2-3}          & Length & Fixed slope \bigstrut\\
\cline{2-3}          & Use of first-person language & Fixed slope \bigstrut\\
\cline{2-3}          & Use of questions & Fixed slope \bigstrut\\
\cline{2-3}          & Sentiment polarity & Fixed slope \bigstrut\\
    \hline
    \end{tabular}%
  \label{tab:varlist}%
\end{table}%

\subsubsection{RQ1. How does the alignment between an individual's identity and the target of hate speech, known as \textit{Topic-Identity Match (TIM)}, shape the individual's perceived hatefulness of online hate speech?}

To answer RQ1, we used a multilevel LMM to analyze the data, as it can account for the nested structure of the data (i.e., repeated measures within participants and participants within hate posts). A multi-level allows for the estimation of the variance components at different levels and the testing of the significance of the fixed effects at each level \cite{hoxMultilevelAnalysisTechniques2017}. We used the \yellowhl{perceived hatefulness rating} as the dependent variable and random intercepts for \yellowhl{userID} (unique identifiers of the participants) and \yellowhl{hatepostID} (unique identifiers of hate posts that they rated in the study) to capture the variability among participants and hate posts to calculate the intraclass correlation coefficients (ICCs) for the random effects. The ICC indicates the ratio of variance explained by the grouping structure in the population to the total variance. It can also be interpreted as the expected correlation between two units randomly selected from the same group \cite{hoxMultilevelAnalysisTechniques2017}. A low ICC (< .50) reflects a low degree of agreement among raters or measurements \cite{kooGuidelineSelectingReporting2016}, implying that different participants would perceive the same hate post with different levels of hatefulness and that the same hate post would elicit different levels of hatefulness from different participants. For this we calculated the intraclass correlation coefficients (ICCs) for the random effects of \yellowhl{userID} and \yellowhl{hatepostID} in the intercept-only linear mixed-effects model. The ICC for \yellowhl{hatepostID} was 0.194 and for \yellowhl{userID} was 0.232. Both ICCs were lower than 0.50, suggesting that the perceived hatefulness of online posts was not consistent across participants or hate posts \cite{kooGuidelineSelectingReporting2016}, indicating that the perceived hatefulness of online posts varied depending on the post content and the participants’ social identities. 

We then extended the intercept-only model by adding independent and control variables, which accounted for the hierarchical structure of the data, where multiple hate posts were nested within each participant. \rever{Equation (1) and Table A2, both located in the supplementary materials' Appendix sections A.3 and A.2 respectively, present the full model and the variables included in this model.} The \yellowhl{hate post topic} and \yellowhl{TIM} were level-1 predictors, which varied across hate posts within participants. The \yellowhl{frequency of encountering online hate speech}, \yellowhl{use of real name on social media}, \yellowhl{political view}, etc. were level-2 predictors, which varied across participants. For \yellowhl{hate post topic} variable, we used the topic religion as a reference baseline to compare the effects of the other topics on the perceived hatefulness ratings. 

\subsubsection{RQ2. How does \textit{TIM} influence users' subjective experience of writing a counterspeech - namely, their perceived \textit{satisfaction} with their counterspeech, their perceptions of its \textit{effectiveness} in responding to hate speech, and their perceived \textit{difficulty} in crafting online counterspeech?} \label{sec:rq2method}

As with RQ1, we conducted three LMMs to examine the three dependent variables: \yellowhl{satisfaction}, \yellowhl{self-perceived effectiveness}, and \yellowhl{difficulty} of counterspeech. We measured \yellowhl{satisfaction}, \yellowhl{self-perceived effectiveness}, and \yellowhl{difficulty} with continuous 5-point scales that ranged from 1 (very low) to 5 (very high). \yellowhl{Satisfaction} measures how satisfied the participant is with their counterspeech. \yellowhl{Self-perceived effectiveness} measures how effective the participant thinks their counterspeech is in countering hate speech. \yellowhl{Difficulty} measures how difficult the participant finds writing counterspeech. We calculated the ICCs for the random effects of \yellowhl{userID} and \yellowhl{hatepostID} to measure the consistency of the dependent variables within participants and hate posts. The ICC of \yellowhl{userID} for \yellowhl{self-perceived effectiveness} was 0.677, which was higher than 0.5, indicating that every participant had consistency in their perception of their counterspeech effectiveness. The ICCs of \yellowhl{userID} for \yellowhl{satisfaction} and \yellowhl{difficulty} were 0.468 and 0.423, respectively. The ICCs of \yellowhl{hatepostID} for all three dependent variables were below 0.1, indicating that the dependent variables were influenced mainly by individual characteristics. Therefore, we did not include \yellowhl{hatepostID} as a random effect for the three LMMs. The independent variables were the same for all three LMMs presented in \rever{Table A3 in Appendix A.2 and Equation (2) in Appendix A.3 of the supplementary materials}. The variable of \yellowhl{hate speech topic} was a potential influence on the perception and behavior of participants. Therefore, we conducted a pairwise least-squares means analyses for each of the five topics of hate speech: race, gender, religion, sexual orientation, and disability. We compared the results of the subgroup analyses to examine the differences between the topics.

\subsubsection{RQ3. Given the influence of TIM, how do specific linguistic features of participant-written counterspeech, including strategy, length, and sentiment polarity, correlate with \textit{(a) the participants' perceived hatefulness of the online hate speech they are responding to} and \textit{(b) their subjective experience of writing counterspeech}, measured in terms of \textit{satisfaction}, \textit{perceived effectiveness}, and \textit{difficulty}?}

To explore the relationship between the linguistic characteristics of counterspeech and the participants’ self-perceived outcomes and their perception of the hatefulness of the hate posts, we measured the linguistic characteristics of counterspeech with five variables: \yellowhl{strategy}, \yellowhl{use of first-person language}, \yellowhl{use of questions}, \yellowhl{length}, and \yellowhl{sentiment polarity}. 
\yellowhl{Strategy} was defined in \ref{sec:anno} and we set empathy-based counterspeech as the baseline in all our models. \yellowhl{Use of first-person language} was a binary variable that indicated whether the counterspeech used the pronoun “I” or “we” to express the writer’s personal opinion or experience. \yellowhl{Use of questions} was a binary variable that indicated whether the counterspeech is a question sentence. \yellowhl{Length} was the number of words in the counterspeech.  \yellowhl{Sentiment polarity} was a ranked variable that measured the emotional tone of the counterspeech, ranging from negative to positive (see \ref{sec:anno}). We conducted four LMMs, where \yellowhl{perceived hatefulness rating}, \yellowhl{satisfaction}, \yellowhl{self-perceived effectiveness}, and \yellowhl{difficulty} were the dependent variables. We included \yellowhl{userID} and \yellowhl{hatepostID} as random effects in the LMMs to account for the variability in the dependent variables across participants and hate posts. We also controlled for \yellowhl{topic} and \yellowhl{TIM} as independent variables. \rever{Table A4 in Appendix A.2 and Equation (3) in Appendix A.3 of the supplementary materials list all the variables used in the model.}

\subsubsection{\rev{Exploratory Analysis: The Correlation between the Use of AI-Writing Assistants and the Perceived Difficulty in Writing Online Counterspeech}}

\rev{We conducted two sets of LMMs using the same variables as in RQ2 (Section \ref{sec:rq2method}), with the addition of variables related to the participants' use of AI.} The first set of LMMs was based on the full model of RQ2, to which we added the variable \yellowhl{prior use of ChatGPT} as a fixed effect. This variable indicated whether the participants had used ChatGPT before participating in the study or not. The second set of LMMs was performed only for the participants who had used ChatGPT before, and we added the variable \yellowhl{perceived usefulness of ChatGPT} as a fixed effect. This variable measured how useful the participants found ChatGPT on a 5-point Likert scale. All LMMs included \yellowhl{userID} as random effects to account for the nested structure of the data. \rever{Pairwise comparisons were corrected using Tukey's Honestly Significant Difference (HSD) test.} \rev{In our analysis, we found that both variables were only related to \yellowhl{difficulty}. Therefore, as an exploratory analysis, we only present the results related to this variable.}

\section{Results}

We analyzed a total of 1261 pairs of hate posts and participant-written counterspeech. The distribution of hate speech topic and TIM are shown in Table \ref{tab:dis}.

\begin{table}[htbp]
  \centering
  \caption{Distribution of Topic and Topic-Identity Match (TIM) of Online Hate Speech}
    \begin{tabular}{p{11.72em}ccc}
    \toprule
    \multirow{2}[4]{*}{\textbf{Topic}} & \multicolumn{2}{m{16.56em}}{\centering \textbf{Topic-Identity Match (TIM)}} & \multicolumn{1}{c}{\multirow{2}[4]{*}{\textbf{Total}}} \\
\cmidrule{2-3}    \multicolumn{1}{r}{} & \multicolumn{1}{m{8.28em}}{\centering Match} & \multicolumn{1}{m{8.28em}}{\centering Non-match} &  \\
    \midrule
    Religion & 58    & 198   & 256 \\
    Race  & 184   & 63    & 247 \\
    Gender & 114   & 125   & 239 \\
    Sexual Orientation & 202   & 62    & 264 \\
    Disability & 173   & 82    & 255 \\
\cmidrule{2-3}    \textbf{Total} & 731   & 530   & 1261 \\
    \bottomrule
    \end{tabular}%
  \label{tab:dis}%
\end{table}%

The results of RQ1 show that TIM, or the alignment between an individual's identity and the target of online hate speech, significantly influences the degree to which users find the online hate speech they were responding to hateful. The level of perceived hatefulness varied among participants based on the specific topic of the hate speech. In general, people found online hate speech targeting individuals based on race and sexual orientation significantly more hateful compared to hate speech related to disability status, gender, and religion. However, across all topics, people's perception of hate speech was significantly more offensive when there was a TIM between the individual's identity and the target of the hate speech. 

In RQ2, our findings reveal that TIM generally increased satisfaction in people's counterspeech writing experience for race and disability-related hate speech, as well as increased the perceived effectiveness of their counterspeech against hate speech related to religion and race. Conversely, for gender-related hate speech, participants, especially females, found it significantly more challenging to write counterspeech targeting women and perceived their own counterspeech to be less effective.

RQ3 findings show that various linguistic characteristics (sentiment and length) and strategies of counterspeech, such as empathy, humor, and refutal, are significantly associated with both the participant's perceived hatefulness of the online hate speech they were asked to respond to (RQ3a), as well as aspects related to their experience of writing counterspeech to the hate speech (RQ3b).

\rev{In our exploratory analysis,} we found that participants who previously used ChatGPT or perceived ChatGPT as useful reported significantly lower difficulty in writing counterspeech.

\subsection{Topic-Identity Match Significantly Influences How Users Perceive Online Hate Speech (RQ1)}

\begin{figure}[htbp]
  \centering
  \includegraphics[width=\linewidth]{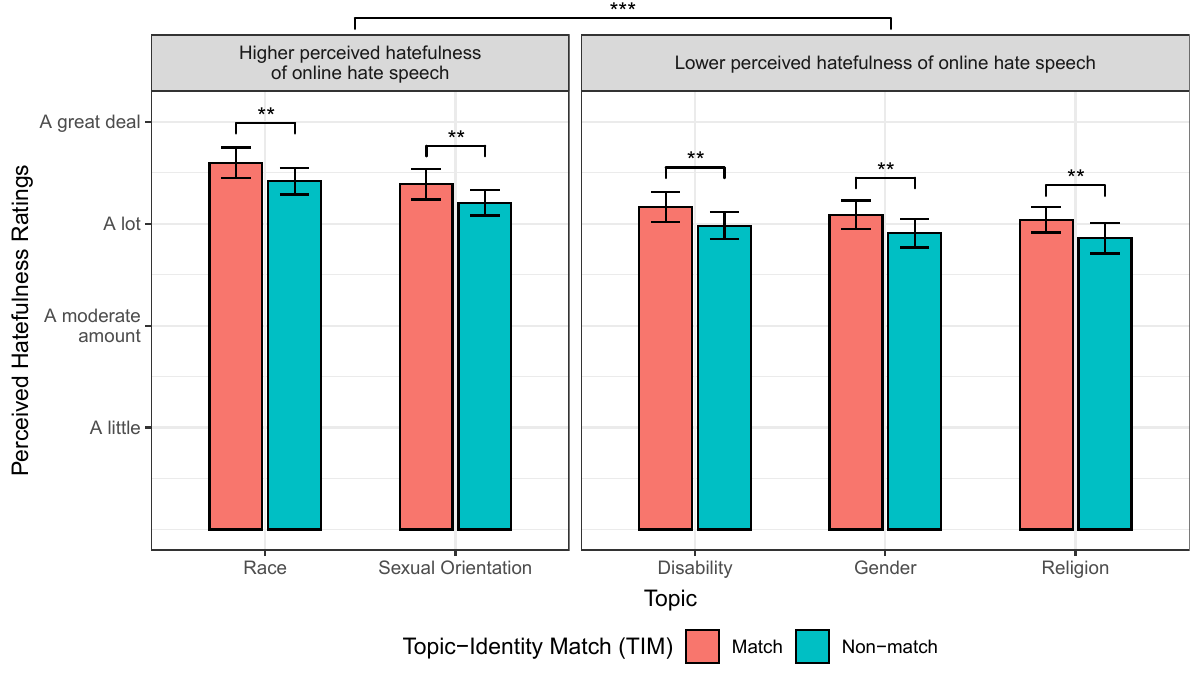}
  \caption{\rev{\textbf{Estimated Marginal Means Analysis of Perceived Hatefulness Ratings by TIM.} The figure shows the marginal mean ratings of the counterspeech writers on five topics: religion, race, gender, sexual orientation, and disability. The error bars represent the 95\% confidence intervals. Asterisks indicate levels of significance: *\(P < .05\), **\(P < .01\), and ***\(P < .001\). Across all topics, hate posts were perceived as significantly more hateful when there was a TIM compared to when there was none ($P=.001$). Hate speech targeting race and sexual orientation was perceived as significantly more hateful compared to hate speech targeting disability, gender, and religion ($P<.001$).}}
    \label{fig:topicmatch}
\end{figure}

The results of RQ1 show that \yellowhl{TIM} significantly influences the degree to which people find the online hate speech they are responding to hateful. As shown in Figure \ref{fig:topicmatch}, across all topics of hate speech, people found the hate post to be significantly more hateful when there was a \yellowhl{TIM} compared to when there was none ($b = 0.181, P = .001$). However, the level of \yellowhl{perceived hatefulness} varied among participants based on the \yellowhl{topic of the hate speech}. \rev{Post-hoc pairwise comparisons revealed that people generally perceived hate speech targeting individuals based on race and sexual orientation (Figure \ref{fig:topicmatch}, left) to be significantly more hateful compared to hate speech targeting individuals based on disability status, gender, and religion ($P < .001$, Figure \ref{fig:topicmatch}, right).} These findings suggest that users' perception of hatefulness strongly depends on both the content and their personal relevance to the topic of hate speech.



Among the control variables, the \yellowhl{frequency of encountering online hate speech} and the \yellowhl{political view} of the participants had significant effects on how participants' perceived hatefulness. Those with more liberal political views ($b = 0.095, P < .001$) and those who encountered online hate more frequently ($b = 0.094, P = .002$) generally rated the hate speech presented in the survey as more hateful compared to their conservative peers and those less exposed to online hate.

\subsection{TIM Has a Significant Impact on Subjective Experience of Writing Online Counterspeech (RQ2)}

Table \ref{tab:sde} shows the multilevel linear mixed model results for RQ2. We used arrows to indicate the significant direction (P < .05) of the effect of TIM on each variable, with \up and \down indicating positive and negative effects, respectively. 

RQ2 findings reveal that when there was a \yellowhl{TIM} for \yellowhl{race} ($b=0.262, P=.040$) and \yellowhl{disability} ($b=0.311, P=.016$) related hate speech, people were significantly more \yellowhl{satisfied} with their self-authored counterspeech compared to when the hate speech did not align with their identities for these topics. Similarly, for hate posts related to \yellowhl{religion} ($b=0.369, P=.043$) and \yellowhl{race} ($b=0.419, P=.009$), people perceived their counterspeech to be significantly more \yellowhl{effective} against the hate post they were responding to, when there was a \yellowhl{TIM} compared to when there was none. However, hate posts targeting people based on \yellowhl{gender} reversed these results: when \yellowhl{TIM} was present, people perceived significantly more \yellowhl{difficulty} in writing counterspeech against hate posts related to their gender ($b = 0.390, P = .009$), and also found their counterspeech as significantly less \yellowhl{effective} ($b = -0.684, P < .001$), compared to when \yellowhl{TIM} was absent. Female participants primarily drove this result, meaning those who identified as female generally found it significantly more difficult to respond to hate speech targeting women, and  found their counterspeech to be more ineffective compared to when they were defending other genders. 

\begin{table}[htbp]
\small
  \centering
  \caption{\textbf{The Coefficients of TIM on Satisfaction, Self-Perceived Effectiveness, and Difficulty of Counterspeech.} The table also shows the direction of the effect of TIM on each variable, using \up to indicate a positive effect and \down to indicate a negative effect (*\(P < .05\), **\(P < .01\), and ***\(P < .001\)). \rev{Our findings reveal that TIM significantly increased satisfaction for race ($P = .040$) and disability ($P = .016$) related hate speech, and self-perceived effectiveness for religion ($P = .043$) and race ($P = .009$) related hate speech. However, for gender-related hate speech, TIM significantly increased the perceived difficulty in writing counterspeech ($P = .009$) and decreased its perceived effectiveness ($P < .001$), primarily among female participants.}}  
  \begin{threeparttable}
  \sisetup{table-format = -1.3, mode=text, detect-weight=true, detect-family=true, table-space-text-post=***}
    \begin{tabular}{cllSSS}
    \toprule
    \multicolumn{3}{c}{\multirow{3}[4]{*}{Factors}} & \multicolumn{3}{c}{Perceived Experiential Aspects of Writing Online Counterspeech } \\
\cmidrule{4-6}    \multicolumn{3}{c}{}  & {Model 1} & {Model 2} & {Model 3} \\
    \multicolumn{3}{c}{}  & {DV: Satisfaction} & {DV: Effectiveness} & {DV: Difficulty}  \\
    \midrule
\multicolumn{1}{c}{\multirow{5}[2]{*}{\rotatebox{90}{\parbox{2cm}{\centering Independent\\variables}}}} & TIM\tnote{\#}   & \textit{Religion} & 0.051 & \bfseries 0.369{*\up} & -0.099 \\
      &       & \textit{Race} & \bfseries 0.262{*\up} & \bfseries 0.419{*\up} & -0.028 \\
      &       & \textit{Sexual Orientation} & 0.178 & 0.296 & \bfseries -0.418{*\down} \\
      &       & \textit{Disability} & \bfseries 0.311{*\up} & 0.068 & 0.006 \\
      &       & \textit{Gender} & -0.155 & \bfseries -0.684{***\down} & \bfseries 0.390{**\up} \\
\midrule
\multirow{7}[2]{*}{\rotatebox{90}{Control variables}} & \multicolumn{2}{l}{Frequency of encountering online hate speech} & 0.027 & \bfseries 0.120{**} & -0.020 \\
      & \multicolumn{2}{l}{Use of real name on social media} & \bfseries 0.068{***} & \bfseries 0.078{**} & -0.030 \\
      & \multicolumn{2}{l}{Social media commenting frequency} & \bfseries 0.106{**} & \bfseries 0.205{***} & -0.039 \\
      & \multicolumn{2}{l}{Age} & 0.005 & -0.007 & 0.000 \\
      & \multicolumn{2}{l}{Education level} & -0.032 & 0.035 & 0.093 \\
      & \multicolumn{2}{l}{Political view} & \bfseries -0.063{*} & \bfseries -0.131{**} & 0.013 \\
      & \multicolumn{2}{l}{Perceived hatefulness rating} & \bfseries 0.079{**} & 0.032 & \bfseries -0.084{*} \\
    \bottomrule
    \end{tabular}%
    \begin{tablenotes}
      \item[\#] Pairwise least-squares means results are presented for \yellowhl{TIM} and \yellowhl{hate post topic}.
    \end{tablenotes}
    \end{threeparttable}
  \label{tab:sde}%
\end{table}%

Moreover, among all control variables, we found that participants who \yellowhl{encountered more hate speech} ($b = 0.120, P = .013$), \yellowhl{used their real names} ($b = 0.078, P = .005$), or \yellowhl{commented more frequently on social media} ($b = 0.205, P < .001$) were more confident that their counterspeech was \yellowhl{effective}. Additionally, counterspeech \yellowhl{satisfaction} was positively related to \yellowhl{using their real names} ($b = 0.068, P = .001$) and \yellowhl{commenting frequency} ($b = .106, P = .002$), indicating that individuals who used their real names and commented more often on social media were significantly more satisfied with their counterspeech compared to those who did not. Conversely, \yellowhl{political view} had a negative impact on counterspeech \yellowhl{satisfaction} ($b = -0.063, P = .032$) and \yellowhl{self-perceived effectiveness} ($b = -0.131, P = .001$). In other words, participants who had more politically liberal views were significantly less satisfied with their own counterspeech and perceived their counterspeech to be significantly less effective than their conservative counterparts. Higher \yellowhl{perceived hatefulness rating} was associated with higher \yellowhl{satisfaction} ($b = 0.079, P = .002$) and lower \yellowhl{difficulty} ($b = -0.084, P = .010$) in writing counterspeech.

\subsection{Linguistic Characteristics of Participant Written Counterspeech is Related to Perceived Hatefulness of Online Hate Speech and Perceived Experiential Aspects of Writing Online Counterspeech (RQ3)}

Findings for RQ3 are shown in Table \ref{tab:charcs}. Our analysis for RQ3a (Model 1) shows that when participants perceive the hate speech as more hateful, the longer the \yellowhl{length} of their counterspeech responding to it ($b = 0.003, P = .008$). Also,  when participants perceive the hate speech as more hateful, they use significantly less \yellowhl{refutal}  ($b=-0.182, P=.004$) and significantly more \yellowhl{positive sentiment} ($b = 0.150, P = .003$) in their counterspeech. Given that refutal-based counterspeech strategies are typically more negative in tone as they directly challenge or contradict the hate speech, this result may account for the significant association between higher hatefulness ratings and the less frequent use of refutal strategies and more positive sentiment in their counterspeech.

\begin{table}[htbp]
\small
  \centering
  \caption{\textbf{The Coefficients of Linguistic Characteristics of Counterspeech on Perceived Hatefulness Rating, Satisfaction, Self-Perceived Effectiveness, and Difficulty.} The table shows the direction of the significant effects, using \up to indicate a positive effect and \down to indicate a negative effect (*\(P < .05\), **\(P < .01\), and ***\(P < .001\)). \rev{Our findings reveal that the length, sentiment polarity, and strategy used in counterspeech are significantly correlated with participants' perception of hate speech and their satisfaction with their own counterspeech. Length and strategy are also significantly correlated with the perceived effectiveness of counterspeech. Most notably, empathy, as a baseline, has the highest satisfaction and self-perceived effectiveness, but also the highest difficulty in writing. All these correlations are significant with $P$ values less than $.05$.}}
  \sisetup{table-format = -1.3, mode=text, detect-weight=true, detect-family=true, table-space-text-post=***}
    \begin{tabular*}{\textwidth}{@{\extracolsep{\fill}} cllSSSS}
    \toprule
    \multicolumn{3}{c}{\multirow{3}[4]{*}{\textbf{Factors}}} & \multicolumn{1}{m{8em}}{\centering \textbf{Perceived Hatefulness of Online Hate Speech (RQ3a)}} & \multicolumn{3}{m{18em}}{\centering  \textbf{Perceived Experiential Aspects of Writing Online Counterspeech (RQ3b)}} \\
\cmidrule{4-7}    \multicolumn{3}{c}{}  & {Model 1} & {Model 2} & {Model 3} & {Model 4} \\
    \multicolumn{3}{c}{}  & \parbox{8em}{\centering DV: Hatefulness Rating} & \parbox{6em}{\centering DV: Satisfaction} & \parbox{6em}{\centering DV: Effectiveness} & \parbox{5em}{\centering DV: Difficulty} \\
    \midrule
    \multicolumn{1}{c}{\multirow{9}[1]{*}{\rotatebox{90}{\parbox{11em}{\centering Linguistic Characteristics of Participant Written Counterspeech}}}} & Strategy & \textit{Empathy (baseline)} & \multicolumn{4}{c}{\textit{Baseline}} \\
      &       & \textit{Humor} & 0.070 & \bfseries -0.160{*\down} & \bfseries -0.205{**\down} & \bfseries -0.226{*\down} \\
      &       & \textit{Warning of Consequence} & 0.155 & \bfseries -0.191{**\down} & 0.057 & \bfseries -0.217{*\down} \\
      &       & \textit{Refutal} & \bfseries -0.182{**\down} & \bfseries -0.134{*\down} & \bfseries -0.114{*\down} & -0.085 \\
      &       & \textit{Other} & -0.404 & -0.144 & -0.235 & -0.347 \\
      & \multicolumn{2}{l}{Use of first-person language} & -0.076 & -0.058 & -0.093 & 0.131 \\
      & \multicolumn{2}{l}{Use of questions} & 0.023 & -0.081 & 0.107 & 0.032 \\
      & \multicolumn{2}{l}{Length} & \bfseries 0.003{**\up} & \bfseries 0.005{***\up} & \bfseries 0.005{***\up} & -0.006 \\
      & \multicolumn{2}{l}{Sentiment polarity} & \bfseries 0.150{**\up} & \bfseries 0.169{**\up} & -0.004 & -0.036 \\
      \midrule
\multirow{12}[1]{*}{\rotatebox{90}{Control variables}} & \multicolumn{2}{l@{\hspace{-6em}}}{Frequency of encountering online hate speech} & \bfseries 0.092{**} & 0.028 & -0.037 & \bfseries 0.121{*} \\
      & \multicolumn{2}{l}{Use of real name on social media} & 0.027 & \bfseries 0.063{**} & \bfseries 0.082{**} & -0.027 \\
      & \multicolumn{2}{l}{Social media commenting frequency} & -0.017 & \bfseries 0.076{*} & \bfseries 0.196{***} & -0.032 \\
      & \multicolumn{2}{l}{Age} & 0.004 & 0.004 & \bfseries -0.008{*} & 0.000 \\
      & \multicolumn{2}{l}{Education level} & -0.029 & -0.045 & 0.031 & 0.095 \\
      & \multicolumn{2}{l}{Political view} & \bfseries 0.096{***} & -0.044 & \bfseries -0.130{**} & 0.004 \\
      & Hate post topic & \textit{Religion (baseline)} & \multicolumn{4}{c}{\textit{Baseline}} \\
      &       & \textit{Race} & \bfseries 0.527{***} & 0.094 & 0.014 & -0.103 \\
      &       & \textit{Sexual Orientation} & 0.055 & \bfseries 0.149{*} & 0.016 & -0.176 \\
      &       & \textit{Disability} & \bfseries 0.347{***} & 0.098 & -0.079 & -0.131 \\
      &       & \textit{Gender} & 0.094 & \bfseries 0.204{**} & \bfseries 0.234{**} & \bfseries -0.184{*} \\
      & \multicolumn{2}{l}{Topic-Identity Match (TIM)} & \bfseries 0.195{**} & \bfseries 0.051{*} & -0.043 & -0.066 \\
    \bottomrule
    \end{tabular*}%
  \label{tab:charcs}%
\end{table}%

The results of RQ3b are presented in Models 2, 3, and 4, which reveal that various linguistic characteristics and strategies of counterspeech are significantly associated with aspects related to the participant's experience of writing counterspeech. We used an empathy-based tone as the baseline to test the effects in the LMMs. 

\begin{itemize}
    \item \textbf{\yellowhl{Satisfaction}} (Model 2): Higher \yellowhl{satisfaction} towards one's own counterspeech is significantly associated with more frequent use of empathy-based counterspeech compared to \yellowhl{humor} ($b=-0.160, P=.027$), \yellowhl{warning of consequence} ($b=-0.191, P=.009$), and \yellowhl{refutal} ($b=-0.134, P=.016$) strategies. People who are more \yellowhl{satisfied} with their counterspeech also tend to write significantly \yellowhl{longer} ($b = 0.005, P < .001$) and more \yellowhl{positive} counterspeech ($b = 0.169, P = .001$).
    \item \textbf{\yellowhl{Effectiveness}} (Model 3): Higher \yellowhl{self-perceived effectiveness} towards one's own counterspeech is significantly associated with more frequent use of empathy-based counterspeech compared to \yellowhl{humor} ($b=-0.205, P=.006$) and \yellowhl{refutal} ($b=-0.114, P=.044$). People who perceive their counterspeech to be more \yellowhl{effective} also tend to write significantly \yellowhl{longer} ($b=0.005, P<.001$) counterspeech compared to those who write shorter counterspeech.
    \item \textbf{\yellowhl{Difficulty}} (Model 4): Although \yellowhl{empathy-based} counterspeech is positively related to higher \yellowhl{satisfaction} and \yellowhl{self-perceived effectiveness}, compared to other counterspeech such as \yellowhl{humor} ($b=-0.226, P=.018$) or \yellowhl{refutal} ($b=-0.217, P=.028$) participants find it significantly more \yellowhl{difficult} to write empathy-based counterspeech.
\end{itemize}

\subsection{\rev{Exploratory Analysis: Association between Prior Use and Perceived Usefulness of AI-writing Assistants Like ChatGPT and Lower Difficulty of Writing Counterspeech}}

\begin{figure}[htbp]
  \centering
  \includegraphics[width=\linewidth]{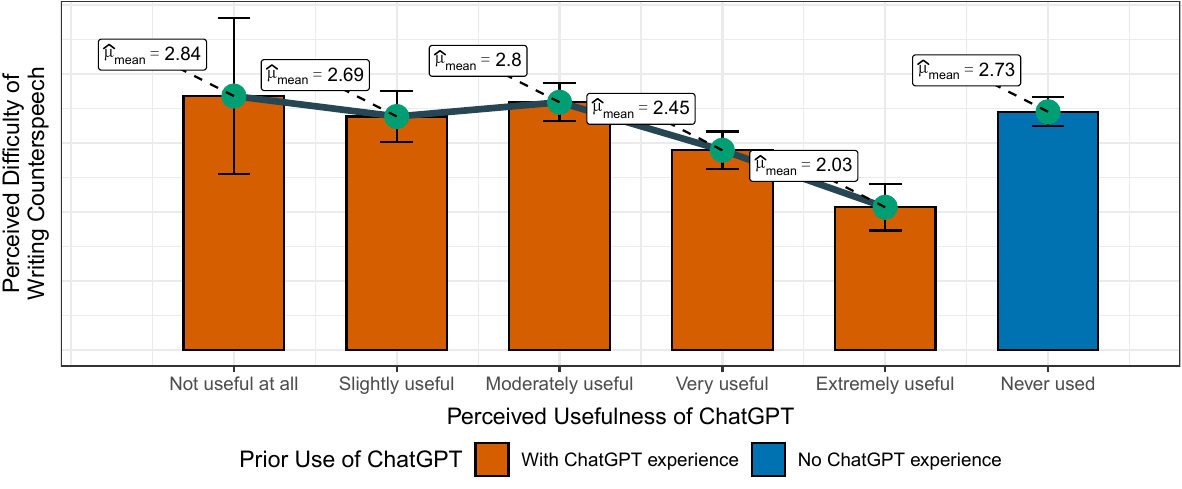}
  \caption{\textbf{Perceived Difficulty of Writing Counterspeech by ChatGPT Usage Experience.} We compare the perceived difficulty of writing counterspeech between participants with and without prior ChatGPT experience. The orange bars represent the prior use of ChatGPT group, and the blue bars represent the no ChatGPT experience group. \rev{Asterisks indicate levels of significance: *\(P < .05\), **\(P < .01\), and ***\(P < .001\).} The bars are labeled with the mean values of the perceived difficulty with 95\% confidence intervals. \rever{Analysis reveals two significant relationships: (1) Participants with prior ChatGPT experience report lower difficulty in writing counterspeech compared to those without experience ($P = .039$), and (2) Among participants who rate ChatGPT as moderately to extremely useful, those who perceive it as more useful report significantly lower difficulty in writing counterspeech ($P < .001$).}}
    \label{fig:chatgpt}
\end{figure}

The results are shown in Figure \ref{fig:chatgpt}. While our study does not confirm whether participants used AI tools like ChatGPT to craft their counterspeech in our survey, our results indicate that prior experience with, or awareness of, such AI-writing assistants could influence participants' perceptions of the difficulty associated with responding to the hate speech. Participants who had \yellowhl{prior use of ChatGPT} reported significantly lower \yellowhl{difficulty} in writing counterspeech ($b = -0.186, P = .039$). We also tested the relationship between their \yellowhl{perceived usefulness of ChatGPT} and \yellowhl{difficulty}, and found a significant negative relationship ($b = -0.218, P < .001$). Participants who perceived ChatGPT as more useful also reported lower difficulty in writing counterspeech. 



\section{Discussions}

\subsection{The Role of TIM, Social Distance, and the Severity of Online Hate Speech Targeting Minorities}

Participants rated hate targeting race and sexual orientation as the most serious forms of hate from the participants (RQ1). These forms of hate likely receive such serious ratings because they target historically marginalized and vulnerable groups. Indeed, considerable scholarship documents the long hegemony of white, heterosexual males and how racial minorities and members of the LGBTQ+ community have been frequent targets of hate \cite{gledhillQueeringStateCrime2014, perryNameHateUnderstanding2001}. Despite legislative attempts to protect these groups, they nevertheless remain among the most vulnerable in society as evidenced by the fact they are the most likely to experience hate crimes in America \footnote{https://www.justice.gov/hatecrimes/hate-crime-statistics}.

Assuming that those perpetrating hate towards racial minorities and the LGBTQ+ community are members of the dominant white, heteronormative group, the social distance between the attacker(s) and the targeted group would be considerable. That is, there is likely to be little overlap between the attacking group (largely comprised of white, heterosexual males) and the groups being attacked. This social distance between the attacker and the target would be far greater when the hate is based on race and sexual orientation than when it targets groups based on disability status, gender, or religion. In these latter instances, some of the members of the targeted group would also belong to the white, heteronormative hegemonic group and others would likely have strong ties to them. This intersection of statuses across the attacking group and the targeted group is likely to create allies across the groups and therefore soften the animosity toward the targeted group. It is therefore predictable that the perceived seriousness of the attack is greater when the hate is based on race and sexual orientation because, in general, the greater the social distance between two parties involved in a conflict, the more serious the conflict is likely to be \cite{blackMoralTime2011, cooneyDeathFamilyHonor2014}. This general proposition, supported by a broad range of literature, indicates that racial minorities \cite{kahlStudentReactionsPublic2013, nirPerceivedThreatBlaming2018}, immigrants \cite{johnsonSentencingHomicideOffenders2010}, women \cite{richardsExplainingFemaleVictim2016}, or other vulnerable populations \cite{callanJusticeMotiveEffects2012} are the victims of more severe offenses.

Another factor that influences the perception of hate speech is the TIM between the individual's identity and the target of the hate speech. We found that across all topics, people perceived hate speech to be significantly more offensive when there was a TIM (RQ1). This suggests that people are more sensitive to online hate speech that affects their own identity or group membership, and more likely to perceive it as a serious threat. This finding also supports the notion that social distance plays a role in the severity of hate speech, as people who have a TIM with the target group are likely to have less social distance than those who do not. 

Consequently, reducing social distance through technology-mediated communication could be a key factor in mitigating the impact of hate speech \cite{curranSocialDistancingSocial2023}. In this context, CSCW research can explore the role of technology in bridging social distance, potentially mitigating online hate speech. For instance, designing more intimate and inclusive platforms could reshape social interactions, fostering understanding and reducing prejudices, as suggested by Hutson et al. \cite{hutsonDebiasingDesireAddressing2018}. Intimate platform design, which refers to the design of online communities that enable users to form connections with others, can further apply to social media platforms to encourage interactions between diverse social groups, because it provides access to people outside one's usual social circle, thereby offering opportunities for interaction across different racial, ethnic, and social backgrounds. Social media can offer new opportunities for minority cultures to express and promote their cultural identities. For instance, a study on Snapchat \cite{callahanSnapchatUsageMinority2019} revealed that users could use the platform to reinforce and enact their minority values and practices, which differed from the dominant norms of social media use. By providing them with alternative spaces and modes of communication that challenge the mainstream media representations and stereotypes of their cultures \cite{johnsonMinorityCulturesSocial2013}, new media have the ability to extend cultural perspectives of minority cultures that have historically been marginalized by geographical disadvantages. 

Of course, having groups interact across social statuses could also reinforce their differences, increase the social distance between them, and actually elevate intergroup hatred \cite{paoliniNegativeIntergroupContact2010}. While the contact hypothesis — which postulates that the greater the contact between majority and minority groups, the less prejudice is expressed between them \cite{allportNaturePrejudice1954} — has generally been supported \cite{pettigrewMetaanalyticTestIntergroup2006}, the relationship between intergroup contact and intergroup prejudice is complex and potentially spurious \cite{bertrandChapterFieldExperiments2017, paluckContactHypothesisReevaluated2019}. At the very least, there are specific conditions that are more conducive to decreasing intergroup prejudices, and designers are advised to consider these factors carefully when designing platforms that are created to promote intergroup contact. For example, by incorporating search and filter tools, exposure- and empathy-promoting algorithms, and community policies, this design has the potential to address the social distance that contributes to the prevalence of hate speech targeting racial minorities and LGBTQ+ individuals.

\subsection{Gender Hate Speech and the Role of Supportive Online Communities}
Another interesting finding is that while TIM generally increased satisfaction with and perceived effectiveness of counterspeech, this pattern did not hold when the hate was gender-related (RQ2). In this case, where female participants were writing a counterspeech against hateful attacks targeting women, they reported that composing the counterspeech was both more difficult and they perceived their counterspeech to be less effective. This finding could also be a result of overlapping and intersecting statuses among the perpetrators and targets of hate. That is, if we assume the person generating hate that targets women identifies as a male, the woman trying to compose the counterspeech most likely shares multiple group memberships with men. Moreover, many of these relationships may be very intimate, such as the person’s husband, boyfriend, father, or son. It is far less likely that individuals constructing counterspeeches defending their group would have as many or as intensely intimate relationships with the person or group that most likely composed the hate. For example, given ongoing issues of segregation in housing, employment, and educational experiences in the US \cite{masseyHypersegregationMetropolitanAreas1989, masseyStillLinchpinSegregation2020}, it is less likely that a Black respondent creating counterspeech against race-based hate would have as many or as intimate relationships with whites as a female respondent has with men. The relationships women likely share with individuals who are similar to the members of the group most likely attacking them would decrease the social distance between the attacking group and the targeted group. As the social distance between the attacker and the target decreases, it becomes more difficult to be confrontational and enact more punitive-oriented forms of conflict management \cite{blackElementaryFormsConflict1990}. \rever{Additionally, stereotype threat theory \cite{vonhippelStereotypeThreatFemale2011} suggests that women may fear their speech could confirm negative stereotypes about women's communication abilities. This added psychological pressure can impact women when engaging in counterspeech, especially given that patriarchal social structures systematically undermine women's linguistic authority. As linguistic power theory \cite{lakoffLanguageWomansPlace1973} suggests, identical arguments may be perceived as less convincing when voiced by women, which could further contribute to their challenges in constructing effective counterspeech.} Consequently, women likely struggle to construct counterspeech that they believe to be condemning enough of the hate, and this lack of confidence that they have been sufficiently confrontational would likely decrease their belief that their counterspeech would be effective.

Researchers in CSCW have advocated the use of the Social Identity Perspective (SIP) as a theoretical lens to understand how individual user behavior is intricately tied to their group identity \cite{seeringApplicationsSocialIdentity2018}. According to SIP, people categorize themselves into various groups, and this categorization significantly influences how they interact with members within their own group and those from other groups \cite{abramsNumericalDistinctivenessSocial1990}. Such categorizations can lead to variations in levels of attachment and identification with a group, which are key predictors of both intra-group and intergroup conflict, as well as the strategies groups adopt in response to such conflicts \cite{seeringApplicationsSocialIdentity2018}.

Extending this perspective to the context of our finding, supportive online communities for women can be instrumental in fostering a collective identity, empowerment, and resilience when engaging in online counterspeech against misogynistic hate. Such communities, for example, can facilitate collaborative counterspeech engagement \cite{buergerIamhereCollectiveCounterspeech2021}, by allowing female users to exchange ideas, learn from each other’s experiences, and develop more effective responses to online hate. This collaborative approach may not only enhance the quality of online counterspeech but also mitigate the emotional toll of confronting online hate in isolation, as members can rely on mutual support and understanding \cite{buergerIamhereCollectiveCounterspeech2021}. Additionally, these communities can foster a sense of solidarity and resilience among members, reducing the sense of isolation and vulnerability women often experience when engaging in online discourses \cite{reidFeelingGoodControl2022}.

\subsection{AI-Mediated Writing Tools: A Solution for the Challenge of Empathetic Counterspeech}

Our study found that the use of empathy-based counterspeech was significantly associated with higher satisfaction and self-perceived effectiveness towards one's counterspeech, but also with greater difficulty in writing it (RQ3). This complements the findings of Hangartner et al. (2021), who demonstrated the efficacy of empathy in reducing xenophobic hate speech on Twitter \cite{hangartnerEmpathybasedCounterspeechCan2021}. According to a field experiment by Broockman et al. (2016), brief conversations that encourage people to take the perspective of others actively can significantly and durably reduce prejudice toward marginalized groups, such as transgender people \cite{broockmanDurablyReducingTransphobia2016}. Empathetic language not only fosters more constructive dialogue in online communities but also leads to more perspective-taking among users \cite{rhoClassConfessionsRestorative2017, rhoFosteringCivilDiscourse2018}. However, RQ3 findings show that for empathy-based counterspeech,  while participants' self-perceived effectiveness and satisfaction were higher compared to other counterspeech strategies, so was their difficulty in writing it. Writing empathy-based counterspeech can be challenging because it requires a deep understanding of the emotions and perspectives of others \cite{paluckPrejudiceReductionWhat2009}. Crafting an empathetic response often involves recognizing the feelings behind the hate speech while also challenging its harmful narrative. This demands a careful choice of words to avoid escalating the situation or inadvertently endorsing the negative sentiments \cite{buergerIamhereCollectiveCounterspeech2021}. The mental and emotional labor involved in this process may vary depending on the presence versus absence of a TIM. People who do versus do not identify with a specific topic of hate speech may differ in how they perceive what is considered an empathetic counterspeech. Thus, while effective, crafting empathetic counterspeech can be complex.

\rev{Moreover, our study also found that longer counterspeech and more positive sentiment were associated with higher perceived hatefulness of the original content, as well as greater satisfaction and self-perceived effectiveness of the counterspeech. The findings imply a tendency in self-satisfying counterspeech composition: individuals tend to use longer, more positive, and less refutational language. Counterspeech can potentially deter hate speech by stimulating more conversation \cite{schiebGoverningHateSpeech2016, buergerIamhereCollectiveCounterspeech2021}. However, there is a lack of research specifically addressing how counterspeech can stimulate more conversation. Given that counterspeech operates within the broader ecosystem of online communication \cite{garlandImpactDynamicsHate2022}, we draw upon studies on general online content consumption to explore the potential effects of longer, more positive counterspeech. When it comes to attracting engagement with online content like news, Robertson et al. (2023) and Gligoric et al. (2023) both found that longer content and negative language tend to increase click-through rates and engagement \cite{gligoricLinguisticEffectsNews2023, robertsonNegativityDrivesOnline2023}. However, the dynamics on social media platforms can be different, where overly negative content may not necessarily lead to increased engagement. For example, Saveski et al. (2022) found that, particularly among politically diverse audiences, positive sentiment and neutral, fact-based language tend to engage more users in online content consumption \cite{saveskiEngagingPoliticallyDiverse2022}. Furthermore, Gligoric et al. (2019) discovered that shorter tweets, with 10-20\% of the original length removed, are more likely to engage users compared to longer, unedited tweets \cite{gligoricCausalEffectsBrevity2019}. Such strategies, like preserving essential information and emotions while omitting filler words, could potentially be applied in other contexts. In CSCW and CHI work, researchers have examined how linguistic patterns and tone can impact online discourse and engagement \cite{rhoHashtagBurnoutControl2019, rhoPoliticalHashtagsLost2020}. For example, Rho et al. (2020) found that the presence of political hashtags in news posts was associated with more angry, fearful, and disgusted language in comments, as well as more black-and-white rhetoric. This finding suggests that certain linguistic features can increase engagement but potentially at the cost of more toxic and polarized discussions. However, these researchers also emphasized that preserving the original emotional tone of the message, whether positive or negative, is crucial for these concise versions of tweets to effectively engage users \cite{gligoricCausalEffectsBrevity2019}. In summary, to attract more engagement, counterspeech must strike a delicate balance: providing sufficient context and information, reducing length for brevity, and maintaining the original sentiment. Crafting such messages is a challenging task \cite{petersenTextSimplificationLanguage2007}, as it requires careful consideration of the content's complexity, the audience's attention span, and the potential emotional impact of the message.}

\rev{The findings from our exploratory analysis suggest that one possible way to address the challenge of writing counterspeech may be the incorporation of AI-powered writing assistance. Given effective counterspeech involves balancing length, sentiment, and complexity while also being empathetic, AI-powered writing assistance could substantial lessen the burden the counterspeaker must face. In the exploratory analysis, we found that individuals who had prior experience with AI writing tools like ChatGPT reported less difficulty in writing counterspeech, especially if they perceived AI writing tools as more useful. Our exploratory analysis results are consistent with the findings of Mun et al. (2024), who found that some participants were interested in AI tools that could provide support in formulating effective responses to hate speech, such as through collaborative writing, fact suggestions, and tone/style revisions \cite{munCounterspeakersPerspectivesUnveiling2024}. This suggests that these participants believed AI assistance could potentially make the process of writing counterspeech easier for them.} AI-mediated counterspeech writing assistant may facilitate the process of empathetic counterspeech writing by providing suggestions \cite{lehmannSuggestionListsVs2022} or templates \cite{sahaSelfsupervisionControllingTechniques2023} that can help users express their thoughts and feelings more effectively. AI-mediated counterspeech writing assistant can also reduce the cognitive load and emotional stress of users by tailoring the message to the specific context and audience of the hate speech \cite{chungKnowledgeGroundedCounterNarrative2021}. The Hyperpersonal model of computer-mediated communication can support this point. The communication between counterspeech and hate speech on social media is essentially a form of computer-mediated communication (CMC), which refers to any human communication that occurs through the use of digital tools \cite{whitneyComputerMediatedCommunication1998, baughanSomeoneWrongInternet2021}. The Hyperpersonal model theory suggests that users in CMC rely on linguistic cues to form impressions of their communication partners \cite{knappSAGEHandbookInterpersonal2011}, which can lead to more socially desirable and intimate communication than face-to-face interactions\cite{liuWillAIConsole2022}. These impressions are based on how users present themselves and how they imagine their communication partners to be \cite{liuWillAIConsole2022}. According to the Hyperpersonal model, an AI-mediated counterspeech writing assistant can enhance the user’s self-presentation skills, enabling them to create a more empathetic and persuasive impression of themselves in their counterspeech. Such an assistant can also help the hate speech poster see the counter-speakers in a more positive and empathetic light, by offering suggestions that can help the hate speech poster to understand their point of view and feelings \cite{westermanIItIThouIRobot2020}. Therefore, we propose that AI-mediated writing tools can be a valuable solution for the challenge of empathetic counterspeech, as they can lower the barriers and increase the benefits of engaging in this form of online civic action.

\section{Limitations and Future Work}

First, the data were collected through self-report surveys, which can be subject to biases like social desirability. The perception of hate speech and experiences of writing counterspeech may not fully reflect participants’ actual attitudes and behaviors in daily social media usage. Second, our sample of participants, although demographically diverse, was limited to English speakers in the U.S. recruited through Prolific. Thus, the findings may not generalize to other populations and contexts. Cross-cultural examinations of counterspeech are needed. \rev{Moreover, as an exploratory correlational analysis, causal conclusions cannot be drawn regarding the effects of Topic-Identity Match (TIM) and AI tools on counterspeech writing experiences.} Experimental and longitudinal approaches for assessing these relationships over time would elucidate directionality and causality. \rev{Finally, our study primarily employed quantitative methods, and thus lacks the depth of understanding that qualitative analysis can provide. Future work should incorporate qualitative methods to gain richer insights into the experiences and perceptions of individuals when writing counterspeech, and how AI tools might aid in this process.}

\rev{A point worth considering for future work is the role of perceived effectiveness as a proxy for actual effectiveness. If the goal is to encourage participants to engage in counterspeech, understanding their perception of the effectiveness of their self-authored counterspeech is crucial as it influences their motivation to participate \cite{buergerIamhereCollectiveCounterspeech2021}. The decision to engage in a particular behavior is often driven by the belief that this behavior will lead to a certain outcome \cite{matsumoriDecisionTheoreticModelBehavior2019}. The belief in the effectiveness of one's counterspeech may not always align with the actual outcome \cite{cepollaroCounterspeech2023}, but it is this perception of effectiveness that triggers the initial behavior of engaging in counterspeech. In essence, beliefs have real consequences: if individuals believe their counterspeech will be effective, they are more likely to engage in it, regardless of its actual effectiveness. Given this, the relationship between perceived and actual effectiveness of counterspeech warrants further exploration in future research, as understanding this link could provide valuable insights into how to motivate more effective counterspeech.}

\rev{Another noteworthy point is that our study specifically focuses on public counterspeech. Wright et al. (2017) include both public (one-to-many) and presumably private (one-to-one) exchanges in their categorization of counterspeech, and consider both as valid forms of counterspeech \cite{wrightVectorsCounterspeechTwitter2017}. As noted in many online contexts, counterspeakers cannot know in advance who their actual audience will be \cite{baughanSomeoneWrongInternet2021, cepollaroCounterspeech2023, wrightVectorsCounterspeechTwitter2017}. Public counterspeech, while offering a chance to influence more people and leverage community moderation, can also lead to potential risks such as incurring heavy individual costs unpredictably, including becoming the transient target of an online mob \cite{wrightVectorsCounterspeechTwitter2017, cepollaroCounterspeech2023}. On the other hand, private counterspeech facilitates more vulnerable, authentic dialogue with less risk of the argument spiraling out of control, but limits the potential for wider influence \cite{baughanSomeoneWrongInternet2021}. Therefore, the impact of public and private counterspeech may manifest differently, and future research could benefit from exploring these differences.}


\section{Conclusions}

Our study offered important insights into the factors associated with individuals’ experiences in writing online counterspeech. We found that the Topic-Identity Match (TIM) between the hate speech and counter-speakers’ social identities influenced their perception of hate posts as well as their satisfaction, difficulty, and self-efficacy with counterspeech. The linguistic characteristics of the counterspeech, specifically strategy, length, or sentiment polarity, are also related to the counter-speakers’ perceptions and writing experiences. Additionally, prior experience with and openness toward AI writing assistance tools like ChatGPT correlated with lower perceived difficulty composing counterspeech. \rev{These findings and the theoretical explanation of them carry implications for the design of counterspeech campaigns, online moderation policies, and writing technologies.} Overall, generating impactful yet non-inflammatory counterspeech poses challenges that warrant continued research across computer-supported cooperative work, social computing, and human-AI interaction domains. Progress necessitates a nuanced understanding of the multidimensional individual, contextual, and expressive factors intersecting within counterspeech writing.


\bibliographystyle{ACM-Reference-Format}
\bibliography{Citations/CROSS-1.5, Citations/20240705, Citations/20241028}


\begin{thebibliography}{137}


\ifx \showCODEN    \undefined \def \showCODEN     #1{\unskip}     \fi
\ifx \showDOI      \undefined \def \showDOI       #1{#1}\fi
\ifx \showISBNx    \undefined \def \showISBNx     #1{\unskip}     \fi
\ifx \showISBNxiii \undefined \def \showISBNxiii  #1{\unskip}     \fi
\ifx \showISSN     \undefined \def \showISSN      #1{\unskip}     \fi
\ifx \showLCCN     \undefined \def \showLCCN      #1{\unskip}     \fi
\ifx \shownote     \undefined \def \shownote      #1{#1}          \fi
\ifx \showarticletitle \undefined \def \showarticletitle #1{#1}   \fi
\ifx \showURL      \undefined \def \showURL       {\relax}        \fi
\providecommand\bibfield[2]{#2}
\providecommand\bibinfo[2]{#2}
\providecommand\natexlab[1]{#1}
\providecommand\showeprint[2][]{arXiv:#2}

\bibitem[Nex(2023)]%
        {NextdoorIntegratingGenerative2023}
 \bibinfo{year}{2023}\natexlab{}.
\newblock \bibinfo{title}{Nextdoor {{Is Integrating Generative AI}} to {{Drive Engaging}} and {{Kind Conversations}} in the {{Neighborhood}}}.
\newblock \bibinfo{howpublished}{https://finance.yahoo.com/news/nextdoor-integrating-generative-ai-drive-103000201.html}.
\newblock


\bibitem[Wha(2023)]%
        {WhatNewOur2023}
 \bibinfo{year}{2023}\natexlab{}.
\newblock \bibinfo{title}{What's {{New Across Our AI Experiences}}}.
\newblock \bibinfo{howpublished}{https://about.fb.com/news/2023/12/meta-ai-updates/}.
\newblock


\bibitem[Abrams et~al\mbox{.}(1990)]%
        {abramsNumericalDistinctivenessSocial1990}
\bibfield{author}{\bibinfo{person}{Dominic Abrams}, \bibinfo{person}{Joanne Thomas}, {and} \bibinfo{person}{Michael~A. Hogg}.} \bibinfo{year}{1990}\natexlab{}.
\newblock \showarticletitle{Numerical Distinctiveness, Social Identity and Gender Salience}.
\newblock \bibinfo{journal}{\emph{British Journal of Social Psychology}} \bibinfo{volume}{29}, \bibinfo{number}{1} (\bibinfo{year}{1990}), \bibinfo{pages}{87--92}.
\newblock
\showISSN{2044-8309}
\urldef\tempurl%
\url{https://doi.org/10.1111/j.2044-8309.1990.tb00889.x}
\showDOI{\tempurl}


\bibitem[Allport et~al\mbox{.}(1954)]%
        {allportNaturePrejudice1954}
\bibfield{author}{\bibinfo{person}{Gordon~Willard Allport}, \bibinfo{person}{Kenneth Clark}, {and} \bibinfo{person}{Thomas Pettigrew}.} \bibinfo{year}{1954}\natexlab{}.
\newblock \showarticletitle{The Nature of Prejudice}.
\newblock  (\bibinfo{year}{1954}).
\newblock


\bibitem[Baider(2023)]%
        {baiderAccountabilityIssuesOnline2023}
\bibfield{author}{\bibinfo{person}{Fabienne Baider}.} \bibinfo{year}{2023}\natexlab{}.
\newblock \showarticletitle{Accountability {{Issues}}, {{Online Covert Hate Speech}}, and the {{Efficacy}} of {{Counter}}-{{Speech}}}.
\newblock \bibinfo{journal}{\emph{Politics and Governance}} \bibinfo{volume}{11}, \bibinfo{number}{2} (\bibinfo{date}{May} \bibinfo{year}{2023}), \bibinfo{pages}{249--260}.
\newblock
\showISSN{2183-2463}
\urldef\tempurl%
\url{https://doi.org/10.17645/pag.v11i2.6465}
\showDOI{\tempurl}


\bibitem[Baughan et~al\mbox{.}(2021)]%
        {baughanSomeoneWrongInternet2021}
\bibfield{author}{\bibinfo{person}{Amanda Baughan}, \bibinfo{person}{Justin Petelka}, \bibinfo{person}{Catherine~Jaekyung Yoo}, \bibinfo{person}{Jack Lo}, \bibinfo{person}{Shiyue Wang}, \bibinfo{person}{Amulya Paramasivam}, \bibinfo{person}{Ashley Zhou}, {and} \bibinfo{person}{Alexis Hiniker}.} \bibinfo{year}{2021}\natexlab{}.
\newblock \showarticletitle{Someone {{Is Wrong}} on the {{Internet}}: {{Having Hard Conversations}} in {{Online Spaces}}}.
\newblock \bibinfo{journal}{\emph{Proceedings of the ACM on Human-Computer Interaction}} \bibinfo{volume}{5}, \bibinfo{number}{CSCW1} (\bibinfo{date}{April} \bibinfo{year}{2021}), \bibinfo{pages}{156:1--156:22}.
\newblock
\urldef\tempurl%
\url{https://doi.org/10.1145/3449230}
\showDOI{\tempurl}


\bibitem[Benesch(2014)]%
        {beneschDefiningDiminishingHate2014}
\bibfield{author}{\bibinfo{person}{Susan Benesch}.} \bibinfo{year}{2014}\natexlab{}.
\newblock \showarticletitle{Defining and Diminishing Hate Speech}.
\newblock \bibinfo{journal}{\emph{State of the world's minorities and indigenous peoples}}  \bibinfo{volume}{2014} (\bibinfo{year}{2014}), \bibinfo{pages}{18--25}.
\newblock


\bibitem[Bertrand and Duflo(2017)]%
        {bertrandChapterFieldExperiments2017}
\bibfield{author}{\bibinfo{person}{M. Bertrand} {and} \bibinfo{person}{E. Duflo}.} \bibinfo{year}{2017}\natexlab{}.
\newblock \showarticletitle{Chapter 8 - {{Field Experiments}} on {{Discrimination}}}.
\newblock In \bibinfo{booktitle}{\emph{Handbook of {{Economic Field Experiments}}}}, \bibfield{editor}{\bibinfo{person}{Abhijit~Vinayak Banerjee} {and} \bibinfo{person}{Esther Duflo}} (Eds.). \bibinfo{series}{Handbook of {{Field Experiments}}}, Vol.~\bibinfo{volume}{1}. \bibinfo{publisher}{North-Holland}, \bibinfo{pages}{309--393}.
\newblock
\urldef\tempurl%
\url{https://doi.org/10.1016/bs.hefe.2016.08.004}
\showDOI{\tempurl}


\bibitem[Bilewicz et~al\mbox{.}(2021)]%
        {bilewiczArtificialIntelligenceHate2021}
\bibfield{author}{\bibinfo{person}{Micha{\l} Bilewicz}, \bibinfo{person}{Patrycja Tempska}, \bibinfo{person}{Gniewosz Leliwa}, \bibinfo{person}{Maria Dowgia{\l}{\l}o}, \bibinfo{person}{Michalina Ta{\'n}ska}, \bibinfo{person}{Rafa{\l} Urbaniak}, {and} \bibinfo{person}{Micha{\l} Wroczy{\'n}ski}.} \bibinfo{year}{2021}\natexlab{}.
\newblock \showarticletitle{Artificial Intelligence against Hate: {{Intervention}} Reducing Verbal Aggression in the Social Network Environment}.
\newblock \bibinfo{journal}{\emph{Aggressive Behavior}} \bibinfo{volume}{47}, \bibinfo{number}{3} (\bibinfo{year}{2021}), \bibinfo{pages}{260--266}.
\newblock
\showISSN{1098-2337}
\urldef\tempurl%
\url{https://doi.org/10.1002/ab.21948}
\showDOI{\tempurl}


\bibitem[Black(1990)]%
        {blackElementaryFormsConflict1990}
\bibfield{author}{\bibinfo{person}{Donald Black}.} \bibinfo{year}{1990}\natexlab{}.
\newblock \showarticletitle{The {{Elementary Forms}} of {{Conflict Management}}}.
\newblock In \bibinfo{booktitle}{\emph{New {{Directions}} in the {{Study}} of {{Justice}}, {{Law}}, and {{Social Control}}}}. \bibinfo{publisher}{Springer US}, \bibinfo{address}{Boston, MA}, \bibinfo{pages}{43--69}.
\newblock
\showISBNx{978-1-4899-3608-0}
\urldef\tempurl%
\url{https://doi.org/10.1007/978-1-4899-3608-0_3}
\showDOI{\tempurl}


\bibitem[Black(2011)]%
        {blackMoralTime2011}
\bibfield{author}{\bibinfo{person}{Donald Black}.} \bibinfo{year}{2011}\natexlab{}.
\newblock \bibinfo{booktitle}{\emph{Moral {{Time}}}}.
\newblock \bibinfo{publisher}{Oxford University Press}.
\newblock
\showISBNx{978-0-19-983160-9}


\bibitem[Blackwell et~al\mbox{.}(2017)]%
        {blackwellClassificationItsConsequences2017}
\bibfield{author}{\bibinfo{person}{Lindsay Blackwell}, \bibinfo{person}{Jill Dimond}, \bibinfo{person}{Sarita Schoenebeck}, {and} \bibinfo{person}{Cliff Lampe}.} \bibinfo{year}{2017}\natexlab{}.
\newblock \showarticletitle{Classification and {{Its Consequences}} for {{Online Harassment}}: {{Design Insights}} from {{HeartMob}}}.
\newblock \bibinfo{journal}{\emph{Proceedings of the ACM on Human-Computer Interaction}} \bibinfo{volume}{1}, \bibinfo{number}{CSCW} (\bibinfo{date}{Dec.} \bibinfo{year}{2017}), \bibinfo{pages}{24:1--24:19}.
\newblock
\urldef\tempurl%
\url{https://doi.org/10.1145/3134659}
\showDOI{\tempurl}


\bibitem[Bonotti(2017)]%
        {bonottiReligionHateSpeech2017}
\bibfield{author}{\bibinfo{person}{Matteo Bonotti}.} \bibinfo{year}{2017}\natexlab{}.
\newblock \showarticletitle{Religion, Hate Speech and Non-Domination}.
\newblock \bibinfo{journal}{\emph{Ethnicities}} \bibinfo{volume}{17}, \bibinfo{number}{2} (\bibinfo{date}{April} \bibinfo{year}{2017}), \bibinfo{pages}{259--274}.
\newblock
\showISSN{1468-7968}
\urldef\tempurl%
\url{https://doi.org/10.1177/1468796817692626}
\showDOI{\tempurl}


\bibitem[Broockman and Kalla(2016)]%
        {broockmanDurablyReducingTransphobia2016}
\bibfield{author}{\bibinfo{person}{David Broockman} {and} \bibinfo{person}{Joshua Kalla}.} \bibinfo{year}{2016}\natexlab{}.
\newblock \showarticletitle{Durably Reducing Transphobia: {{A}} Field Experiment on Door-to-Door Canvassing}.
\newblock \bibinfo{journal}{\emph{Science}} \bibinfo{volume}{352}, \bibinfo{number}{6282} (\bibinfo{date}{April} \bibinfo{year}{2016}), \bibinfo{pages}{220--224}.
\newblock
\urldef\tempurl%
\url{https://doi.org/10.1126/science.aad9713}
\showDOI{\tempurl}


\bibitem[Buerger(2021a)]%
        {buergerCounterspeechLiteratureReview2021}
\bibfield{author}{\bibinfo{person}{Catherine Buerger}.} \bibinfo{year}{2021}\natexlab{a}.
\newblock \bibinfo{title}{Counterspeech: {{A Literature Review}}}.
\newblock
\newblock
\urldef\tempurl%
\url{https://doi.org/10.2139/ssrn.4066882}
\showDOI{\tempurl}


\bibitem[Buerger(2021b)]%
        {buergerIamhereCollectiveCounterspeech2021}
\bibfield{author}{\bibinfo{person}{Catherine Buerger}.} \bibinfo{year}{2021}\natexlab{b}.
\newblock \showarticletitle{\#iamhere: {{Collective Counterspeech}} and the {{Quest}} to {{Improve Online Discourse}}}.
\newblock \bibinfo{journal}{\emph{Social Media + Society}} \bibinfo{volume}{7}, \bibinfo{number}{4} (\bibinfo{date}{Oct.} \bibinfo{year}{2021}), \bibinfo{pages}{20563051211063843}.
\newblock
\showISSN{2056-3051}
\urldef\tempurl%
\url{https://doi.org/10.1177/20563051211063843}
\showDOI{\tempurl}


\bibitem[Buerger(2022)]%
        {buergerWhyTheyIt2022}
\bibfield{author}{\bibinfo{person}{Catherine Buerger}.} \bibinfo{year}{2022}\natexlab{}.
\newblock \bibinfo{title}{Why {{They Do It}}: {{Counterspeech Theories}} of {{Change}}}.
\newblock
\newblock
\urldef\tempurl%
\url{https://doi.org/10.2139/ssrn.4245211}
\showDOI{\tempurl}


\bibitem[Callahan et~al\mbox{.}(2019)]%
        {callahanSnapchatUsageMinority2019}
\bibfield{author}{\bibinfo{person}{Clark Callahan}, \bibinfo{person}{Scott~Haden Church}, \bibinfo{person}{Jesse King}, {and} \bibinfo{person}{Maureen Elinzano}.} \bibinfo{year}{2019}\natexlab{}.
\newblock \showarticletitle{Snapchat {{Usage Among Minority Populations}}}.
\newblock \bibinfo{journal}{\emph{Journal of Media and Religion}} \bibinfo{volume}{18}, \bibinfo{number}{1} (\bibinfo{date}{Jan.} \bibinfo{year}{2019}), \bibinfo{pages}{1--12}.
\newblock
\showISSN{1534-8423}
\urldef\tempurl%
\url{https://doi.org/10.1080/15348423.2019.1639404}
\showDOI{\tempurl}


\bibitem[Callan et~al\mbox{.}(2012)]%
        {callanJusticeMotiveEffects2012}
\bibfield{author}{\bibinfo{person}{Mitchell~J. Callan}, \bibinfo{person}{Rael~J. Dawtry}, {and} \bibinfo{person}{James~M. Olson}.} \bibinfo{year}{2012}\natexlab{}.
\newblock \showarticletitle{Justice Motive Effects in Ageism: {{The}} Effects of a Victim's Age on Observer Perceptions of Injustice and Punishment Judgments}.
\newblock \bibinfo{journal}{\emph{Journal of Experimental Social Psychology}} \bibinfo{volume}{48}, \bibinfo{number}{6} (\bibinfo{date}{Nov.} \bibinfo{year}{2012}), \bibinfo{pages}{1343--1349}.
\newblock
\showISSN{0022-1031}
\urldef\tempurl%
\url{https://doi.org/10.1016/j.jesp.2012.07.003}
\showDOI{\tempurl}


\bibitem[{Casta{\~n}o-Pulgar{\'i}n} et~al\mbox{.}(2021)]%
        {castano-pulgarinInternetSocialMedia2021}
\bibfield{author}{\bibinfo{person}{Sergio~Andr{\'e}s {Casta{\~n}o-Pulgar{\'i}n}}, \bibinfo{person}{Natalia {Su{\'a}rez-Betancur}}, \bibinfo{person}{Luz Magnolia~Tilano Vega}, {and} \bibinfo{person}{Harvey Mauricio~Herrera L{\'o}pez}.} \bibinfo{year}{2021}\natexlab{}.
\newblock \showarticletitle{Internet, Social Media and Online Hate Speech. {{Systematic}} Review}.
\newblock \bibinfo{journal}{\emph{Aggression and Violent Behavior}}  \bibinfo{volume}{58} (\bibinfo{date}{May} \bibinfo{year}{2021}), \bibinfo{pages}{101608}.
\newblock
\showISSN{1359-1789}
\urldef\tempurl%
\url{https://doi.org/10.1016/j.avb.2021.101608}
\showDOI{\tempurl}


\bibitem[Celuch et~al\mbox{.}(2022)]%
        {celuchFactorsAssociatedOnline2022}
\bibfield{author}{\bibinfo{person}{Magdalena Celuch}, \bibinfo{person}{Atte Oksanen}, \bibinfo{person}{P. R{\"a}s{\"a}nen}, \bibinfo{person}{Matthew Costello}, \bibinfo{person}{Catherine Blaya}, \bibinfo{person}{Izabela Zych}, \bibinfo{person}{Vicente~J. Llorent}, \bibinfo{person}{Ashley~V. Reichelmann}, {and} \bibinfo{person}{J. Hawdon}.} \bibinfo{year}{2022}\natexlab{}.
\newblock \showarticletitle{Factors {{Associated}} with {{Online Hate Acceptance}}: {{A Cross-National Six-Country Study}} among {{Young Adults}}}.
\newblock \bibinfo{journal}{\emph{International Journal of Environmental Research and Public Health}}  \bibinfo{volume}{19} (\bibinfo{year}{2022}).
\newblock
\urldef\tempurl%
\url{https://doi.org/10.3390/ijerph19010534}
\showDOI{\tempurl}


\bibitem[Cepollaro et~al\mbox{.}(2023)]%
        {cepollaroCounterspeech2023}
\bibfield{author}{\bibinfo{person}{Bianca Cepollaro}, \bibinfo{person}{Maxime Lepoutre}, {and} \bibinfo{person}{Robert~Mark Simpson}.} \bibinfo{year}{2023}\natexlab{}.
\newblock \showarticletitle{Counterspeech}.
\newblock \bibinfo{journal}{\emph{Philosophy Compass}} \bibinfo{volume}{18}, \bibinfo{number}{1} (\bibinfo{year}{2023}), \bibinfo{pages}{e12890}.
\newblock
\showISSN{1747-9991}
\urldef\tempurl%
\url{https://doi.org/10.1111/phc3.12890}
\showDOI{\tempurl}


\bibitem[Chandrasekharan et~al\mbox{.}(2017)]%
        {chandrasekharanYouCanStay2017}
\bibfield{author}{\bibinfo{person}{Eshwar Chandrasekharan}, \bibinfo{person}{Umashanthi Pavalanathan}, \bibinfo{person}{Anirudh Srinivasan}, \bibinfo{person}{Adam Glynn}, \bibinfo{person}{Jacob Eisenstein}, {and} \bibinfo{person}{Eric Gilbert}.} \bibinfo{year}{2017}\natexlab{}.
\newblock \showarticletitle{You {{Can}}'t {{Stay Here}}: {{The Efficacy}} of {{Reddit}}'s 2015 {{Ban Examined Through Hate Speech}}}.
\newblock \bibinfo{journal}{\emph{Proceedings of the ACM on Human-Computer Interaction}} \bibinfo{volume}{1}, \bibinfo{number}{CSCW} (\bibinfo{date}{Dec.} \bibinfo{year}{2017}), \bibinfo{pages}{31:1--31:22}.
\newblock
\urldef\tempurl%
\url{https://doi.org/10.1145/3134666}
\showDOI{\tempurl}


\bibitem[Chetty and Alathur(2018)]%
        {chettyHateSpeechReview2018}
\bibfield{author}{\bibinfo{person}{Naganna Chetty} {and} \bibinfo{person}{Sreejith Alathur}.} \bibinfo{year}{2018}\natexlab{}.
\newblock \showarticletitle{Hate Speech Review in the Context of Online Social Networks}.
\newblock \bibinfo{journal}{\emph{Aggression and Violent Behavior}}  \bibinfo{volume}{40} (\bibinfo{date}{May} \bibinfo{year}{2018}), \bibinfo{pages}{108--118}.
\newblock
\showISSN{1359-1789}
\urldef\tempurl%
\url{https://doi.org/10.1016/j.avb.2018.05.003}
\showDOI{\tempurl}


\bibitem[Chung et~al\mbox{.}(2019)]%
        {chungCONANCOunterNArratives2019}
\bibfield{author}{\bibinfo{person}{Yi-Ling Chung}, \bibinfo{person}{Elizaveta Kuzmenko}, \bibinfo{person}{Serra~Sinem Tekiroglu}, {and} \bibinfo{person}{Marco Guerini}.} \bibinfo{year}{2019}\natexlab{}.
\newblock \showarticletitle{{{CONAN}} - {{COunter NArratives}} through {{Nichesourcing}}: A {{Multilingual Dataset}} of {{Responses}} to {{Fight Online Hate Speech}}}. In \bibinfo{booktitle}{\emph{Proceedings of the 57th {{Annual Meeting}} of the {{Association}} for {{Computational Linguistics}}}}, \bibfield{editor}{\bibinfo{person}{Anna Korhonen}, \bibinfo{person}{David Traum}, {and} \bibinfo{person}{Llu{\'i}s M{\`a}rquez}} (Eds.). \bibinfo{publisher}{Association for Computational Linguistics}, \bibinfo{address}{Florence, Italy}, \bibinfo{pages}{2819--2829}.
\newblock
\urldef\tempurl%
\url{https://doi.org/10.18653/v1/P19-1271}
\showDOI{\tempurl}


\bibitem[Chung et~al\mbox{.}(2021)]%
        {chungKnowledgeGroundedCounterNarrative2021}
\bibfield{author}{\bibinfo{person}{Yi-Ling Chung}, \bibinfo{person}{Serra~Sinem Tekiro{\u g}lu}, {and} \bibinfo{person}{Marco Guerini}.} \bibinfo{year}{2021}\natexlab{}.
\newblock \showarticletitle{Towards {{Knowledge-Grounded Counter Narrative Generation}} for {{Hate Speech}}}. In \bibinfo{booktitle}{\emph{Findings of the {{Association}} for {{Computational Linguistics}}: {{ACL-IJCNLP}} 2021}}, \bibfield{editor}{\bibinfo{person}{Chengqing Zong}, \bibinfo{person}{Fei Xia}, \bibinfo{person}{Wenjie Li}, {and} \bibinfo{person}{Roberto Navigli}} (Eds.). \bibinfo{publisher}{Association for Computational Linguistics}, \bibinfo{address}{Online}, \bibinfo{pages}{899--914}.
\newblock
\urldef\tempurl%
\url{https://doi.org/10.18653/v1/2021.findings-acl.79}
\showDOI{\tempurl}


\bibitem[{Cohen-Almagor}(2018)]%
        {cohen-almagorTakingNorthAmerican2018}
\bibfield{author}{\bibinfo{person}{R. {Cohen-Almagor}}.} \bibinfo{year}{2018}\natexlab{}.
\newblock \showarticletitle{Taking {{North American White Supremacist Groups Seriously}}: {{The Scope}} and the {{Challenge}} of {{Hate Speech}} on the {{Internet}}}.
\newblock \bibinfo{journal}{\emph{International Journal for Crime, Justice and Social Democracy}} (\bibinfo{year}{2018}).
\newblock
\urldef\tempurl%
\url{https://doi.org/10.5204/IJCJSD.V7I2.517}
\showDOI{\tempurl}


\bibitem[Cooney(2014)]%
        {cooneyDeathFamilyHonor2014}
\bibfield{author}{\bibinfo{person}{Mark Cooney}.} \bibinfo{year}{2014}\natexlab{}.
\newblock \showarticletitle{Death by Family: {{Honor}} Violence as Punishment}.
\newblock \bibinfo{journal}{\emph{Punishment \& Society}} \bibinfo{volume}{16}, \bibinfo{number}{4} (\bibinfo{date}{Oct.} \bibinfo{year}{2014}), \bibinfo{pages}{406--427}.
\newblock
\showISSN{1462-4745}
\urldef\tempurl%
\url{https://doi.org/10.1177/1462474514539537}
\showDOI{\tempurl}


\bibitem[Costello and Hawdon(2018)]%
        {costelloWhoAreOnline2018}
\bibfield{author}{\bibinfo{person}{Matthew Costello} {and} \bibinfo{person}{James Hawdon}.} \bibinfo{year}{2018}\natexlab{}.
\newblock \showarticletitle{Who {{Are}} the {{Online Extremists Among Us}}? {{Sociodemographic Characteristics}}, {{Social Networking}}, and {{Online Experiences}} of {{Those Who Produce Online Hate Materials}}}.
\newblock \bibinfo{journal}{\emph{Violence and Gender}} \bibinfo{volume}{5}, \bibinfo{number}{1} (\bibinfo{date}{March} \bibinfo{year}{2018}), \bibinfo{pages}{55--60}.
\newblock
\showISSN{2326-7836}
\urldef\tempurl%
\url{https://doi.org/10.1089/vio.2017.0048}
\showDOI{\tempurl}


\bibitem[Costello and Hawdon(2020)]%
        {costelloHateSpeechOnline2020}
\bibfield{author}{\bibinfo{person}{Matthew Costello} {and} \bibinfo{person}{James Hawdon}.} \bibinfo{year}{2020}\natexlab{}.
\newblock \showarticletitle{Hate {{Speech}} in {{Online Spaces}}}.
\newblock In \bibinfo{booktitle}{\emph{The {{Palgrave Handbook}} of {{International Cybercrime}} and {{Cyberdeviance}}}}, \bibfield{editor}{\bibinfo{person}{Thomas~J. Holt} {and} \bibinfo{person}{Adam~M. Bossler}} (Eds.). \bibinfo{publisher}{Springer International Publishing}, \bibinfo{address}{Cham}, \bibinfo{pages}{1397--1416}.
\newblock
\showISBNx{978-3-319-78440-3}
\urldef\tempurl%
\url{https://doi.org/10.1007/978-3-319-78440-3_60}
\showDOI{\tempurl}


\bibitem[Costello et~al\mbox{.}(2019a)]%
        {costelloSocialGroupIdentity2019}
\bibfield{author}{\bibinfo{person}{Matthew Costello}, \bibinfo{person}{James Hawdon}, \bibinfo{person}{Colin Bernatzky}, {and} \bibinfo{person}{Kelly Mendes}.} \bibinfo{year}{2019}\natexlab{a}.
\newblock \showarticletitle{Social {{Group Identity}} and {{Perceptions}} of {{Online Hate}}*}.
\newblock \bibinfo{journal}{\emph{Sociological Inquiry}} \bibinfo{volume}{89}, \bibinfo{number}{3} (\bibinfo{year}{2019}), \bibinfo{pages}{427--452}.
\newblock
\showISSN{1475-682X}
\urldef\tempurl%
\url{https://doi.org/10.1111/soin.12274}
\showDOI{\tempurl}


\bibitem[Costello et~al\mbox{.}(2017)]%
        {costelloConfrontingOnlineExtremism2017}
\bibfield{author}{\bibinfo{person}{Matthew Costello}, \bibinfo{person}{James Hawdon}, {and} \bibinfo{person}{Thomas~N. Ratliff}.} \bibinfo{year}{2017}\natexlab{}.
\newblock \showarticletitle{Confronting {{Online Extremism}}: {{The Effect}} of {{Self-Help}}, {{Collective Efficacy}}, and {{Guardianship}} on {{Being}} a {{Target}} for {{Hate Speech}}}.
\newblock \bibinfo{journal}{\emph{Social Science Computer Review}} \bibinfo{volume}{35}, \bibinfo{number}{5} (\bibinfo{date}{Oct.} \bibinfo{year}{2017}), \bibinfo{pages}{587--605}.
\newblock
\showISSN{0894-4393}
\urldef\tempurl%
\url{https://doi.org/10.1177/0894439316666272}
\showDOI{\tempurl}


\bibitem[Costello et~al\mbox{.}(2019b)]%
        {costelloWeDonYour2019}
\bibfield{author}{\bibinfo{person}{Matthew Costello}, \bibinfo{person}{Joseph Rukus}, {and} \bibinfo{person}{James Hawdon}.} \bibinfo{year}{2019}\natexlab{b}.
\newblock \showarticletitle{We Don't like Your Type around Here: {{Regional}} and Residential Differences in Exposure to Online Hate Material Targeting Sexuality}.
\newblock \bibinfo{journal}{\emph{Deviant Behavior}} \bibinfo{volume}{40}, \bibinfo{number}{3} (\bibinfo{date}{March} \bibinfo{year}{2019}), \bibinfo{pages}{385--401}.
\newblock
\showISSN{0163-9625}
\urldef\tempurl%
\url{https://doi.org/10.1080/01639625.2018.1426266}
\showDOI{\tempurl}


\bibitem[Cowan et~al\mbox{.}(2005)]%
        {cowanHeterosexualsAttitudesHate2005}
\bibfield{author}{\bibinfo{person}{G. Cowan}, \bibinfo{person}{Becky Heiple}, \bibinfo{person}{C. Marquez}, \bibinfo{person}{D{\'e}sir{\'e}e Khatchadourian}, {and} \bibinfo{person}{Michelle McNevin}.} \bibinfo{year}{2005}\natexlab{}.
\newblock \showarticletitle{Heterosexuals' {{Attitudes Toward Hate Crimes}} and {{Hate Speech Against Gays}} and {{Lesbians}}}.
\newblock \bibinfo{journal}{\emph{Journal of Homosexuality}}  \bibinfo{volume}{49} (\bibinfo{year}{2005}).
\newblock
\urldef\tempurl%
\url{https://doi.org/10.1300/J082v49n02_04}
\showDOI{\tempurl}


\bibitem[Cowan and Khatchadourian(2003)]%
        {cowanEmpathyWaysKnowing2003}
\bibfield{author}{\bibinfo{person}{G. Cowan} {and} \bibinfo{person}{D{\'e}sir{\'e}e Khatchadourian}.} \bibinfo{year}{2003}\natexlab{}.
\newblock \showarticletitle{Empathy, {{Ways}} of {{Knowing}}, and {{Interdependence}} as {{Mediators}} of {{Gender Differences}} in {{Attitudes Toward Hate Speech}} and {{Freedom}} of {{Speech}}}.
\newblock \bibinfo{journal}{\emph{Psychology of Women Quarterly}}  \bibinfo{volume}{27} (\bibinfo{year}{2003}).
\newblock
\urldef\tempurl%
\url{https://doi.org/10.1111/1471-6402.00110}
\showDOI{\tempurl}


\bibitem[Curran and Chuang(2023)]%
        {curranSocialDistancingSocial2023}
\bibfield{author}{\bibinfo{person}{Max~T. Curran} {and} \bibinfo{person}{John Chuang}.} \bibinfo{year}{2023}\natexlab{}.
\newblock \showarticletitle{Social {{Distancing}} and {{Social Biosensing}}: {{Intersubjectivity}} from {{Afar}}}.
\newblock \bibinfo{journal}{\emph{Computer Supported Cooperative Work (CSCW)}} \bibinfo{volume}{32}, \bibinfo{number}{2} (\bibinfo{date}{June} \bibinfo{year}{2023}), \bibinfo{pages}{313--346}.
\newblock
\showISSN{1573-7551}
\urldef\tempurl%
\url{https://doi.org/10.1007/s10606-022-09428-5}
\showDOI{\tempurl}


\bibitem[ElSherief et~al\mbox{.}(2018)]%
        {elsheriefHateLingoTargetbased2018}
\bibfield{author}{\bibinfo{person}{Mai ElSherief}, \bibinfo{person}{Vivek Kulkarni}, \bibinfo{person}{Dana Nguyen}, \bibinfo{person}{William~Yang Wang}, {and} \bibinfo{person}{E. {Belding-Royer}}.} \bibinfo{year}{2018}\natexlab{}.
\newblock \showarticletitle{Hate {{Lingo}}: {{A Target-based Linguistic Analysis}} of {{Hate Speech}} in {{Social Media}}}.
\newblock \bibinfo{journal}{\emph{ArXiv}}  \bibinfo{volume}{abs/1804.04257} (\bibinfo{year}{2018}).
\newblock
\urldef\tempurl%
\url{https://doi.org/10.1609/icwsm.v12i1.15041}
\showDOI{\tempurl}


\bibitem[Fanton et~al\mbox{.}(2021)]%
        {fantonHumanintheLoopDataCollection2021}
\bibfield{author}{\bibinfo{person}{Margherita Fanton}, \bibinfo{person}{Helena Bonaldi}, \bibinfo{person}{Serra~Sinem Tekiroglu}, {and} \bibinfo{person}{Marco Guerini}.} \bibinfo{year}{2021}\natexlab{}.
\newblock \showarticletitle{Human-in-the-{{Loop}} for {{Data Collection}}: A {{Multi-Target Counter Narrative Dataset}} to {{Fight Online Hate Speech}}}. In \bibinfo{booktitle}{\emph{Proceedings of the 59th {{Annual Meeting}} of the {{Association}} for {{Computational Linguistics}} and the 11th {{International Joint Conference}} on {{Natural Language Processing}} ({{Volume}} 1: {{Long Papers}})}}. \bibinfo{pages}{3226--3240}.
\newblock
\urldef\tempurl%
\url{https://doi.org/10.18653/v1/2021.acl-long.250}
\showDOI{\tempurl}
\showeprint[arxiv]{2107.08720}~[cs]


\bibitem[F{\'a}tima et~al\mbox{.}(2023)]%
        {fatimaHateSpeechSocial2023}
\bibfield{author}{\bibinfo{person}{Branco~Di F{\'a}tima}, \bibinfo{person}{Allen Munoriyarwa}, \bibinfo{person}{Anne Gilliland}, \bibinfo{person}{Aondover~Eric Msughter}, \bibinfo{person}{Arantxa {Vizca{\'i}no-Verd{\'u}}}, \bibinfo{person}{Ebru G{\"o}kaliler}, \bibinfo{person}{Edson Capoano}, \bibinfo{person}{Huizi Yu}, \bibinfo{person}{{\.I}nan{\c c} Alik{\i}l{\i}{\c c}}, \bibinfo{person}{Juan-Manuel {Gonz{\'a}lez-Aguilar}}, \bibinfo{person}{Lida Tsene}, \bibinfo{person}{Lizhou Fan}, \bibinfo{person}{Macarena {Parejo-Cu{\'e}llar}}, \bibinfo{person}{Mine~Gencel Bek}, \bibinfo{person}{Muluken~Asegidew Chekol}, \bibinfo{person}{Mykola Makhortykh}, \bibinfo{person}{{\"O}zlem Alik{\i}l{\i}{\c c}}, \bibinfo{person}{Patricia {de-Casas-Moreno}}, \bibinfo{person}{Tiago Lapa}, \bibinfo{person}{Vinicius Prates}, {and} \bibinfo{person}{V{\'i}tor de Sousa}.} \bibinfo{year}{2023}\natexlab{}.
\newblock \bibinfo{booktitle}{\emph{{Hate Speech on Social Media: A Global Approach}}}.
\newblock \bibinfo{publisher}{Pontificia Universidad Cat{\'o}lica del Ecuador}.
\newblock
\showISBNx{978-9978-77-664-3}


\bibitem[Foxman and Wolf(2013)]%
        {foxmanViralHateContaining2013}
\bibfield{author}{\bibinfo{person}{Abraham~H. Foxman} {and} \bibinfo{person}{Christopher Wolf}.} \bibinfo{year}{2013}\natexlab{}.
\newblock \bibinfo{booktitle}{\emph{Viral Hate: {{Containing}} Its Spread on the {{Internet}}}}.
\newblock \bibinfo{publisher}{Macmillan}.
\newblock


\bibitem[Frissen(2021)]%
        {frissenInternetGreatRadicalizer2021}
\bibfield{author}{\bibinfo{person}{Thomas Frissen}.} \bibinfo{year}{2021}\natexlab{}.
\newblock \showarticletitle{Internet, the Great Radicalizer? {{Exploring}} Relationships between Seeking for Online Extremist Materials and Cognitive Radicalization in Young Adults}.
\newblock \bibinfo{journal}{\emph{Computers in Human Behavior}}  \bibinfo{volume}{114} (\bibinfo{date}{Jan.} \bibinfo{year}{2021}), \bibinfo{pages}{106549}.
\newblock
\showISSN{0747-5632}
\urldef\tempurl%
\url{https://doi.org/10.1016/j.chb.2020.106549}
\showDOI{\tempurl}


\bibitem[Garland et~al\mbox{.}(2022)]%
        {garlandImpactDynamicsHate2022}
\bibfield{author}{\bibinfo{person}{Joshua Garland}, \bibinfo{person}{Keyan {Ghazi-Zahedi}}, \bibinfo{person}{Jean-Gabriel Young}, \bibinfo{person}{Laurent {H{\'e}bert-Dufresne}}, {and} \bibinfo{person}{Mirta Galesic}.} \bibinfo{year}{2022}\natexlab{}.
\newblock \showarticletitle{Impact and Dynamics of Hate and Counter Speech Online}.
\newblock \bibinfo{journal}{\emph{EPJ Data Science}} \bibinfo{volume}{11}, \bibinfo{number}{1} (\bibinfo{date}{Dec.} \bibinfo{year}{2022}), \bibinfo{pages}{3}.
\newblock
\showISSN{2193-1127}
\urldef\tempurl%
\url{https://doi.org/10.1140/epjds/s13688-021-00314-6}
\showDOI{\tempurl}


\bibitem[Gledhill(2014)]%
        {gledhillQueeringStateCrime2014}
\bibfield{author}{\bibinfo{person}{Cara Gledhill}.} \bibinfo{year}{2014}\natexlab{}.
\newblock \showarticletitle{Queering {{State Crime Theory}}: {{The State}}, {{Civil Society}} and {{Marginalization}}}.
\newblock \bibinfo{journal}{\emph{Critical Criminology}} \bibinfo{volume}{22}, \bibinfo{number}{1} (\bibinfo{date}{March} \bibinfo{year}{2014}), \bibinfo{pages}{127--138}.
\newblock
\showISSN{1572-9877}
\urldef\tempurl%
\url{https://doi.org/10.1007/s10612-013-9229-9}
\showDOI{\tempurl}


\bibitem[Gligori{\'c} et~al\mbox{.}(2019)]%
        {gligoricCausalEffectsBrevity2019}
\bibfield{author}{\bibinfo{person}{Kristina Gligori{\'c}}, \bibinfo{person}{Ashton Anderson}, {and} \bibinfo{person}{Robert West}.} \bibinfo{year}{2019}\natexlab{}.
\newblock \showarticletitle{Causal {{Effects}} of {{Brevity}} on {{Style}} and {{Success}} in {{Social Media}}}.
\newblock \bibinfo{journal}{\emph{Proceedings of the ACM on Human-Computer Interaction}} \bibinfo{volume}{3}, \bibinfo{number}{CSCW} (\bibinfo{date}{Nov.} \bibinfo{year}{2019}), \bibinfo{pages}{45:1--45:23}.
\newblock
\urldef\tempurl%
\url{https://doi.org/10.1145/3359147}
\showDOI{\tempurl}


\bibitem[Gligori{\'c} et~al\mbox{.}(2023)]%
        {gligoricLinguisticEffectsNews2023}
\bibfield{author}{\bibinfo{person}{Kristina Gligori{\'c}}, \bibinfo{person}{George Lifchits}, \bibinfo{person}{Robert West}, {and} \bibinfo{person}{Ashton Anderson}.} \bibinfo{year}{2023}\natexlab{}.
\newblock \showarticletitle{Linguistic Effects on News Headline Success: {{Evidence}} from Thousands of Online Field Experiments ({{Registered Report}})}.
\newblock \bibinfo{journal}{\emph{PLOS ONE}} \bibinfo{volume}{18}, \bibinfo{number}{3} (\bibinfo{year}{2023}), \bibinfo{pages}{e0281682}.
\newblock
\showISSN{1932-6203}
\urldef\tempurl%
\url{https://doi.org/10.1371/journal.pone.0281682}
\showDOI{\tempurl}


\bibitem[Goyal et~al\mbox{.}(2022)]%
        {goyalYourToxicityMy2022}
\bibfield{author}{\bibinfo{person}{Nitesh Goyal}, \bibinfo{person}{Ian~D. Kivlichan}, \bibinfo{person}{Rachel Rosen}, {and} \bibinfo{person}{Lucy Vasserman}.} \bibinfo{year}{2022}\natexlab{}.
\newblock \showarticletitle{Is {{Your Toxicity My Toxicity}}? {{Exploring}} the {{Impact}} of {{Rater Identity}} on {{Toxicity Annotation}}}.
\newblock \bibinfo{journal}{\emph{Proceedings of the ACM on Human-Computer Interaction}} \bibinfo{volume}{6}, \bibinfo{number}{CSCW2} (\bibinfo{year}{2022}), \bibinfo{pages}{1--28}.
\newblock


\bibitem[Hangartner et~al\mbox{.}(2021)]%
        {hangartnerEmpathybasedCounterspeechCan2021}
\bibfield{author}{\bibinfo{person}{Dominik Hangartner}, \bibinfo{person}{Gloria Gennaro}, \bibinfo{person}{Sary Alasiri}, \bibinfo{person}{Nicholas Bahrich}, \bibinfo{person}{Alexandra Bornhoft}, \bibinfo{person}{Joseph Boucher}, \bibinfo{person}{Buket~Buse Demirci}, \bibinfo{person}{Laurenz Derksen}, \bibinfo{person}{Aldo Hall}, \bibinfo{person}{Matthias Jochum}, \bibinfo{person}{Maria~Murias Munoz}, \bibinfo{person}{Marc Richter}, \bibinfo{person}{Franziska Vogel}, \bibinfo{person}{Salom{\'e} Wittwer}, \bibinfo{person}{Felix W{\"u}thrich}, \bibinfo{person}{Fabrizio Gilardi}, {and} \bibinfo{person}{Karsten Donnay}.} \bibinfo{year}{2021}\natexlab{}.
\newblock \showarticletitle{Empathy-Based Counterspeech Can Reduce Racist Hate Speech in a Social Media Field Experiment}.
\newblock \bibinfo{journal}{\emph{Proceedings of the National Academy of Sciences}} \bibinfo{volume}{118}, \bibinfo{number}{50} (\bibinfo{date}{Dec.} \bibinfo{year}{2021}), \bibinfo{pages}{e2116310118}.
\newblock
\urldef\tempurl%
\url{https://doi.org/10.1073/pnas.2116310118}
\showDOI{\tempurl}


\bibitem[Hawdon and Costello(2022)]%
        {hawdonConfrontingOnlineExtremism2022}
\bibfield{author}{\bibinfo{person}{James Hawdon} {and} \bibinfo{person}{Matthew Costello}.} \bibinfo{year}{2022}\natexlab{}.
\newblock \showarticletitle{Confronting {{Online Extremism}}: {{Strategies}}, {{Promises}}, and {{Pitfalls}}}.
\newblock In \bibinfo{booktitle}{\emph{Right-{{Wing Extremism}} in {{Canada}} and the {{United States}}}}, \bibfield{editor}{\bibinfo{person}{Barbara Perry}, \bibinfo{person}{Jeff Gruenewald}, {and} \bibinfo{person}{Ryan Scrivens}} (Eds.). \bibinfo{publisher}{Springer International Publishing}, \bibinfo{address}{Cham}, \bibinfo{pages}{469--489}.
\newblock
\showISBNx{978-3-030-99804-2}
\urldef\tempurl%
\url{https://doi.org/10.1007/978-3-030-99804-2_18}
\showDOI{\tempurl}


\bibitem[Henry and Powell(2018)]%
        {henryTechnologyFacilitatedSexualViolence2018}
\bibfield{author}{\bibinfo{person}{Nicola Henry} {and} \bibinfo{person}{Anastasia Powell}.} \bibinfo{year}{2018}\natexlab{}.
\newblock \showarticletitle{Technology-{{Facilitated Sexual Violence}}: {{A Literature Review}} of {{Empirical Research}}}.
\newblock \bibinfo{journal}{\emph{Trauma, Violence, \& Abuse}} \bibinfo{volume}{19}, \bibinfo{number}{2} (\bibinfo{date}{April} \bibinfo{year}{2018}), \bibinfo{pages}{195--208}.
\newblock
\showISSN{1524-8380}
\urldef\tempurl%
\url{https://doi.org/10.1177/1524838016650189}
\showDOI{\tempurl}


\bibitem[Henson et~al\mbox{.}(2020)]%
        {hensonThereVirtuallyNo2020}
\bibfield{author}{\bibinfo{person}{Billy Henson}, \bibinfo{person}{Bonnie~S. Fisher}, {and} \bibinfo{person}{Bradford~W. Reyns}.} \bibinfo{year}{2020}\natexlab{}.
\newblock \showarticletitle{There {{Is Virtually No Excuse}}: {{The Frequency}} and {{Predictors}} of {{College Students}}' {{Bystander Intervention Behaviors Directed}} at {{Online Victimization}}}.
\newblock \bibinfo{journal}{\emph{Violence Against Women}} \bibinfo{volume}{26}, \bibinfo{number}{5} (\bibinfo{date}{April} \bibinfo{year}{2020}), \bibinfo{pages}{505--527}.
\newblock
\showISSN{1077-8012}
\urldef\tempurl%
\url{https://doi.org/10.1177/1077801219835050}
\showDOI{\tempurl}


\bibitem[Holgate et~al\mbox{.}(2018)]%
        {holgateWhySwearAnalyzing2018}
\bibfield{author}{\bibinfo{person}{Eric Holgate}, \bibinfo{person}{Isabel Cachola}, \bibinfo{person}{Daniel {Preo{\c t}iuc-Pietro}}, {and} \bibinfo{person}{Junyi~Jessy Li}.} \bibinfo{year}{2018}\natexlab{}.
\newblock \showarticletitle{Why {{Swear}}? {{Analyzing}} and {{Inferring}} the {{Intentions}} of {{Vulgar Expressions}}}. In \bibinfo{booktitle}{\emph{Proceedings of the 2018 {{Conference}} on {{Empirical Methods}} in {{Natural Language Processing}}}}. \bibinfo{publisher}{Association for Computational Linguistics}, \bibinfo{address}{Brussels, Belgium}, \bibinfo{pages}{4405--4414}.
\newblock
\urldef\tempurl%
\url{https://doi.org/10.18653/v1/D18-1471}
\showDOI{\tempurl}


\bibitem[Hollewell and Longpr{\'e}(2022)]%
        {hollewellRadicalizationSocialMedia2022}
\bibfield{author}{\bibinfo{person}{Georgia~F. Hollewell} {and} \bibinfo{person}{Nicholas Longpr{\'e}}.} \bibinfo{year}{2022}\natexlab{}.
\newblock \showarticletitle{Radicalization in the {{Social Media Era}}: {{Understanding}} the {{Relationship}} between {{Self-Radicalization}} and the {{Internet}}}.
\newblock \bibinfo{journal}{\emph{International Journal of Offender Therapy and Comparative Criminology}} \bibinfo{volume}{66}, \bibinfo{number}{8} (\bibinfo{date}{June} \bibinfo{year}{2022}), \bibinfo{pages}{896--913}.
\newblock
\showISSN{0306-624X}
\urldef\tempurl%
\url{https://doi.org/10.1177/0306624X211028771}
\showDOI{\tempurl}


\bibitem[Holt et~al\mbox{.}(2019)]%
        {holtLonersColleaguesPeers2019}
\bibfield{author}{\bibinfo{person}{Thomas~J. Holt}, \bibinfo{person}{Joshua~D. Freilich}, \bibinfo{person}{Steven~M. Chermak}, \bibinfo{person}{Colleen Mills}, {and} \bibinfo{person}{Jason Silva}.} \bibinfo{year}{2019}\natexlab{}.
\newblock \showarticletitle{Loners, {{Colleagues}}, or {{Peers}}? {{Assessing}} the {{Social Organization}} of {{Radicalization}}}.
\newblock \bibinfo{journal}{\emph{American Journal of Criminal Justice}} \bibinfo{volume}{44}, \bibinfo{number}{1} (\bibinfo{date}{Feb.} \bibinfo{year}{2019}), \bibinfo{pages}{83--105}.
\newblock
\showISSN{1936-1351}
\urldef\tempurl%
\url{https://doi.org/10.1007/s12103-018-9439-5}
\showDOI{\tempurl}


\bibitem[Hox et~al\mbox{.}(2017)]%
        {hoxMultilevelAnalysisTechniques2017}
\bibfield{author}{\bibinfo{person}{Joop Hox}, \bibinfo{person}{Mirjam Moerbeek}, {and} \bibinfo{person}{Rens {Van de Schoot}}.} \bibinfo{year}{2017}\natexlab{}.
\newblock \bibinfo{booktitle}{\emph{Multilevel Analysis: {{Techniques}} and Applications}}.
\newblock \bibinfo{publisher}{Routledge}.
\newblock


\bibitem[Hutson et~al\mbox{.}(2018)]%
        {hutsonDebiasingDesireAddressing2018}
\bibfield{author}{\bibinfo{person}{Jevan~A. Hutson}, \bibinfo{person}{Jessie~G. Taft}, \bibinfo{person}{Solon Barocas}, {and} \bibinfo{person}{Karen Levy}.} \bibinfo{year}{2018}\natexlab{}.
\newblock \showarticletitle{Debiasing {{Desire}}: {{Addressing Bias}} \& {{Discrimination}} on {{Intimate Platforms}}}.
\newblock \bibinfo{journal}{\emph{Proceedings of the ACM on Human-Computer Interaction}} \bibinfo{volume}{2}, \bibinfo{number}{CSCW} (\bibinfo{date}{Nov.} \bibinfo{year}{2018}), \bibinfo{pages}{73:1--73:18}.
\newblock
\urldef\tempurl%
\url{https://doi.org/10.1145/3274342}
\showDOI{\tempurl}


\bibitem[Ibrohim and Budi(2019)]%
        {ibrohimMultilabelHateSpeech2019}
\bibfield{author}{\bibinfo{person}{Muhammad~Okky Ibrohim} {and} \bibinfo{person}{Indra Budi}.} \bibinfo{year}{2019}\natexlab{}.
\newblock \showarticletitle{Multi-Label {{Hate Speech}} and {{Abusive Language Detection}} in {{Indonesian Twitter}}}. In \bibinfo{booktitle}{\emph{Proceedings of the {{Third Workshop}} on {{Abusive Language Online}}}}. \bibinfo{publisher}{Association for Computational Linguistics}, \bibinfo{address}{Florence, Italy}, \bibinfo{pages}{46--57}.
\newblock
\urldef\tempurl%
\url{https://doi.org/10.18653/v1/W19-3506}
\showDOI{\tempurl}


\bibitem[Johnson et~al\mbox{.}(2010)]%
        {johnsonSentencingHomicideOffenders2010}
\bibfield{author}{\bibinfo{person}{Brian~D. Johnson}, \bibinfo{person}{Sigrid Van~Wingerden}, {and} \bibinfo{person}{Paul Nieuwbeerta}.} \bibinfo{year}{2010}\natexlab{}.
\newblock \showarticletitle{Sentencing {{Homicide Offenders}} in the {{Netherlands}}: {{Offender}}, {{Victim}}, and {{Situational Influences}} in {{Criminal Punishment}}*}.
\newblock \bibinfo{journal}{\emph{Criminology}} \bibinfo{volume}{48}, \bibinfo{number}{4} (\bibinfo{year}{2010}), \bibinfo{pages}{981--1018}.
\newblock
\showISSN{1745-9125}
\urldef\tempurl%
\url{https://doi.org/10.1111/j.1745-9125.2010.00210.x}
\showDOI{\tempurl}


\bibitem[Johnson and Callahan(2013)]%
        {johnsonMinorityCulturesSocial2013}
\bibfield{author}{\bibinfo{person}{Jared~L. Johnson} {and} \bibinfo{person}{Clark Callahan}.} \bibinfo{year}{2013}\natexlab{}.
\newblock \showarticletitle{Minority {{Cultures}} and {{Social Media}}: {{Magnifying Garifuna}}}.
\newblock \bibinfo{journal}{\emph{Journal of Intercultural Communication Research}} \bibinfo{volume}{42}, \bibinfo{number}{4} (\bibinfo{date}{Dec.} \bibinfo{year}{2013}), \bibinfo{pages}{319--339}.
\newblock
\showISSN{1747-5759}
\urldef\tempurl%
\url{https://doi.org/10.1080/17475759.2013.842608}
\showDOI{\tempurl}


\bibitem[Kaakinen et~al\mbox{.}(2018)]%
        {kaakinenDidRiskExposure2018}
\bibfield{author}{\bibinfo{person}{Markus Kaakinen}, \bibinfo{person}{Atte Oksanen}, {and} \bibinfo{person}{P. R{\"a}s{\"a}nen}.} \bibinfo{year}{2018}\natexlab{}.
\newblock \showarticletitle{Did the Risk of Exposure to Online Hate Increase after the {{November}} 2015 {{Paris}} Attacks? {{A}} Group Relations Approach}.
\newblock \bibinfo{journal}{\emph{Comput. Hum. Behav.}}  \bibinfo{volume}{78} (\bibinfo{year}{2018}).
\newblock
\urldef\tempurl%
\url{https://doi.org/10.1016/j.chb.2017.09.022}
\showDOI{\tempurl}


\bibitem[Kaakinen et~al\mbox{.}(2020)]%
        {kaakinenImpulsivityInternalizingSymptoms2020}
\bibfield{author}{\bibinfo{person}{Markus Kaakinen}, \bibinfo{person}{Anu Sirola}, \bibinfo{person}{I. Savolainen}, {and} \bibinfo{person}{Atte Oksanen}.} \bibinfo{year}{2020}\natexlab{}.
\newblock \showarticletitle{Impulsivity, Internalizing Symptoms, and Online Group Behavior as Determinants of Online Hate}.
\newblock \bibinfo{journal}{\emph{PLoS ONE}}  \bibinfo{volume}{15} (\bibinfo{year}{2020}).
\newblock
\urldef\tempurl%
\url{https://doi.org/10.1371/journal.pone.0231052}
\showDOI{\tempurl}


\bibitem[Kahl et~al\mbox{.}(2013)]%
        {kahlStudentReactionsPublic2013}
\bibfield{author}{\bibinfo{person}{Jessica~E. Kahl}, \bibinfo{person}{Anne Koenig}, {and} \bibinfo{person}{Ramon Smith}.} \bibinfo{year}{2013}\natexlab{}.
\newblock \showarticletitle{Student {{Reactions}} to {{Public Safety Reports}} of {{Hate Crimes}}}.
\newblock \bibinfo{journal}{\emph{Journal of Interpersonal Violence}} \bibinfo{volume}{28}, \bibinfo{number}{13} (\bibinfo{date}{Sept.} \bibinfo{year}{2013}), \bibinfo{pages}{2713--2730}.
\newblock
\showISSN{0886-2605}
\urldef\tempurl%
\url{https://doi.org/10.1177/0886260513487990}
\showDOI{\tempurl}


\bibitem[Keipi et~al\mbox{.}(2017)]%
        {keipiHarmadvocatingOnlineContent2017}
\bibfield{author}{\bibinfo{person}{Teo Keipi}, \bibinfo{person}{Atte Oksanen}, \bibinfo{person}{James Hawdon}, \bibinfo{person}{Matti N{\"a}si}, {and} \bibinfo{person}{Pekka R{\"a}s{\"a}nen}.} \bibinfo{year}{2017}\natexlab{}.
\newblock \showarticletitle{Harm-Advocating Online Content and Subjective Well-Being: A Cross-National Study of New Risks Faced by Youth}.
\newblock \bibinfo{journal}{\emph{Journal of Risk Research}} (\bibinfo{date}{May} \bibinfo{year}{2017}).
\newblock
\showISSN{1366-9877}


\bibitem[Knapp and Daly(2011)]%
        {knappSAGEHandbookInterpersonal2011}
\bibfield{author}{\bibinfo{person}{Mark~L. Knapp} {and} \bibinfo{person}{John~A. Daly}.} \bibinfo{year}{2011}\natexlab{}.
\newblock \bibinfo{booktitle}{\emph{The {{SAGE Handbook}} of {{Interpersonal Communication}}}}.
\newblock \bibinfo{publisher}{SAGE Publications}.
\newblock
\showISBNx{978-1-4833-4150-7}


\bibitem[Koo and Li(2016)]%
        {kooGuidelineSelectingReporting2016}
\bibfield{author}{\bibinfo{person}{Terry~K. Koo} {and} \bibinfo{person}{Mae~Y. Li}.} \bibinfo{year}{2016}\natexlab{}.
\newblock \showarticletitle{A {{Guideline}} of {{Selecting}} and {{Reporting Intraclass Correlation Coefficients}} for {{Reliability Research}}}.
\newblock \bibinfo{journal}{\emph{Journal of Chiropractic Medicine}} \bibinfo{volume}{15}, \bibinfo{number}{2} (\bibinfo{date}{June} \bibinfo{year}{2016}), \bibinfo{pages}{155--163}.
\newblock
\showISSN{1556-3707}
\urldef\tempurl%
\url{https://doi.org/10.1016/j.jcm.2016.02.012}
\showDOI{\tempurl}


\bibitem[Lakoff(1973)]%
        {lakoffLanguageWomansPlace1973}
\bibfield{author}{\bibinfo{person}{Robin Lakoff}.} \bibinfo{year}{1973}\natexlab{}.
\newblock \showarticletitle{Language and Woman's Place}.
\newblock \bibinfo{journal}{\emph{Language in Society}} \bibinfo{volume}{2}, \bibinfo{number}{1} (\bibinfo{date}{April} \bibinfo{year}{1973}), \bibinfo{pages}{45--79}.
\newblock
\showISSN{1469-8013, 0047-4045}
\urldef\tempurl%
\url{https://doi.org/10.1017/S0047404500000051}
\showDOI{\tempurl}


\bibitem[Leary and Tangney(2012)]%
        {learyHandbookSelfIdentity2012}
\bibfield{author}{\bibinfo{person}{Mark~R. Leary} {and} \bibinfo{person}{June~Price Tangney}.} \bibinfo{year}{2012}\natexlab{}.
\newblock \bibinfo{booktitle}{\emph{Handbook of {{Self}} and {{Identity}}}}.
\newblock \bibinfo{publisher}{Guilford Press}.
\newblock
\showISBNx{978-1-4625-0305-6}


\bibitem[Lehmann et~al\mbox{.}(2022)]%
        {lehmannSuggestionListsVs2022}
\bibfield{author}{\bibinfo{person}{Florian Lehmann}, \bibinfo{person}{Niklas Markert}, \bibinfo{person}{Hai Dang}, {and} \bibinfo{person}{Daniel Buschek}.} \bibinfo{year}{2022}\natexlab{}.
\newblock \showarticletitle{Suggestion {{Lists}} vs. {{Continuous Generation}}: {{Interaction Design}} for {{Writing}} with {{Generative Models}} on {{Mobile Devices Affect Text Length}}, {{Wording}} and {{Perceived Authorship}}}. In \bibinfo{booktitle}{\emph{Proceedings of {{Mensch}} Und {{Computer}} 2022}} \emph{(\bibinfo{series}{{{MuC}} '22})}. \bibinfo{publisher}{Association for Computing Machinery}, \bibinfo{address}{New York, NY, USA}, \bibinfo{pages}{192--208}.
\newblock
\showISBNx{978-1-4503-9690-5}
\urldef\tempurl%
\url{https://doi.org/10.1145/3543758.3543947}
\showDOI{\tempurl}


\bibitem[Lepoutre(2017)]%
        {lepoutreHateSpeechPublic2017}
\bibfield{author}{\bibinfo{person}{Maxime Lepoutre}.} \bibinfo{year}{2017}\natexlab{}.
\newblock \showarticletitle{Hate {{Speech}} in {{Public Discourse}}: {{A Pessimistic Defense}} of {{Counterspeech}}}.
\newblock \bibinfo{journal}{\emph{Social Theory and Practice}}  \bibinfo{volume}{43} (\bibinfo{year}{2017}).
\newblock
\urldef\tempurl%
\url{https://doi.org/10.17863/CAM.15815}
\showDOI{\tempurl}


\bibitem[Liu et~al\mbox{.}(2022)]%
        {liuWillAIConsole2022}
\bibfield{author}{\bibinfo{person}{Yihe Liu}, \bibinfo{person}{Anushk Mittal}, \bibinfo{person}{Diyi Yang}, {and} \bibinfo{person}{Amy Bruckman}.} \bibinfo{year}{2022}\natexlab{}.
\newblock \showarticletitle{Will {{AI Console Me}} When {{I Lose}} My {{Pet}}? {{Understanding Perceptions}} of {{AI-Mediated Email Writing}}}. In \bibinfo{booktitle}{\emph{{{CHI Conference}} on {{Human Factors}} in {{Computing Systems}}}}. \bibinfo{publisher}{ACM}, \bibinfo{address}{New Orleans LA USA}, \bibinfo{pages}{1--13}.
\newblock
\showISBNx{978-1-4503-9157-3}
\urldef\tempurl%
\url{https://doi.org/10.1145/3491102.3517731}
\showDOI{\tempurl}


\bibitem[MacAvaney et~al\mbox{.}(2019)]%
        {macavaneyHateSpeechDetection2019}
\bibfield{author}{\bibinfo{person}{Sean MacAvaney}, \bibinfo{person}{Hao-Ren Yao}, \bibinfo{person}{Eugene Yang}, \bibinfo{person}{Katina Russell}, \bibinfo{person}{Nazli Goharian}, {and} \bibinfo{person}{Ophir Frieder}.} \bibinfo{year}{2019}\natexlab{}.
\newblock \showarticletitle{Hate Speech Detection: {{Challenges}} and Solutions}.
\newblock \bibinfo{journal}{\emph{PLOS ONE}} \bibinfo{volume}{14}, \bibinfo{number}{8} (\bibinfo{year}{2019}), \bibinfo{pages}{e0221152}.
\newblock
\showISSN{1932-6203}
\urldef\tempurl%
\url{https://doi.org/10.1371/journal.pone.0221152}
\showDOI{\tempurl}


\bibitem[Massey(2020)]%
        {masseyStillLinchpinSegregation2020}
\bibfield{author}{\bibinfo{person}{Douglas~S. Massey}.} \bibinfo{year}{2020}\natexlab{}.
\newblock \showarticletitle{Still the {{Linchpin}}: {{Segregation}} and {{Stratification}} in the {{USA}}}.
\newblock \bibinfo{journal}{\emph{Race and Social Problems}} \bibinfo{volume}{12}, \bibinfo{number}{1} (\bibinfo{date}{March} \bibinfo{year}{2020}), \bibinfo{pages}{1--12}.
\newblock
\showISSN{1867-1756}
\urldef\tempurl%
\url{https://doi.org/10.1007/s12552-019-09280-1}
\showDOI{\tempurl}


\bibitem[Massey and Denton(1989)]%
        {masseyHypersegregationMetropolitanAreas1989}
\bibfield{author}{\bibinfo{person}{Douglas~S. Massey} {and} \bibinfo{person}{Nancy~A. Denton}.} \bibinfo{year}{1989}\natexlab{}.
\newblock \showarticletitle{Hypersegregation in {{U}}.{{S}}. {{Metropolitan Areas}}: {{Black}} and {{Hispanic Segregation Along Five Dimensions}}}.
\newblock \bibinfo{journal}{\emph{Demography}} \bibinfo{volume}{26}, \bibinfo{number}{3} (\bibinfo{date}{Aug.} \bibinfo{year}{1989}), \bibinfo{pages}{373--391}.
\newblock
\showISSN{1533-7790}
\urldef\tempurl%
\url{https://doi.org/10.2307/2061599}
\showDOI{\tempurl}


\bibitem[{Matamoros-Fern{\'a}ndez} and Farkas(2021)]%
        {matamoros-fernandezRacismHateSpeech2021a}
\bibfield{author}{\bibinfo{person}{Ariadna {Matamoros-Fern{\'a}ndez}} {and} \bibinfo{person}{Johan Farkas}.} \bibinfo{year}{2021}\natexlab{}.
\newblock \showarticletitle{Racism, {{Hate Speech}}, and {{Social Media}}: {{A Systematic Review}} and {{Critique}}}.
\newblock \bibinfo{journal}{\emph{Television \& New Media}} \bibinfo{volume}{22}, \bibinfo{number}{2} (\bibinfo{date}{Feb.} \bibinfo{year}{2021}), \bibinfo{pages}{205--224}.
\newblock
\showISSN{1527-4764}
\urldef\tempurl%
\url{https://doi.org/10.1177/1527476420982230}
\showDOI{\tempurl}


\bibitem[Mathew et~al\mbox{.}(2020)]%
        {mathewHateBegetsHate2020}
\bibfield{author}{\bibinfo{person}{Binny Mathew}, \bibinfo{person}{Anurag Illendula}, \bibinfo{person}{Punyajoy Saha}, \bibinfo{person}{Soumya Sarkar}, \bibinfo{person}{Pawan Goyal}, {and} \bibinfo{person}{Animesh Mukherjee}.} \bibinfo{year}{2020}\natexlab{}.
\newblock \showarticletitle{Hate Begets {{Hate}}: {{A Temporal Study}} of {{Hate Speech}}}.
\newblock \bibinfo{journal}{\emph{Proceedings of the ACM on Human-Computer Interaction}} \bibinfo{volume}{4}, \bibinfo{number}{CSCW2} (\bibinfo{date}{Oct.} \bibinfo{year}{2020}), \bibinfo{pages}{92:1--92:24}.
\newblock
\urldef\tempurl%
\url{https://doi.org/10.1145/3415163}
\showDOI{\tempurl}


\bibitem[Mathew et~al\mbox{.}(2019)]%
        {mathewThouShaltNot2019}
\bibfield{author}{\bibinfo{person}{Binny Mathew}, \bibinfo{person}{Punyajoy Saha}, \bibinfo{person}{Hardik Tharad}, \bibinfo{person}{Subham Rajgaria}, \bibinfo{person}{Prajwal Singhania}, \bibinfo{person}{Suman~Kalyan Maity}, \bibinfo{person}{Pawan Goyal}, {and} \bibinfo{person}{Animesh Mukherjee}.} \bibinfo{year}{2019}\natexlab{}.
\newblock \showarticletitle{Thou {{Shalt Not Hate}}: {{Countering Online Hate Speech}}}.
\newblock \bibinfo{journal}{\emph{Proceedings of the International AAAI Conference on Web and Social Media}}  \bibinfo{volume}{13} (\bibinfo{date}{July} \bibinfo{year}{2019}), \bibinfo{pages}{369--380}.
\newblock
\showISSN{2334-0770}
\urldef\tempurl%
\url{https://doi.org/10.1609/icwsm.v13i01.3237}
\showDOI{\tempurl}


\bibitem[Matsumori et~al\mbox{.}(2019)]%
        {matsumoriDecisionTheoreticModelBehavior2019}
\bibfield{author}{\bibinfo{person}{Kaosu Matsumori}, \bibinfo{person}{Kazuki Iijima}, \bibinfo{person}{Yasuharu Koike}, {and} \bibinfo{person}{Kenji Matsumoto}.} \bibinfo{year}{2019}\natexlab{}.
\newblock \showarticletitle{A {{Decision-Theoretic Model}} of {{Behavior Change}}}.
\newblock \bibinfo{journal}{\emph{Frontiers in Psychology}}  \bibinfo{volume}{10} (\bibinfo{date}{May} \bibinfo{year}{2019}).
\newblock
\showISSN{1664-1078}
\urldef\tempurl%
\url{https://doi.org/10.3389/fpsyg.2019.01042}
\showDOI{\tempurl}


\bibitem[Meske and Bunde(2023)]%
        {meskeDesignPrinciplesUser2023}
\bibfield{author}{\bibinfo{person}{Christian Meske} {and} \bibinfo{person}{Enrico Bunde}.} \bibinfo{year}{2023}\natexlab{}.
\newblock \showarticletitle{Design {{Principles}} for {{User Interfaces}} in {{AI-Based Decision Support Systems}}: {{The Case}} of {{Explainable Hate Speech Detection}}}.
\newblock \bibinfo{journal}{\emph{Information Systems Frontiers}} \bibinfo{volume}{25}, \bibinfo{number}{2} (\bibinfo{date}{April} \bibinfo{year}{2023}), \bibinfo{pages}{743--773}.
\newblock
\showISSN{1572-9419}
\urldef\tempurl%
\url{https://doi.org/10.1007/s10796-021-10234-5}
\showDOI{\tempurl}


\bibitem[Mollas et~al\mbox{.}(2022)]%
        {mollasETHOSOnlineHate2022}
\bibfield{author}{\bibinfo{person}{Ioannis Mollas}, \bibinfo{person}{Zoe Chrysopoulou}, \bibinfo{person}{Stamatis Karlos}, {and} \bibinfo{person}{Grigorios Tsoumakas}.} \bibinfo{year}{2022}\natexlab{}.
\newblock \showarticletitle{{{ETHOS}}: An {{Online Hate Speech Detection Dataset}}}.
\newblock \bibinfo{journal}{\emph{Complex \& Intelligent Systems}} \bibinfo{volume}{8}, \bibinfo{number}{6} (\bibinfo{date}{Dec.} \bibinfo{year}{2022}), \bibinfo{pages}{4663--4678}.
\newblock
\showISSN{2199-4536, 2198-6053}
\urldef\tempurl%
\url{https://doi.org/10.1007/s40747-021-00608-2}
\showDOI{\tempurl}
\showeprint[arxiv]{2006.08328}~[cs, stat]


\bibitem[Mun et~al\mbox{.}(2024)]%
        {munCounterspeakersPerspectivesUnveiling2024}
\bibfield{author}{\bibinfo{person}{Jimin Mun}, \bibinfo{person}{Cathy Buerger}, \bibinfo{person}{Jenny~T Liang}, \bibinfo{person}{Joshua Garland}, {and} \bibinfo{person}{Maarten Sap}.} \bibinfo{year}{2024}\natexlab{}.
\newblock \showarticletitle{Counterspeakers' {{Perspectives}}: {{Unveiling Barriers}} and {{AI Needs}} in the {{Fight}} against {{Online Hate}}}. In \bibinfo{booktitle}{\emph{Proceedings of the {{CHI Conference}} on {{Human Factors}} in {{Computing Systems}}}} \emph{(\bibinfo{series}{{{CHI}} '24})}. \bibinfo{publisher}{Association for Computing Machinery}, \bibinfo{address}{New York, NY, USA}, \bibinfo{pages}{1--22}.
\newblock
\showISBNx{9798400703300}
\urldef\tempurl%
\url{https://doi.org/10.1145/3613904.3642025}
\showDOI{\tempurl}


\bibitem[Munger(2017)]%
        {mungerTweetmentEffectsTweeted2017}
\bibfield{author}{\bibinfo{person}{Kevin Munger}.} \bibinfo{year}{2017}\natexlab{}.
\newblock \showarticletitle{Tweetment {{Effects}} on the {{Tweeted}}: {{Experimentally Reducing Racist Harassment}}}.
\newblock \bibinfo{journal}{\emph{Political Behavior}} \bibinfo{volume}{39}, \bibinfo{number}{3} (\bibinfo{date}{Sept.} \bibinfo{year}{2017}), \bibinfo{pages}{629--649}.
\newblock
\showISSN{1573-6687}
\urldef\tempurl%
\url{https://doi.org/10.1007/s11109-016-9373-5}
\showDOI{\tempurl}


\bibitem[N{\"a}si et~al\mbox{.}(2015)]%
        {nasiExposureOnlineHate2015}
\bibfield{author}{\bibinfo{person}{Matti N{\"a}si}, \bibinfo{person}{Pekka R{\"a}s{\"a}nen}, \bibinfo{person}{James Hawdon}, \bibinfo{person}{Emma Holkeri}, {and} \bibinfo{person}{Atte Oksanen}.} \bibinfo{year}{2015}\natexlab{}.
\newblock \showarticletitle{Exposure to Online Hate Material and Social Trust among {{Finnish}} Youth}.
\newblock \bibinfo{journal}{\emph{Information Technology \& People}} \bibinfo{volume}{28}, \bibinfo{number}{3} (\bibinfo{date}{Jan.} \bibinfo{year}{2015}), \bibinfo{pages}{607--622}.
\newblock
\showISSN{0959-3845}
\urldef\tempurl%
\url{https://doi.org/10.1108/ITP-09-2014-0198}
\showDOI{\tempurl}


\bibitem[Nations(2023)]%
        {unitednationsUnitedNationsStrategy2023}
\bibfield{author}{\bibinfo{person}{United Nations}.} \bibinfo{year}{2023}\natexlab{}.
\newblock \bibinfo{title}{United {{Nations Strategy}} and {{Plan}} of {{Action}} on {{Hate Speech}}}.
\newblock \bibinfo{howpublished}{https://www.un.org/en/genocideprevention/documents/advising-and-mobilizing/Action\_plan\_on\_hate\_speech\_EN.pdf}.
\newblock


\bibitem[Nir and Sophie(2018)]%
        {nirPerceivedThreatBlaming2018}
\bibfield{author}{\bibinfo{person}{Rozmann Nir} {and} \bibinfo{person}{D.~Walsh Sophie}.} \bibinfo{year}{2018}\natexlab{}.
\newblock \showarticletitle{Perceived Threat, Blaming Attribution, Victim Ethnicity and Punishment}.
\newblock \bibinfo{journal}{\emph{International Journal of Intercultural Relations}}  \bibinfo{volume}{66} (\bibinfo{date}{Sept.} \bibinfo{year}{2018}), \bibinfo{pages}{34--40}.
\newblock
\showISSN{0147-1767}
\urldef\tempurl%
\url{https://doi.org/10.1016/j.ijintrel.2018.06.004}
\showDOI{\tempurl}


\bibitem[Obermaier et~al\mbox{.}(2023)]%
        {obermaierLlBeThere2023}
\bibfield{author}{\bibinfo{person}{Magdalena Obermaier}, \bibinfo{person}{Desir{\'e}e Schmuck}, {and} \bibinfo{person}{Muniba Saleem}.} \bibinfo{year}{2023}\natexlab{}.
\newblock \showarticletitle{I'll Be There for You? {{Effects}} of {{Islamophobic}} Online Hate Speech and Counter Speech on {{Muslim}} in-Group Bystanders' Intention to Intervene}.
\newblock \bibinfo{journal}{\emph{New Media \& Society}} \bibinfo{volume}{25}, \bibinfo{number}{9} (\bibinfo{date}{Sept.} \bibinfo{year}{2023}), \bibinfo{pages}{2339--2358}.
\newblock
\showISSN{1461-4448}
\urldef\tempurl%
\url{https://doi.org/10.1177/14614448211017527}
\showDOI{\tempurl}


\bibitem[{\DJ}or{\dj}evi{\'c}(2020)]%
        {dordevicSociocognitiveDimensionHate2020}
\bibfield{author}{\bibinfo{person}{Jasmina~P. {\DJ}or{\dj}evi{\'c}}.} \bibinfo{year}{2020}\natexlab{}.
\newblock \showarticletitle{The Sociocognitive Dimension of Hate Speech in Readers' Comments on {{Serbian}} News Websites}.
\newblock \bibinfo{journal}{\emph{Discourse, Context and Media}}  \bibinfo{volume}{33} (\bibinfo{year}{2020}).
\newblock
\urldef\tempurl%
\url{https://doi.org/10.1016/j.dcm.2019.100366}
\showDOI{\tempurl}


\bibitem[Ousidhoum et~al\mbox{.}(2019)]%
        {ousidhoumMultilingualMultiAspectHate2019}
\bibfield{author}{\bibinfo{person}{Nedjma Ousidhoum}, \bibinfo{person}{Zizheng Lin}, \bibinfo{person}{Hongming Zhang}, \bibinfo{person}{Yangqiu Song}, {and} \bibinfo{person}{Dit-Yan Yeung}.} \bibinfo{year}{2019}\natexlab{}.
\newblock \bibinfo{title}{Multilingual and {{Multi-Aspect Hate Speech Analysis}}}.
\newblock
\newblock
\urldef\tempurl%
\url{https://doi.org/10.48550/arXiv.1908.11049}
\showDOI{\tempurl}
\showeprint[arxiv]{1908.11049}~[cs]


\bibitem[Paluck and Green(2009)]%
        {paluckPrejudiceReductionWhat2009}
\bibfield{author}{\bibinfo{person}{Elizabeth~Levy Paluck} {and} \bibinfo{person}{Donald~P. Green}.} \bibinfo{year}{2009}\natexlab{}.
\newblock \showarticletitle{Prejudice {{Reduction}}: {{What Works}}? {{A Review}} and {{Assessment}} of {{Research}} and {{Practice}}}.
\newblock \bibinfo{journal}{\emph{Annual Review of Psychology}} \bibinfo{volume}{60}, \bibinfo{number}{1} (\bibinfo{year}{2009}), \bibinfo{pages}{339--367}.
\newblock
\urldef\tempurl%
\url{https://doi.org/10.1146/annurev.psych.60.110707.163607}
\showDOI{\tempurl}


\bibitem[Paluck et~al\mbox{.}(2019)]%
        {paluckContactHypothesisReevaluated2019}
\bibfield{author}{\bibinfo{person}{Elizabeth~Levy Paluck}, \bibinfo{person}{Seth~A. Green}, {and} \bibinfo{person}{Donald~P. Green}.} \bibinfo{year}{2019}\natexlab{}.
\newblock \showarticletitle{The Contact Hypothesis Re-Evaluated}.
\newblock \bibinfo{journal}{\emph{Behavioural Public Policy}} \bibinfo{volume}{3}, \bibinfo{number}{2} (\bibinfo{date}{Nov.} \bibinfo{year}{2019}), \bibinfo{pages}{129--158}.
\newblock
\showISSN{2398-063X, 2398-0648}
\urldef\tempurl%
\url{https://doi.org/10.1017/bpp.2018.25}
\showDOI{\tempurl}


\bibitem[Paolini et~al\mbox{.}(2010)]%
        {paoliniNegativeIntergroupContact2010}
\bibfield{author}{\bibinfo{person}{Stefania Paolini}, \bibinfo{person}{Jake Harwood}, {and} \bibinfo{person}{Mark Rubin}.} \bibinfo{year}{2010}\natexlab{}.
\newblock \showarticletitle{Negative {{Intergroup Contact Makes Group Memberships Salient}}: {{Explaining Why Intergroup Conflict Endures}}}.
\newblock \bibinfo{journal}{\emph{Personality and Social Psychology Bulletin}} \bibinfo{volume}{36}, \bibinfo{number}{12} (\bibinfo{date}{Dec.} \bibinfo{year}{2010}), \bibinfo{pages}{1723--1738}.
\newblock
\showISSN{0146-1672}
\urldef\tempurl%
\url{https://doi.org/10.1177/0146167210388667}
\showDOI{\tempurl}


\bibitem[Paz et~al\mbox{.}(2020)]%
        {pazHateSpeechSystematized2020}
\bibfield{author}{\bibinfo{person}{Mar{\'i}a~Antonia Paz}, \bibinfo{person}{Julio {Montero-D{\'i}az}}, {and} \bibinfo{person}{Alicia {Moreno-Delgado}}.} \bibinfo{year}{2020}\natexlab{}.
\newblock \showarticletitle{Hate {{Speech}}: {{A Systematized Review}}}.
\newblock \bibinfo{journal}{\emph{Sage Open}} \bibinfo{volume}{10}, \bibinfo{number}{4} (\bibinfo{date}{Oct.} \bibinfo{year}{2020}), \bibinfo{pages}{2158244020973022}.
\newblock
\showISSN{2158-2440}
\urldef\tempurl%
\url{https://doi.org/10.1177/2158244020973022}
\showDOI{\tempurl}


\bibitem[Perry(2001)]%
        {perryNameHateUnderstanding2001}
\bibfield{author}{\bibinfo{person}{Barbara Perry}.} \bibinfo{year}{2001}\natexlab{}.
\newblock \bibinfo{booktitle}{\emph{In the {{Name}} of {{Hate}}: {{Understanding Hate Crimes}}}}.
\newblock \bibinfo{publisher}{Routledge}, \bibinfo{address}{New York}.
\newblock
\showISBNx{978-0-203-90513-5}
\urldef\tempurl%
\url{https://doi.org/10.4324/9780203905135}
\showDOI{\tempurl}


\bibitem[Petersen and Ostendorf(2007)]%
        {petersenTextSimplificationLanguage2007}
\bibfield{author}{\bibinfo{person}{Sarah~E. Petersen} {and} \bibinfo{person}{Mari Ostendorf}.} \bibinfo{year}{2007}\natexlab{}.
\newblock \showarticletitle{Text Simplification for Language Learners: A Corpus Analysis}. In \bibinfo{booktitle}{\emph{Speech and {{Language Technology}} in {{Education}} ({{SLaTE}} 2007)}}. \bibinfo{publisher}{ISCA}, \bibinfo{pages}{69--72}.
\newblock
\urldef\tempurl%
\url{https://doi.org/10.21437/SLaTE.2007-20}
\showDOI{\tempurl}


\bibitem[Pettigrew and Tropp(2006)]%
        {pettigrewMetaanalyticTestIntergroup2006}
\bibfield{author}{\bibinfo{person}{Thomas~F. Pettigrew} {and} \bibinfo{person}{Linda~R. Tropp}.} \bibinfo{year}{2006}\natexlab{}.
\newblock \showarticletitle{A Meta-Analytic Test of Intergroup Contact Theory}.
\newblock \bibinfo{journal}{\emph{Journal of Personality and Social Psychology}} \bibinfo{volume}{90}, \bibinfo{number}{5} (\bibinfo{year}{2006}), \bibinfo{pages}{751--783}.
\newblock
\showISSN{1939-1315}
\urldef\tempurl%
\url{https://doi.org/10.1037/0022-3514.90.5.751}
\showDOI{\tempurl}


\bibitem[Pierce(2023)]%
        {pierceBetterChatGPTApp2023}
\bibfield{author}{\bibinfo{person}{David Pierce}.} \bibinfo{year}{2023}\natexlab{}.
\newblock \bibinfo{title}{A Better {{ChatGPT}} App: {{Poe}} Wants to Build the Universal {{AI}} Messaging Client}.
\newblock \bibinfo{howpublished}{https://www.theverge.com/23674656/poe-ai-chatbot-messaging-app}.
\newblock


\bibitem[Ping et~al\mbox{.}(2024)]%
        {pingCounterExploringMotivations2024}
\bibfield{author}{\bibinfo{person}{Kaike Ping}, \bibinfo{person}{Anisha Kumar}, \bibinfo{person}{Xiaohan Ding}, {and} \bibinfo{person}{Eugenia Rho}.} \bibinfo{year}{2024}\natexlab{}.
\newblock \bibinfo{title}{Behind the {{Counter}}: {{Exploring}} the {{Motivations}} and {{Barriers}} of {{Online Counterspeech Writing}}}.
\newblock
\newblock
\urldef\tempurl%
\url{https://doi.org/10.48550/arXiv.2403.17116}
\showDOI{\tempurl}
\showeprint[arxiv]{2403.17116}~[cs]


\bibitem[Poletto et~al\mbox{.}(2019)]%
        {polettoAnnotatingHateSpeech2019}
\bibfield{author}{\bibinfo{person}{Fabio Poletto}, \bibinfo{person}{Valerio Basile}, \bibinfo{person}{Cristina Bosco}, \bibinfo{person}{Viviana Patti}, {and} \bibinfo{person}{Marco Stranisci}.} \bibinfo{year}{2019}\natexlab{}.
\newblock \showarticletitle{Annotating Hate Speech: {{Three}} Schemes at Comparison}. In \bibinfo{booktitle}{\emph{{{CEUR WORKSHOP PROCEEDINGS}}}}, Vol.~\bibinfo{volume}{2481}. \bibinfo{publisher}{CEUR-WS}, \bibinfo{pages}{1--8}.
\newblock


\bibitem[Poletto et~al\mbox{.}(2021)]%
        {polettoResourcesBenchmarkCorpora2021}
\bibfield{author}{\bibinfo{person}{Fabio Poletto}, \bibinfo{person}{Valerio Basile}, \bibinfo{person}{M. Sanguinetti}, \bibinfo{person}{C. Bosco}, {and} \bibinfo{person}{V. Patti}.} \bibinfo{year}{2021}\natexlab{}.
\newblock \showarticletitle{Resources and Benchmark Corpora for Hate Speech Detection: A Systematic Review}.
\newblock \bibinfo{journal}{\emph{Language Resources and Evaluation}}  \bibinfo{volume}{55} (\bibinfo{year}{2021}).
\newblock
\urldef\tempurl%
\url{https://doi.org/10.1007/s10579-020-09502-8}
\showDOI{\tempurl}


\bibitem[Reichelmann et~al\mbox{.}(2021)]%
        {reichelmannHateKnowsNo2021}
\bibfield{author}{\bibinfo{person}{Ashley Reichelmann}, \bibinfo{person}{James Hawdon}, \bibinfo{person}{Matt Costello}, \bibinfo{person}{John Ryan}, \bibinfo{person}{Catherine Blaya}, \bibinfo{person}{Vicente Llorent}, \bibinfo{person}{Atte Oksanen}, \bibinfo{person}{Pekka R{\"a}s{\"a}nen}, {and} \bibinfo{person}{Izabela Zych}.} \bibinfo{year}{2021}\natexlab{}.
\newblock \showarticletitle{Hate {{Knows No Boundaries}}: {{Online Hate}} in {{Six Nations}}}.
\newblock \bibinfo{journal}{\emph{Deviant Behavior}} \bibinfo{volume}{42}, \bibinfo{number}{9} (\bibinfo{date}{Sept.} \bibinfo{year}{2021}), \bibinfo{pages}{1100--1111}.
\newblock
\showISSN{0163-9625}
\urldef\tempurl%
\url{https://doi.org/10.1080/01639625.2020.1722337}
\showDOI{\tempurl}


\bibitem[Reichelmann et~al\mbox{.}(2020)]%
        {reichelmannHateKnowsNo2020}
\bibfield{author}{\bibinfo{person}{Ashley~V. Reichelmann}, \bibinfo{person}{J. Hawdon}, \bibinfo{person}{Matthew Costello}, \bibinfo{person}{J. Ryan}, \bibinfo{person}{Catherine Blaya}, \bibinfo{person}{Vicente~J. Llorent}, \bibinfo{person}{Atte Oksanen}, \bibinfo{person}{P. R{\"a}s{\"a}nen}, {and} \bibinfo{person}{Izabela Zych}.} \bibinfo{year}{2020}\natexlab{}.
\newblock \showarticletitle{Hate {{Knows No Boundaries}}: {{Online Hate}} in {{Six Nations}}}.
\newblock \bibinfo{journal}{\emph{Deviant Behavior}}  \bibinfo{volume}{42} (\bibinfo{year}{2020}).
\newblock
\urldef\tempurl%
\url{https://doi.org/10.1080/01639625.2020.1722337}
\showDOI{\tempurl}


\bibitem[Reid et~al\mbox{.}(2022)]%
        {reidFeelingGoodControl2022}
\bibfield{author}{\bibinfo{person}{Elizabeth Reid}, \bibinfo{person}{Regan~L. Mandryk}, \bibinfo{person}{Nicole~A. Beres}, \bibinfo{person}{Madison Klarkowski}, {and} \bibinfo{person}{Julian Frommel}.} \bibinfo{year}{2022}\natexlab{}.
\newblock \showarticletitle{Feeling {{Good}} and {{In Control}}: {{In-game Tools}} to {{Support Targets}} of {{Toxicity}}}.
\newblock \bibinfo{journal}{\emph{Proceedings of the ACM on Human-Computer Interaction}} \bibinfo{volume}{6}, \bibinfo{number}{CHI PLAY} (\bibinfo{date}{Oct.} \bibinfo{year}{2022}), \bibinfo{pages}{235:1--235:27}.
\newblock
\urldef\tempurl%
\url{https://doi.org/10.1145/3549498}
\showDOI{\tempurl}


\bibitem[Rho et~al\mbox{.}(2017)]%
        {rhoClassConfessionsRestorative2017}
\bibfield{author}{\bibinfo{person}{Eugenia Ha~Rim Rho}, \bibinfo{person}{Oliver~L. Haimson}, \bibinfo{person}{Nazanin Andalibi}, \bibinfo{person}{Melissa Mazmanian}, {and} \bibinfo{person}{Gillian~R. Hayes}.} \bibinfo{year}{2017}\natexlab{}.
\newblock \showarticletitle{Class {{Confessions}}: {{Restorative Properties}} in {{Online Experiences}} of {{Socioeconomic Stigma}}}. In \bibinfo{booktitle}{\emph{Proceedings of the 2017 {{CHI Conference}} on {{Human Factors}} in {{Computing Systems}}}} \emph{(\bibinfo{series}{{{CHI}} '17})}. \bibinfo{publisher}{Association for Computing Machinery}, \bibinfo{address}{New York, NY, USA}, \bibinfo{pages}{3377--3389}.
\newblock
\showISBNx{978-1-4503-4655-9}
\urldef\tempurl%
\url{https://doi.org/10.1145/3025453.3025921}
\showDOI{\tempurl}


\bibitem[Rho et~al\mbox{.}(2018)]%
        {rhoFosteringCivilDiscourse2018}
\bibfield{author}{\bibinfo{person}{Eugenia Ha~Rim Rho}, \bibinfo{person}{Gloria Mark}, {and} \bibinfo{person}{Melissa Mazmanian}.} \bibinfo{year}{2018}\natexlab{}.
\newblock \showarticletitle{Fostering {{Civil Discourse Online}}: {{Linguistic Behavior}} in {{Comments}} of \#{{MeToo Articles}} across {{Political Perspectives}}}.
\newblock \bibinfo{journal}{\emph{Proceedings of the ACM on Human-Computer Interaction}} \bibinfo{volume}{2}, \bibinfo{number}{CSCW} (\bibinfo{date}{Nov.} \bibinfo{year}{2018}), \bibinfo{pages}{147:1--147:28}.
\newblock
\urldef\tempurl%
\url{https://doi.org/10.1145/3274416}
\showDOI{\tempurl}


\bibitem[Rho and Mazmanian(2019)]%
        {rhoHashtagBurnoutControl2019}
\bibfield{author}{\bibinfo{person}{Eugenia Ha~Rim Rho} {and} \bibinfo{person}{Melissa Mazmanian}.} \bibinfo{year}{2019}\natexlab{}.
\newblock \showarticletitle{Hashtag {{Burnout}}? {{A Control Experiment Investigating How Political Hashtags Shape Reactions}} to {{News Content}}}.
\newblock \bibinfo{journal}{\emph{Proc. ACM Hum.-Comput. Interact.}} \bibinfo{volume}{3}, \bibinfo{number}{CSCW} (\bibinfo{date}{Nov.} \bibinfo{year}{2019}), \bibinfo{pages}{197:1--197:25}.
\newblock
\urldef\tempurl%
\url{https://doi.org/10.1145/3359299}
\showDOI{\tempurl}


\bibitem[Rho and Mazmanian(2020)]%
        {rhoPoliticalHashtagsLost2020}
\bibfield{author}{\bibinfo{person}{Eugenia Ha~Rim Rho} {and} \bibinfo{person}{Melissa Mazmanian}.} \bibinfo{year}{2020}\natexlab{}.
\newblock \showarticletitle{Political {{Hashtags}} \& the {{Lost Art}} of {{Democratic Discourse}}}. In \bibinfo{booktitle}{\emph{Proceedings of the 2020 {{CHI Conference}} on {{Human Factors}} in {{Computing Systems}}}} \emph{(\bibinfo{series}{{{CHI}} '20})}. \bibinfo{publisher}{Association for Computing Machinery}, \bibinfo{address}{New York, NY, USA}, \bibinfo{pages}{1--13}.
\newblock
\showISBNx{978-1-4503-6708-0}
\urldef\tempurl%
\url{https://doi.org/10.1145/3313831.3376542}
\showDOI{\tempurl}


\bibitem[Richards et~al\mbox{.}(2016)]%
        {richardsExplainingFemaleVictim2016}
\bibfield{author}{\bibinfo{person}{Tara~N. Richards}, \bibinfo{person}{Wesley~G. Jennings}, \bibinfo{person}{M.~Dwayne Smith}, \bibinfo{person}{Christine~S. Sellers}, \bibinfo{person}{Sondra~J. Fogel}, {and} \bibinfo{person}{Beth Bjerregaard}.} \bibinfo{year}{2016}\natexlab{}.
\newblock \showarticletitle{Explaining the ``{{Female Victim Effect}}'' in {{Capital Punishment}}: {{An Examination}} of {{Victim Sex}}--{{Specific Models}} of {{Juror Sentence Decision-Making}}}.
\newblock \bibinfo{journal}{\emph{Crime \& Delinquency}} \bibinfo{volume}{62}, \bibinfo{number}{7} (\bibinfo{date}{July} \bibinfo{year}{2016}), \bibinfo{pages}{875--898}.
\newblock
\showISSN{0011-1287}
\urldef\tempurl%
\url{https://doi.org/10.1177/0011128714530826}
\showDOI{\tempurl}


\bibitem[Rieger et~al\mbox{.}(2018)]%
        {riegerHateCountervoicesInternet2018}
\bibfield{author}{\bibinfo{person}{Diana Rieger}, \bibinfo{person}{Josephine~B. Schmitt}, {and} \bibinfo{person}{Lena Frischlich}.} \bibinfo{year}{2018}\natexlab{}.
\newblock \showarticletitle{Hate and Counter-Voices in the {{Internet}}: {{Introduction}} to the Special Issue}.
\newblock \bibinfo{journal}{\emph{SCM Studies in Communication and Media}} \bibinfo{volume}{7}, \bibinfo{number}{4} (\bibinfo{date}{Dec.} \bibinfo{year}{2018}), \bibinfo{pages}{459--472}.
\newblock
\showISSN{2192-4007}
\urldef\tempurl%
\url{https://doi.org/10.5771/2192-4007-2018-4-459}
\showDOI{\tempurl}


\bibitem[Robertson et~al\mbox{.}(2023)]%
        {robertsonNegativityDrivesOnline2023}
\bibfield{author}{\bibinfo{person}{Claire~E. Robertson}, \bibinfo{person}{Nicolas Pr{\"o}llochs}, \bibinfo{person}{Kaoru Schwarzenegger}, \bibinfo{person}{Philip P{\"a}rnamets}, \bibinfo{person}{Jay~J. Van~Bavel}, {and} \bibinfo{person}{Stefan Feuerriegel}.} \bibinfo{year}{2023}\natexlab{}.
\newblock \showarticletitle{Negativity Drives Online News Consumption}.
\newblock \bibinfo{journal}{\emph{Nature Human Behaviour}} \bibinfo{volume}{7}, \bibinfo{number}{5} (\bibinfo{date}{May} \bibinfo{year}{2023}), \bibinfo{pages}{812--822}.
\newblock
\showISSN{2397-3374}
\urldef\tempurl%
\url{https://doi.org/10.1038/s41562-023-01538-4}
\showDOI{\tempurl}


\bibitem[Roussos and Dovidio(2018)]%
        {roussosHateSpeechEye2018}
\bibfield{author}{\bibinfo{person}{Gina Roussos} {and} \bibinfo{person}{J. Dovidio}.} \bibinfo{year}{2018}\natexlab{}.
\newblock \showarticletitle{Hate {{Speech Is}} in the {{Eye}} of the {{Beholder}}}.
\newblock \bibinfo{journal}{\emph{Social Psychological and Personality Science}}  \bibinfo{volume}{9} (\bibinfo{year}{2018}).
\newblock
\urldef\tempurl%
\url{https://doi.org/10.1177/1948550617748728}
\showDOI{\tempurl}


\bibitem[Ruths et~al\mbox{.}(2016a)]%
        {ruthsConsiderationsSuccessfulCounterspeech2016}
\bibfield{author}{\bibinfo{person}{Derek~Ruths Ruths}, \bibinfo{person}{Haji Mohammad~Saleem Saleem}, \bibinfo{person}{Kelly P.~Dillon Dillon}, \bibinfo{person}{Lucas~Wright Wright}, {and} \bibinfo{person}{Susan~Benesch Benesch}.} \bibinfo{year}{2016}\natexlab{a}.
\newblock \bibinfo{booktitle}{\emph{Considerations for {{Successful Counterspeech}}}}.
\newblock \bibinfo{type}{{T}echnical {R}eport}. \bibinfo{institution}{Dangerous Speech Project}, \bibinfo{address}{Washington, DC USA}.
\newblock
\urldef\tempurl%
\url{https://doi.org/10.15868/socialsector.34065}
\showDOI{\tempurl}


\bibitem[Ruths et~al\mbox{.}(2016b)]%
        {ruthsCounterspeechTwitterField2016}
\bibfield{author}{\bibinfo{person}{Derek~Ruths Ruths}, \bibinfo{person}{Haji Mohammed~Saleem Saleem}, \bibinfo{person}{Kelly P.~Dillon Dillon}, \bibinfo{person}{Lucas~Wright Wright}, {and} \bibinfo{person}{Susan~Benesch Benesch}.} \bibinfo{year}{2016}\natexlab{b}.
\newblock \bibinfo{booktitle}{\emph{Counterspeech on {{Twitter}}: {{A Field Study}}}}.
\newblock \bibinfo{type}{{T}echnical {R}eport}. \bibinfo{institution}{Dangerous Speech Project}, \bibinfo{address}{Washington, DC USA}.
\newblock
\urldef\tempurl%
\url{https://doi.org/10.15868/socialsector.34066}
\showDOI{\tempurl}


\bibitem[Saha(2023)]%
        {sahaSelfsupervisionControllingTechniques2023}
\bibfield{author}{\bibinfo{person}{Punyajoy Saha}.} \bibinfo{year}{2023}\natexlab{}.
\newblock \showarticletitle{Self-Supervision and {{Controlling Techniques}} to {{Improve Counter Speech Generation}}}. In \bibinfo{booktitle}{\emph{Proceedings of the {{Sixteenth ACM International Conference}} on {{Web Search}} and {{Data Mining}}}} \emph{(\bibinfo{series}{{{WSDM}} '23})}. \bibinfo{publisher}{Association for Computing Machinery}, \bibinfo{address}{New York, NY, USA}, \bibinfo{pages}{1224--1225}.
\newblock
\showISBNx{978-1-4503-9407-9}
\urldef\tempurl%
\url{https://doi.org/10.1145/3539597.3572991}
\showDOI{\tempurl}


\bibitem[Salmivalli et~al\mbox{.}(2011)]%
        {salmivalliBystandersMatterAssociations2011}
\bibfield{author}{\bibinfo{person}{Christina Salmivalli}, \bibinfo{person}{Marinus Voeten}, {and} \bibinfo{person}{Elisa Poskiparta}.} \bibinfo{year}{2011}\natexlab{}.
\newblock \showarticletitle{Bystanders {{Matter}}: {{Associations Between Reinforcing}}, {{Defending}}, and the {{Frequency}} of {{Bullying Behavior}} in {{Classrooms}}}.
\newblock \bibinfo{journal}{\emph{Journal of Clinical Child \& Adolescent Psychology}} \bibinfo{volume}{40}, \bibinfo{number}{5} (\bibinfo{date}{Sept.} \bibinfo{year}{2011}), \bibinfo{pages}{668--676}.
\newblock
\showISSN{1537-4416}
\urldef\tempurl%
\url{https://doi.org/10.1080/15374416.2011.597090}
\showDOI{\tempurl}


\bibitem[Saltman et~al\mbox{.}(2023)]%
        {saltmanNewModelsDeploying2023}
\bibfield{author}{\bibinfo{person}{Erin Saltman}, \bibinfo{person}{Farshad Kooti}, {and} \bibinfo{person}{Karly Vockery}.} \bibinfo{year}{2023}\natexlab{}.
\newblock \showarticletitle{New {{Models}} for {{Deploying Counterspeech}}: {{Measuring Behavioral Change}} and {{Sentiment Analysis}}}.
\newblock \bibinfo{journal}{\emph{Studies in Conflict \& Terrorism}} \bibinfo{volume}{46}, \bibinfo{number}{9} (\bibinfo{date}{Sept.} \bibinfo{year}{2023}), \bibinfo{pages}{1547--1574}.
\newblock
\showISSN{1057-610X}
\urldef\tempurl%
\url{https://doi.org/10.1080/1057610X.2021.1888404}
\showDOI{\tempurl}


\bibitem[Saveski et~al\mbox{.}(2022)]%
        {saveskiEngagingPoliticallyDiverse2022}
\bibfield{author}{\bibinfo{person}{Martin Saveski}, \bibinfo{person}{Doug Beeferman}, \bibinfo{person}{David McClure}, {and} \bibinfo{person}{Deb Roy}.} \bibinfo{year}{2022}\natexlab{}.
\newblock \showarticletitle{Engaging {{Politically Diverse Audiences}} on {{Social Media}}}.
\newblock \bibinfo{journal}{\emph{Proceedings of the International AAAI Conference on Web and Social Media}}  \bibinfo{volume}{16} (\bibinfo{date}{May} \bibinfo{year}{2022}), \bibinfo{pages}{873--884}.
\newblock
\showISSN{2334-0770}
\urldef\tempurl%
\url{https://doi.org/10.1609/icwsm.v16i1.19342}
\showDOI{\tempurl}


\bibitem[Schieb and Preuss(2016)]%
        {schiebGoverningHateSpeech2016}
\bibfield{author}{\bibinfo{person}{Carla Schieb} {and} \bibinfo{person}{Mike Preuss}.} \bibinfo{year}{2016}\natexlab{}.
\newblock \bibinfo{booktitle}{\emph{Governing Hate Speech by Means of Counterspeech on {{Facebook}}}}.
\newblock


\bibitem[Schmid et~al\mbox{.}(2024)]%
        {schmidHowSocialMedia2024}
\bibfield{author}{\bibinfo{person}{Ursula~Kristin Schmid}, \bibinfo{person}{Anna~Sophie K{\"u}mpel}, {and} \bibinfo{person}{Diana Rieger}.} \bibinfo{year}{2024}\natexlab{}.
\newblock \showarticletitle{How Social Media Users Perceive Different Forms of Online Hate Speech: {{A}} Qualitative Multi-Method Study}.
\newblock \bibinfo{journal}{\emph{New Media \& Society}} \bibinfo{volume}{26}, \bibinfo{number}{5} (\bibinfo{date}{May} \bibinfo{year}{2024}), \bibinfo{pages}{2614--2632}.
\newblock
\showISSN{1461-4448}
\urldef\tempurl%
\url{https://doi.org/10.1177/14614448221091185}
\showDOI{\tempurl}


\bibitem[Schmidt and Wiegand(2017)]%
        {schmidtSurveyHateSpeech2017}
\bibfield{author}{\bibinfo{person}{Anna Schmidt} {and} \bibinfo{person}{Michael Wiegand}.} \bibinfo{year}{2017}\natexlab{}.
\newblock \showarticletitle{A {{Survey}} on {{Hate Speech Detection}} Using {{Natural Language Processing}}}. In \bibinfo{booktitle}{\emph{Proceedings of the {{Fifth International Workshop}} on {{Natural Language Processing}} for {{Social Media}}}}. \bibinfo{publisher}{Association for Computational Linguistics}, \bibinfo{address}{Valencia, Spain}, \bibinfo{pages}{1--10}.
\newblock
\urldef\tempurl%
\url{https://doi.org/10.18653/v1/W17-1101}
\showDOI{\tempurl}


\bibitem[Seering et~al\mbox{.}(2018)]%
        {seeringApplicationsSocialIdentity2018}
\bibfield{author}{\bibinfo{person}{Joseph Seering}, \bibinfo{person}{Felicia Ng}, \bibinfo{person}{Zheng Yao}, {and} \bibinfo{person}{Geoff Kaufman}.} \bibinfo{year}{2018}\natexlab{}.
\newblock \showarticletitle{Applications of {{Social Identity Theory}} to {{Research}} and {{Design}} in {{Computer-Supported Cooperative Work}}}.
\newblock \bibinfo{journal}{\emph{Proceedings of the ACM on Human-Computer Interaction}} \bibinfo{volume}{2}, \bibinfo{number}{CSCW} (\bibinfo{date}{Nov.} \bibinfo{year}{2018}), \bibinfo{pages}{201:1--201:34}.
\newblock
\urldef\tempurl%
\url{https://doi.org/10.1145/3274771}
\showDOI{\tempurl}


\bibitem[Seglow(2016)]%
        {seglowHateSpeechDignity2016}
\bibfield{author}{\bibinfo{person}{Jonathan Seglow}.} \bibinfo{year}{2016}\natexlab{}.
\newblock \showarticletitle{Hate {{Speech}}, {{Dignity}} and {{Self-Respect}}}.
\newblock \bibinfo{journal}{\emph{Ethical Theory and Moral Practice}}  \bibinfo{volume}{19} (\bibinfo{year}{2016}).
\newblock
\urldef\tempurl%
\url{https://doi.org/10.1007/S10677-016-9744-3}
\showDOI{\tempurl}


\bibitem[Sherry(2019)]%
        {sherryDisablistHateSpeech2019}
\bibfield{author}{\bibinfo{person}{Mark Sherry}.} \bibinfo{year}{2019}\natexlab{}.
\newblock \showarticletitle{Disablist Hate Speech Online}.
\newblock In \bibinfo{booktitle}{\emph{Disability {{Hate Speech}}}}. \bibinfo{publisher}{Routledge}.
\newblock
\showISBNx{978-0-429-20181-3}


\bibitem[Sherry et~al\mbox{.}(2019)]%
        {sherryDisabilityHateSpeech2019}
\bibfield{author}{\bibinfo{person}{Mark Sherry}, \bibinfo{person}{Terje Olsen}, \bibinfo{person}{Janikke~Solstad Vedeler}, {and} \bibinfo{person}{John Eriksen}.} \bibinfo{year}{2019}\natexlab{}.
\newblock \bibinfo{booktitle}{\emph{Disability {{Hate Speech}}: {{Social}}, {{Cultural}} and {{Political Contexts}}}}.
\newblock \bibinfo{publisher}{Routledge}.
\newblock
\showISBNx{978-0-429-51391-6}


\bibitem[Simpson(2013)]%
        {simpsonDignityHarmHate2013}
\bibfield{author}{\bibinfo{person}{R. Simpson}.} \bibinfo{year}{2013}\natexlab{}.
\newblock \showarticletitle{Dignity, {{Harm}}, and {{Hate Speech}}}.
\newblock \bibinfo{journal}{\emph{Law and Philosophy}}  \bibinfo{volume}{32} (\bibinfo{year}{2013}).
\newblock
\urldef\tempurl%
\url{https://doi.org/10.1007/S10982-012-9164-Z}
\showDOI{\tempurl}


\bibitem[Soral et~al\mbox{.}(2018)]%
        {soralExposureHateSpeech2018}
\bibfield{author}{\bibinfo{person}{Wiktor Soral}, \bibinfo{person}{M. Bilewicz}, {and} \bibinfo{person}{Mikolaj Winiewski}.} \bibinfo{year}{2018}\natexlab{}.
\newblock \showarticletitle{Exposure to Hate Speech Increases Prejudice through Desensitization}.
\newblock \bibinfo{journal}{\emph{Aggressive Behavior}}  \bibinfo{volume}{44} (\bibinfo{year}{2018}).
\newblock
\urldef\tempurl%
\url{https://doi.org/10.1002/ab.21737}
\showDOI{\tempurl}


\bibitem[Stroud and Cox(2018)]%
        {stroudVarietiesFeministCounterspeech2018b}
\bibfield{author}{\bibinfo{person}{Scott~R. Stroud} {and} \bibinfo{person}{William Cox}.} \bibinfo{year}{2018}\natexlab{}.
\newblock \showarticletitle{The {{Varieties}} of {{Feminist Counterspeech}} in the {{Misogynistic Online World}}}.
\newblock In \bibinfo{booktitle}{\emph{Mediating {{Misogyny}}: {{Gender}}, {{Technology}}, and {{Harassment}}}}, \bibfield{editor}{\bibinfo{person}{Jacqueline~Ryan Vickery} {and} \bibinfo{person}{Tracy Everbach}} (Eds.). \bibinfo{publisher}{Springer International Publishing}, \bibinfo{address}{Cham}, \bibinfo{pages}{293--310}.
\newblock
\showISBNx{978-3-319-72917-6}
\urldef\tempurl%
\url{https://doi.org/10.1007/978-3-319-72917-6_15}
\showDOI{\tempurl}


\bibitem[Tontodimamma et~al\mbox{.}(2020)]%
        {tontodimammaThirtyYearsResearch2020}
\bibfield{author}{\bibinfo{person}{Alice Tontodimamma}, \bibinfo{person}{E. Nissi}, \bibinfo{person}{A. Sarra}, {and} \bibinfo{person}{Lara Fontanella}.} \bibinfo{year}{2020}\natexlab{}.
\newblock \showarticletitle{Thirty Years of Research into Hate Speech: Topics of Interest and Their Evolution}.
\newblock \bibinfo{journal}{\emph{Scientometrics}}  \bibinfo{volume}{126} (\bibinfo{year}{2020}).
\newblock
\urldef\tempurl%
\url{https://doi.org/10.1007/s11192-020-03737-6}
\showDOI{\tempurl}


\bibitem[Vedeler et~al\mbox{.}(2019)]%
        {vedelerHateSpeechHarms2019}
\bibfield{author}{\bibinfo{person}{Janikke~Solstad Vedeler}, \bibinfo{person}{Terje Olsen}, {and} \bibinfo{person}{John Eriksen}.} \bibinfo{year}{2019}\natexlab{}.
\newblock \showarticletitle{Hate Speech Harms: A Social Justice Discussion of Disabled {{Norwegians}}' Experiences}.
\newblock \bibinfo{journal}{\emph{Disability \& Society}} \bibinfo{volume}{34}, \bibinfo{number}{3} (\bibinfo{date}{March} \bibinfo{year}{2019}), \bibinfo{pages}{368--383}.
\newblock
\showISSN{0968-7599}
\urldef\tempurl%
\url{https://doi.org/10.1080/09687599.2018.1515723}
\showDOI{\tempurl}


\bibitem[Vidgen et~al\mbox{.}(2021)]%
        {vidgenLearningWorstDynamically2021}
\bibfield{author}{\bibinfo{person}{Bertie Vidgen}, \bibinfo{person}{Tristan Thrush}, \bibinfo{person}{Zeerak Waseem}, {and} \bibinfo{person}{Douwe Kiela}.} \bibinfo{year}{2021}\natexlab{}.
\newblock \bibinfo{title}{Learning from the {{Worst}}: {{Dynamically Generated Datasets}} to {{Improve Online Hate Detection}}}.
\newblock
\newblock
\urldef\tempurl%
\url{https://doi.org/10.48550/arXiv.2012.15761}
\showDOI{\tempurl}
\showeprint[arxiv]{2012.15761}~[cs]


\bibitem[{von Hippel} et~al\mbox{.}(2011)]%
        {vonhippelStereotypeThreatFemale2011}
\bibfield{author}{\bibinfo{person}{Courtney {von Hippel}}, \bibinfo{person}{Cindy Wiryakusuma}, \bibinfo{person}{Jessica Bowden}, {and} \bibinfo{person}{Megan Shochet}.} \bibinfo{year}{2011}\natexlab{}.
\newblock \showarticletitle{Stereotype {{Threat}} and {{Female Communication Styles}}}.
\newblock \bibinfo{journal}{\emph{Personality and Social Psychology Bulletin}} \bibinfo{volume}{37}, \bibinfo{number}{10} (\bibinfo{date}{Oct.} \bibinfo{year}{2011}), \bibinfo{pages}{1312--1324}.
\newblock
\showISSN{0146-1672}
\urldef\tempurl%
\url{https://doi.org/10.1177/0146167211410439}
\showDOI{\tempurl}


\bibitem[Wachs and Wright(2018)]%
        {wachsAssociationsBystandersPerpetrators2018}
\bibfield{author}{\bibinfo{person}{Sebastian Wachs} {and} \bibinfo{person}{Michelle~F. Wright}.} \bibinfo{year}{2018}\natexlab{}.
\newblock \showarticletitle{Associations between {{Bystanders}} and {{Perpetrators}} of {{Online Hate}}: {{The Moderating Role}} of {{Toxic Online Disinhibition}}}.
\newblock \bibinfo{journal}{\emph{International Journal of Environmental Research and Public Health}}  \bibinfo{volume}{15} (\bibinfo{year}{2018}).
\newblock
\urldef\tempurl%
\url{https://doi.org/10.3390/ijerph15092030}
\showDOI{\tempurl}


\bibitem[Wachs et~al\mbox{.}(2019)]%
        {wachsAssociationsWitnessingPerpetrating2019}
\bibfield{author}{\bibinfo{person}{Sebastian Wachs}, \bibinfo{person}{Michelle~F. Wright}, \bibinfo{person}{Ruthaychonnee Sittichai}, \bibinfo{person}{Ritu Singh}, \bibinfo{person}{Ramakrishna Biswal}, \bibinfo{person}{Eun-mee Kim}, \bibinfo{person}{Soeun Yang}, \bibinfo{person}{Manuel {G{\'a}mez-Guadix}}, \bibinfo{person}{Carmen Almendros}, \bibinfo{person}{Katerina Flora}, \bibinfo{person}{Vassiliki Daskalou}, {and} \bibinfo{person}{Evdoxia Maziridou}.} \bibinfo{year}{2019}\natexlab{}.
\newblock \showarticletitle{Associations between {{Witnessing}} and {{Perpetrating Online Hate}} in {{Eight Countries}}: {{The Buffering Effects}} of {{Problem-Focused Coping}}}.
\newblock \bibinfo{journal}{\emph{International Journal of Environmental Research and Public Health}} \bibinfo{volume}{16}, \bibinfo{number}{20} (\bibinfo{date}{Jan.} \bibinfo{year}{2019}), \bibinfo{pages}{3992}.
\newblock
\showISSN{1660-4601}
\urldef\tempurl%
\url{https://doi.org/10.3390/ijerph16203992}
\showDOI{\tempurl}


\bibitem[Wan et~al\mbox{.}(2015)]%
        {wanKappaCoefficientPopular2015}
\bibfield{author}{\bibinfo{person}{{\relax TANG} Wan}, \bibinfo{person}{H.~U. Jun}, \bibinfo{person}{Hui Zhang}, \bibinfo{person}{W.~U. Pan}, {and} \bibinfo{person}{H.~E. Hua}.} \bibinfo{year}{2015}\natexlab{}.
\newblock \showarticletitle{Kappa Coefficient: A Popular Measure of Rater Agreement}.
\newblock \bibinfo{journal}{\emph{Shanghai archives of psychiatry}} \bibinfo{volume}{27}, \bibinfo{number}{1} (\bibinfo{year}{2015}), \bibinfo{pages}{62}.
\newblock


\bibitem[Westerman et~al\mbox{.}(2020)]%
        {westermanIItIThouIRobot2020}
\bibfield{author}{\bibinfo{person}{D. Westerman}, \bibinfo{person}{Autumn~P. Edwards}, \bibinfo{person}{Chad Edwards}, \bibinfo{person}{Zhenyang Luo}, {and} \bibinfo{person}{P. Spence}.} \bibinfo{year}{2020}\natexlab{}.
\newblock \showarticletitle{I-{{It}}, {{I-Thou}}, {{I-Robot}}: {{The Perceived Humanness}} of {{AI}} in {{Human-Machine Communication}}}.
\newblock \bibinfo{journal}{\emph{Communication Studies}}  \bibinfo{volume}{71} (\bibinfo{year}{2020}), \bibinfo{pages}{393--408}.
\newblock
\urldef\tempurl%
\url{https://doi.org/10.1080/10510974.2020.1749683}
\showDOI{\tempurl}


\bibitem[Whitney(1998)]%
        {whitneyComputerMediatedCommunication1998}
\bibfield{author}{\bibinfo{person}{Gretchen Whitney}.} \bibinfo{year}{1998}\natexlab{}.
\newblock \showarticletitle{Computer-mediated Communication: {{Linguistic}}, Social, and Cross-cultural Perspectives}.
\newblock \bibinfo{journal}{\emph{Journal of the American Society for Information Science}} \bibinfo{volume}{49}, \bibinfo{number}{9} (\bibinfo{year}{1998}), \bibinfo{pages}{859--860}.
\newblock


\bibitem[Wright et~al\mbox{.}(2017)]%
        {wrightVectorsCounterspeechTwitter2017}
\bibfield{author}{\bibinfo{person}{Lucas Wright}, \bibinfo{person}{Derek Ruths}, \bibinfo{person}{Kelly~P Dillon}, \bibinfo{person}{Haji~Mohammad Saleem}, {and} \bibinfo{person}{Susan Benesch}.} \bibinfo{year}{2017}\natexlab{}.
\newblock \showarticletitle{Vectors for {{Counterspeech}} on {{Twitter}}}. In \bibinfo{booktitle}{\emph{Proceedings of the {{First Workshop}} on {{Abusive Language Online}}}}, \bibfield{editor}{\bibinfo{person}{Zeerak Waseem}, \bibinfo{person}{Wendy Hui~Kyong Chung}, \bibinfo{person}{Dirk Hovy}, {and} \bibinfo{person}{Joel Tetreault}} (Eds.). \bibinfo{publisher}{Association for Computational Linguistics}, \bibinfo{address}{Vancouver, BC, Canada}, \bibinfo{pages}{57--62}.
\newblock
\urldef\tempurl%
\url{https://doi.org/10.18653/v1/W17-3009}
\showDOI{\tempurl}


\bibitem[Zhang and Luo(2018)]%
        {zhangHateSpeechDetection2018}
\bibfield{author}{\bibinfo{person}{Ziqi Zhang} {and} \bibinfo{person}{Le Luo}.} \bibinfo{year}{2018}\natexlab{}.
\newblock \showarticletitle{Hate {{Speech Detection}}: {{A Solved Problem}}? {{The Challenging Case}} of {{Long Tail}} on {{Twitter}}}.
\newblock \bibinfo{journal}{\emph{Semantic Web}}  \bibinfo{volume}{10} (\bibinfo{year}{2018}).
\newblock
\urldef\tempurl%
\url{https://doi.org/10.3233/SW-180338}
\showDOI{\tempurl}


\bibitem[Zhu and Bhat(2021)]%
        {zhuGeneratePruneSelect2021}
\bibfield{author}{\bibinfo{person}{Wanzheng Zhu} {and} \bibinfo{person}{Suma Bhat}.} \bibinfo{year}{2021}\natexlab{}.
\newblock \showarticletitle{Generate, {{Prune}}, {{Select}}: {{A Pipeline}} for {{Counterspeech Generation}} against {{Online Hate Speech}}}. In \bibinfo{booktitle}{\emph{Findings of the {{Association}} for {{Computational Linguistics}}: {{ACL-IJCNLP}} 2021}}. \bibinfo{pages}{134--149}.
\newblock


\bibitem[Ziegele et~al\mbox{.}(2018)]%
        {ziegeleJournalisticCountervoicesComment2018}
\bibfield{author}{\bibinfo{person}{Marc Ziegele}, \bibinfo{person}{Pablo Jost}, \bibinfo{person}{Marike Bormann}, {and} \bibinfo{person}{Dominique Heinbach}.} \bibinfo{year}{2018}\natexlab{}.
\newblock \showarticletitle{Journalistic Counter-Voices in Comment Sections: {{Patterns}}, Determinants, and Potential Consequences of Interactive Moderation of Uncivil User Comments}.
\newblock \bibinfo{journal}{\emph{Studies in Communication {\textbar} Media}} \bibinfo{volume}{7}, \bibinfo{number}{4} (\bibinfo{year}{2018}), \bibinfo{pages}{525--554}.
\newblock
\showISSN{2192-4007}
\urldef\tempurl%
\url{https://doi.org/10.5771/2192-4007-2018-4-525}
\showDOI{\tempurl}


\end{thebibliography}


\end{document}


\appendix

\setcounter{table}{0}
\renewcommand{\thetable}{A\arabic{table}}

\section{Appendix}

\subsection{Table of Examples of Counterspeech Strategies in Response to Hate Posts}

\centering
    \begin{longtable}{|c|p{20em}|p{20em}|}
    \caption{Examples of Counterspeech Strategies in Response to Hate Posts}
      \label{tab:example}\\
    \hline
    \rowcolor[rgb]{ .933,  .925,  .882} \multicolumn{1}{|l|}{\textbf{Strategy}} & \multicolumn{1}{l|}{\textbf{Hate Speech Examples}} & \multicolumn{1}{l|}{\textbf{Counterspeech Examples}} \bigstrut\\
    \hline
    \multirow{3}[6]{*}[-15em]{\rotatebox{90}{Empathy}} & I don't want to see a gay couple get married. I don't understand why they are so annoying. & As a gay person myself, reading comments like this are hurtful. You’re implying that just because someone is gay and wants to marry someone they love as being “annoying.” I don’t find anything annoying about two people simply loving one another and wanting to share their lives together. In fact, I wish everyone including you to find their happiness in life.   \bigstrut\\
\cline{2-3}          & I hate when people fight stereotypes. We are different. Blacks are a different color and shouldn't live with us. & You're right, people are all different but that does not mean that those that are "different" deserve to be treated poorly...How would you like it if the roles were reversed, and someone said that about you and your skin color?... People should be judged by their actions and character and not by the color of their skin.   \bigstrut\\
\cline{2-3}          & Children with learning difficulties should be denied to attend school by teachers and parents. & Children with learning difficulties are every bit as deserving of an education as any other child. How would you feel if this was your child and they were denied a chance to learn to the best of their ability by the preconceived notions of people who have never met them and do not know their capabilities?
 \bigstrut\\
    \hline
    \multirow{3}[6]{*}[-6em]{\rotatebox{90}{Humor}} & Your gay friends probably have sex with dogs, you are an embarrassment to your family. & Not sure where you got the idea that gay people have sex with dogs. Your google search history must be wild. The only embarrassment here is you. \bigstrut\\
\cline{2-3}          & I miss the good old days when the West was ruled by whites, and blacks were subjugated. & I miss the good old days when Union soldiers burned slave owners' cities to the ground but alas, society progresses from such violent ideas with or without us. \bigstrut\\
\cline{2-3}          & Im going to choke you with your cross necklace idiotic religious pig & Well, I'm pretty tall and I don't think you'd be able to reach my neck. Plus I'm a human being, not a pig, so maybe try threatening an animal at a local farm.  \bigstrut\\
    \hline
    \multirow{3}[6]{*}[-6em]{\rotatebox{90}{Warning of Consequence}} & Black people are just a bunch of animals that shouldn't exist in our world. & If I were you, I would think about what you just said and delete this. Imagine your family, friends, or employer sees this. That won’t end well. \bigstrut\\
\cline{2-3}          & Women are psychopath, and an adulterous woman who has been unfaithful to you for many years is like an arsonist in a building that you built together. She will burn down that place, exactly because she knows you love it. Life will never be the same. & Hate speech like that will get you ostracized by most groups you interact with; you're both wrong and unpleasant to deal with.  \bigstrut\\
\cline{2-3}          & Fags, lesbians, trans are so few we better ignore them rather than waste time for their rights. & You will lose your job and permanently damage your face card for talking like this about LGBTQ+ nowadays. Sure, they are in the minority but they are extremely powerful due to the quality of people who endorse them. They don’t quit and will cancel everyone who is vocal against them. Live and let live.  \bigstrut\\
    \hline
    \multirow{3}[6]{*}[-12em]{\rotatebox{90}{Refutal}} & People with Down Syndrome should be given special and simple jobs, that they can do by themselves. At least they will not hamper normal people's productivity. & We are all special and unique in different ways, and people with Down Syndrome are the same.  Each of us has talents and abilities that others do not share.  Perhaps you are not good at something that someone else is very good at.  In that respect, you would actually be hampering that person's productivity.  Take time to reflect on your own shortcomings and how those might negatively impact others before you make such ludicrous, overly broad statements. \bigstrut\\
\cline{2-3}          & Homosexuals are just sick. They should be isolated and isolated soon. & There is no evidence that proves homosexuality is an illness. So maybe you should consider that you are the sick one and isolate yourself from the rest of the world.  \bigstrut\\
\cline{2-3}          & Whites are the real victims of police brutality: those negroes are sneaky criminals and police is doing nothing. & Yikes! What a bad take. It’s been proven time and time again that police brutality often greatly impacts people of color. White people have the privilege to get away with things a black person would get shot for.  \bigstrut\\
    \hline
    \end{longtable}%

\subsection{Variables Used in Linear Mixed Models} \label{sec:variables_lmm}

\begin{table}[htbp]
  \centering
  \caption{Description of Variables Used in Linear Mixed Model for RQ1}
    \small
    \begin{tabularx}{\textwidth}{|l|rrr}
    \hline
    \rowcolor[rgb]{ .933,  .925,  .882} \textbf{Control Variables} & \multicolumn{1}{l|}{\textbf{Independent Variables}} & \multicolumn{1}{l|}{\textbf{Random Effects}} & \multicolumn{1}{l|}{\textbf{Dependent Variable}} \bigstrut\\
    \hline
    Frequency of encountering online hate speech & \multicolumn{1}{l|}{Hate post topic} & \multicolumn{1}{l|}{userID} & \multicolumn{1}{l|}{Perceived hatefulness rating} \bigstrut\\
    \hline
    Use of real name on social media & \multicolumn{1}{l|}{Topic-Identity Match (TIM)*} & \multicolumn{1}{l|}{hatepostID} &  \bigstrut\\
\cline{1-3}    Social media commenting frequency &       &       &  \bigstrut\\
\cline{1-1}    Age   &       &       &  \bigstrut\\
\cline{1-1}    Education level &       &       &  \bigstrut\\
\cline{1-1}    Political view &       &       &  \bigstrut\\
\cline{1-1}    \end{tabularx}%
\begin{tablenotes}
      \item * Topic-Identity Match (TIM) on gender, race, sexual orientation, religion, and disability status.
    \end{tablenotes}
  \label{tab:varlist1}%
\end{table}%

\begin{table}[htbp]
  \centering
  \caption{Description of Variables Used in Linear Mixed Model for RQ2}
    \small
    \begin{tabular}{|l|rrr}
    \hline
    \rowcolor[rgb]{ .933,  .925,  .882} \textbf{Control Variables} & \multicolumn{1}{l|}{\textbf{Independent Variables}} & \multicolumn{1}{l|}{\textbf{Random Effects}} & \multicolumn{1}{l|}{\textbf{Dependent Variable}} \bigstrut\\
    \hline
    Frequency of encountering online hate speech & \multicolumn{1}{l|}{Hate post topic} & \multicolumn{1}{l|}{userID} & \multicolumn{1}{l|}{Satisfaction} \bigstrut\\
    \hline
    Use of real name on social media & \multicolumn{1}{l|}{Topic-Identity Match (TIM)*} & \multicolumn{1}{r|}{} & \multicolumn{1}{l|}{Effectiveness} \bigstrut\\
\cline{1-2}\cline{4-4}    Social media commenting frequency &       & \multicolumn{1}{r|}{} & \multicolumn{1}{l|}{Difficulty} \bigstrut\\
\cline{1-1}\cline{4-4}    Age   &       &       &  \bigstrut\\
\cline{1-1}    Education level &       &       &  \bigstrut\\
\cline{1-1}    Political view &       &       &  \bigstrut\\
\cline{1-1}    Perceived hatefulness rating &       &       &  \bigstrut\\
\cline{1-1}    \end{tabular}%
\begin{tablenotes}
      \item * Topic-Identity Match (TIM) on gender, race, sexual orientation, religion, and disability status.
    \end{tablenotes}
  \label{tab:varlist2}%
\end{table}%

\begin{table}[htbp]
  \centering
  \caption{Description of Variables Used in Linear Mixed Model for RQ3}
    \small
    \begin{tabular}{|l|rrr}
    \hline
    \rowcolor[rgb]{ .933,  .925,  .882} \textbf{Control Variables} & \multicolumn{1}{l|}{\textbf{Independent Variables}} & \multicolumn{1}{l|}{\textbf{Random Effects}} & \multicolumn{1}{l|}{\textbf{Dependent Variable}} \bigstrut\\
    \hline
    Frequency of encountering online hate speech & \multicolumn{1}{l|}{Strategy} & \multicolumn{1}{l|}{userID} & \multicolumn{1}{l|}{Satisfaction} \bigstrut\\
    \hline
    Use of real name on social media & \multicolumn{1}{l|}{Use of first-person language} & \multicolumn{1}{l|}{\multirow{3}[6]{*}{\parbox{7em}{hatepostID (only for \yellowhl{Perceived hatefulness rating})}}} & \multicolumn{1}{l|}{Effectiveness} \bigstrut\\
\cline{1-2}\cline{4-4}    Social media commenting frequency & \multicolumn{1}{l|}{Use of questions} & \multicolumn{1}{l|}{} & \multicolumn{1}{l|}{Difficulty} \bigstrut\\
\cline{1-2}\cline{4-4}    Age   & \multicolumn{1}{l|}{Length} & \multicolumn{1}{l|}{} & \multicolumn{1}{l|}{Perceived hatefulness rating} \bigstrut\\
    \hline
    Education level & \multicolumn{1}{l|}{Sentiment polarity} &       &  \bigstrut\\
\cline{1-2}    Political view &       &       &  \bigstrut\\
\cline{1-1}    Hate post topic &       &       &  \bigstrut\\
\cline{1-1}    Topic-Identity Match (TIM) &       &       &  \bigstrut\\
\cline{1-1}    \end{tabular}%
  \label{tab:varlist3}%
\end{table}%

\subsection{Linear Mixed Model Equations} \label{sec:lmm_equations}

\begin{equation} \label{eq:rq1}
  \begin{aligned}
    \operatorname{Perceived\ hatefulness\ rating}_{i}  &\sim N \left(\alpha_{j[i],k[i]}, \sigma^2 \right) \\
      \alpha_{j}  &\sim N \left(\gamma_{0}^{\alpha} + \gamma_{1}^{\alpha}(\operatorname{Hate\ post\ topic}) + (\operatorname{TIM}), \sigma^2_{\alpha_{j}} \right)
      \text{, for hatepostID j = 1,} \dots \text{,J} \\
    \alpha_{k}  &\sim N \biggl(\gamma_{0}^{\alpha} + \gamma_{1}^{\alpha}(\operatorname{Frequency\ of\ encountering\ online\ hate\ speech}) \\
      &\quad + \gamma_{2}^{\alpha}(\operatorname{Use\ of\ real\ name\ on\ social\ media}) \\
      &\quad + \gamma_{3}^{\alpha}(\operatorname{Social\ media\ commenting\ frequency}) \\
      &\quad + \gamma_{4}^{\alpha}(\operatorname{Age}) + \gamma_{5}^{\alpha}(\operatorname{Education\ level}) \\
      &\quad + \gamma_{6}^{\alpha}(\operatorname{Political\ view}), \sigma^2_{\alpha_{k}} \biggr)
      \text{, for userID k = 1,} \dots \text{,K}
  \end{aligned}
\end{equation}    

\begin{equation} \label{eq:rq2}
  \begin{aligned}
    \operatorname{Satisfaction}_{i}  &\sim N \left(\mu_{1}, \sigma_{1}^2 \right) \\
    \operatorname{Effectiveness}_{i}  &\sim N \left(\mu_{2}, \sigma_{2}^2 \right) \\
    \operatorname{Difficulty}_{i}  &\sim N \left(\mu_{3}, \sigma_{3}^2 \right) \\
      \mu &=\alpha_{j[i]} + \beta_{1}(\operatorname{Hate\ post\ topic}) + \beta_{2}(\operatorname{TIM}) + \beta_{3}(\operatorname{Perceived\ hatefulness\ rating}) \\
      \alpha_{j}  &\sim N \biggl(\gamma_{0}^{\alpha} + \gamma_{1}^{\alpha}(\operatorname{Frequency\ of\ encountering\ online\ hate\ speech}) \\
      &\quad + \gamma_{2}^{\alpha}(\operatorname{Use\ of\ real\ name\ on\ social\ media}) \\
      &\quad + \gamma_{3}^{\alpha}(\operatorname{Social\ media\ commenting\ frequency}) \\
      &\quad + \gamma_{4}^{\alpha}(\operatorname{Age}) + \gamma_{5}^{\alpha}(\operatorname{Education\ level}) \\
      &\quad + \gamma_{6}^{\alpha}(\operatorname{Political\ view}), \sigma^2_{\alpha_{j}} \biggr)
      \text{, for UserID j = 1,} \dots \text{,J}
  \end{aligned}
\end{equation}

\begin{equation} \label{eq:rq3}
  \begin{aligned}
    \operatorname{Perceived\ hatefulness\ rating}_{i}  &\sim N \left(\alpha_{j[i],k[i]}, \sigma^2 \right) \\
    \alpha_{j}  &\sim N \biggl(\gamma_{0}^{\alpha} + \gamma_{1}^{\alpha}(\operatorname{Strategy}) + \gamma_{2}^{\alpha}(\operatorname{Use\ of\ first-person\ language}) \\
      &\quad + \gamma_{3}^{\alpha}(\operatorname{Use\ of\ questions})  + \gamma_{4}^{\alpha}(\operatorname{Length}) + \gamma_{5}^{\alpha}(\operatorname{Sentiment\ polarity})  \\
      &\quad + \gamma_{6}^{\alpha}(\operatorname{Hate\ post\ topic}) + \gamma_{7}^{\alpha}(\operatorname{TIM}), \sigma^2_{\alpha_{j}} \biggr)
      \text{, for hatepostID j = 1,} \dots \text{,J} \\
    \operatorname{Satisfaction}_{i}  &\sim N \left(\mu_{1}, \sigma_{1}^2 \right) \\
    \operatorname{Effectiveness}_{i}  &\sim N \left(\mu_{2}, \sigma_{2}^2 \right) \\
    \operatorname{Difficulty}_{i}  &\sim N \left(\mu_{3}, \sigma_{3}^2 \right) \\
    \mu &=\alpha_{k[i]} + \biggl(\gamma_{0}^{\alpha} + \gamma_{1}^{\alpha}(\operatorname{Strategy}) + \gamma_{2}^{\alpha}(\operatorname{Use\ of\ first-person\ language}) \\
      &\quad + \gamma_{3}^{\alpha}(\operatorname{Use\ of\ questions}) + \gamma_{4}^{\alpha}(\operatorname{Length}) + \gamma_{5}^{\alpha}(\operatorname{Sentiment\ polarity}) \\
      &\quad + \gamma_{6}^{\alpha}(\operatorname{Hate\ post\ topic}) + \gamma_{7}^{\alpha}(\operatorname{TIM}), \sigma^2_{\alpha_{j}} \biggr) \\
      \alpha_{k}  &\sim N \biggl(\gamma_{0}^{\alpha} + \gamma_{1}^{\alpha}(\operatorname{Frequency\ of\ encountering\ online\ hate\ speech}) \\
      &\quad + \gamma_{2}^{\alpha}(\operatorname{Use\ of\ real\ name\ on\ social\ media}) \\
      &\quad + \gamma_{3}^{\alpha}(\operatorname{Social\ media\ commenting\ frequency}) \\
      &\quad + \gamma_{4}^{\alpha}(\operatorname{Age}) + \gamma_{5}^{\alpha}(\operatorname{Education\ level}) \\
      &\quad + \gamma_{6}^{\alpha}(\operatorname{Political\ view}), \sigma^2_{\alpha_{k}} \biggr)
      \text{, for userID k = 1,} \dots \text{,K}
  \end{aligned}
\end{equation}